\documentclass{article} 
\usepackage[final]{colm2025_conference}

\PassOptionsToPackage{dvipsnames}{xcolor}
\usepackage[T1]{fontenc}
\usepackage{capt-of}
\usepackage{microtype}
\usepackage{hyperref}
\usepackage{url}
\usepackage{booktabs}
\usepackage{graphicx}
\usepackage{amsmath}
\usepackage{lineno}
\usepackage{tikz}
\usepackage{tcolorbox}
\usepackage{xcolor} 
\usepackage{enumitem}
\usepackage{caption}
\definecolor{chart}{HTML}{1f77b4}

\newtcolorbox{example}[1][]{
  colback=chart!5!white,
  colframe=chart,
  floatplacement=floating,
  title=\centering \textsf{#1}
}
\usepackage{tabularx, booktabs}
\usepackage{listings}
\lstdefinestyle{mystyle}{
  basicstyle=\ttfamily\footnotesize,
  numbers=left,
  numberstyle=\tiny,
  frame=lines,
  breaklines=true,
  keywordstyle=\color{blue},
  commentstyle=\itshape\color{teal!60!black},
  stringstyle=\color{purple!70!black}
}

\definecolor{darkblue}{rgb}{0, 0, 0.5}
\hypersetup{colorlinks=true, citecolor=darkblue, linkcolor=darkblue, urlcolor=darkblue}

\tcbuselibrary{listings}
\tcbuselibrary{breakable}

\lstdefinestyle{extraCodeStyle}{
  basicstyle=\ttfamily\scriptsize,
  breaklines=true,
  keywordstyle=\color{blue},
  commentstyle=\itshape\color{teal!60!black},
  stringstyle=\color{purple!70!black}
}

\newtcblisting{extraCodeExample}[1][]{
  colback=chart!5!white,
  colframe=chart,
  listing only,
  listing options={style=extraCodeStyle, frame=none},
  title=\centering \textsf{#1},
  breakable
}

\definecolor{elorankingHighlightBarColor}{HTML}{74cc86}

\title{CodeClash: Benchmarking Goal-Oriented\\Software Engineering}

\newcommand{\stanford}{\textcolor{red}{\textsuperscript{1}}}
\newcommand{\princeton}{\textcolor{orange}
{\textsuperscript{2}}}
\newcommand{\cornell}{\textcolor{blue}{\textsuperscript{3}}}
\newcommand{\tum}{\textcolor{teal}{\textsuperscript{4}}}

\author{
John Yang*\stanford,
Kilian Lieret*\princeton,
Joyce Yang\cornell,
Carlos E. Jimenez\princeton, \\[0.25em]
\textbf{
Muhtasham Oblokulov\tum,
Aryan Siddiqui\stanford,
Ofir Press\princeton,
Ludwig Schmidt\stanford,
Diyi Yang\stanford
} \\[0.75em]
\stanford Stanford University\quad
\princeton Princeton University\quad
\cornell Cornell University\quad
\tum Technical University of Munich
}

\newcommand{\clash}{CodeClash}

\newif\ifshowcomments
\showcommentsfalse   

\begin{document}

\maketitle

\begin{abstract}
Current benchmarks for coding evaluate language models (LMs) on concrete, well-specified tasks such as fixing specific bugs or writing targeted tests.
However, human programmers do not spend all day addressing isolated tasks.
Instead, real-world software development is grounded in the pursuit of high-level goals, like improving user retention or reducing costs.
Evaluating whether LMs can also iteratively develop code to better accomplish open-ended objectives without any explicit guidance remains an open challenge.
To address this, we introduce \clash{}, a benchmark where LMs compete in multi-round tournaments to build the best codebase for achieving a competitive objective.
Each round proceeds in two phases: agents edit their code, then their codebases compete head-to-head in a code arena that determines winners based on objectives like score maximization, resource acquisition, or survival.
Whether it's writing notes, scrutinizing documentation, analyzing competition logs, or creating test suites, models must decide for themselves how to improve their codebases both absolutely and against their opponents.
We run $1680$ tournaments ($25$,$200$ rounds total) to evaluate $8$ LMs across $6$ arenas.
Our results reveal that while models exhibit diverse development styles, they share fundamental limitations in strategic reasoning.
Models also struggle with long-term codebase maintenance, as repositories become progressively messy and redundant.
These limitations are stark: top models lose every round against expert human programmers.
We open-source \clash{} to advance the study of autonomous, goal-oriented code development.

\let\thefootnote\relax\footnote{
$^*$Equal contribution. Correspondence to \texttt{\href{mailto:johnby@stanford.edu}{johnby@stanford.edu},\href{mailto:kl5675@princeton.edu}{kl5675@princeton.edu}}.
}
\end{abstract}
\section{Introduction}
Existing coding benchmarks challenge language models (LMs) to complete small, focused tasks, such as implementing an algorithm~\citep{jain2024livecodebenchholisticcontaminationfree}, fixing a specific bug in a single function~\citep{jimenez2024swebenchlanguagemodelsresolve}, or writing a test for a target class~\citep{mundler2024swtbench}.
Problem statements are straightforward and fine-grained in their description of a task. 
Given explicit instructions, models are evaluated on their ability to execute them correctly. 

On the contrary, real world software development demands a much broader scope of agency.
Instead of maintenance tasks, developers are driven by high-level goals like improving user retention, increasing revenue, or reducing costs. 
This requires fundamentally different capabilities; engineers must recursively decompose these objectives into actionable steps, prioritize them, and make strategic decisions about which solutions to pursue.
The process is a continuous loop -- propose changes, deploy them, analyze real-world feedback (e.g., metrics, user behavior, A/B test results), and repeat to inform the next move.
Evaluating how models fare under such conditions remains an unaddressed challenge in benchmarking.

Therefore, we introduce \clash{}, a benchmark for goal-oriented software engineering.
Specifically, multiple LM systems compete to build the best codebase for achieving a high-level objective over the course of a multi-round tournament.
These codebases implement solutions that compete in a code arena, such as BattleSnake (grid-based survival), Poker (no-limit Texas Hold'em), and RoboCode (tank combat).
Crucially, LMs do not play the games directly, unlike existing game-based benchmarks~\citep{silver2016mastering,openai2019dota2largescale,zhang2025videogamebenchvisionlanguagemodelscomplete}.
Instead, they iteratively refine code that competes as their proxy.

\begin{figure}[t]
    \centering
    \includegraphics[width=\textwidth]{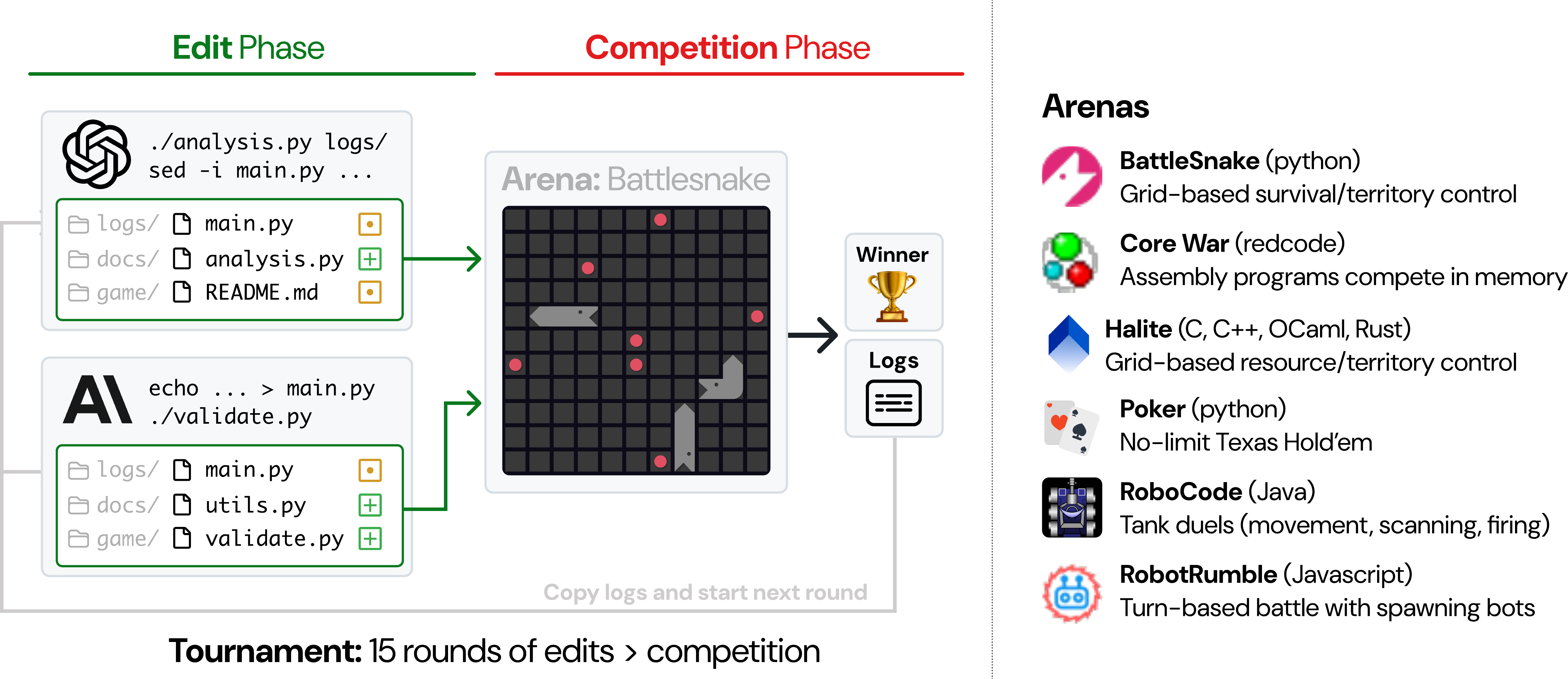}
    \caption{
\clash{} is a benchmark where players (LMs as SWE-agents) compete in programming tournaments spanning multiple rounds.
Per round, models edit their codebases (\textit{edit} phase) before the codebases face off in a code arena (\textit{competition} phase). Then, the competition logs are copied back into the codebases and the next round begins.
    }
    \label{fig:preview}
\end{figure}

As shown in Figure~\ref{fig:preview}, each round proceeds in two phases: agents edit their code, then their codebases compete head-to-head in a code arena.
The code arena then executes multiple implementations against one another and determines winners based on objectives like score maximization, resource acquisition, or survival.

Success in \clash{} requires models to determine their own improvement strategies.
From the outset, LM agents receive only a brief description of the setting.
While information like arena mechanics, example bots, and recommended strategies are available in the starter codebase, models must take initiative to proactively discover them.
Each round, LMs receive gigabytes of logs from past rounds, which they can parse to extract insights about outcomes and opponents -- or ignore entirely.
Across the span of a tournament, \clash{} reveals whether and how models populate their codebases with notes, tests, and analyses.

We evaluate $8$ frontier LMs across $6$ arenas.
We find \clash{} elicits substantial creativity from models; across $1680$ tournaments, we observe that a model's solutions become increasingly dissimilar round over round, even when facing the same opponent in the same arena.
However, our results reveal that while models exhibit diverse development styles, they share common limitations in interpreting competitive feedback, validating changes, and maintaining organized codebases over time.
Even top models hallucinate reasons for failure or modify code without confirming if these changes meaningfully improve performance.
A substantial gap remains between model and human performance; the best model (\texttt{Claude Sonnet 4.5}) \textit{fails to win a single round} against an expert human-written bot.

We release \clash{} as an open source toolkit, including the code, arena logs, and a leaderboard, to further the study of self-evolving, LM-based SWE-agents.

\section{CodeClash}
\label{sec:codeclash}

\subsection{Formulation}
\label{sec:codeclash:formulation}

\clash{} formalizes competitive coding as a tournament, where two or more players compete in a code arena for multiple rounds.
\textit{Player} refers to an LM equipped with an Agent Computer Interface (ACI) or scaffold that enables it to interact with a codebase~\citep{yang2024sweagentagentcomputerinterfacesenable}.
Each player maintains their own codebase for the entire tournament.
A \textit{code arena} is any competition platform that takes in multiple codebases and executes them against one another, producing measurable outcomes about relative performance on a designated objective (e.g., eliminating opponents, acquiring resources, maximizing profit).

Each round proceeds in two phases.
In the \textit{edit} phase, each player independently modifies their codebase using whatever strategies they deem appropriate within a fixed budget of turns.
During the \textit{competition} phase, all codebases are compiled and executed within the code arena, where they interact and compete directly against each other.
The arena determines a winner (or declares a tie) based on the codebases' performance.

\clash{}'s formulation makes several key design decisions.
\textit{Codebase-as-memory}: players have no explicit memory of actions from previous rounds.
Their information is limited to whatever they chose to record in the codebase.
\textit{Log-based feedback}: after each competition phase, the results and logs are copied into each player's codebase as the sole source of new information.
\textit{Strategic opacity}: players cannot see each other's codebases, though we explore lifting this restriction in Section~\ref{sec:results:ablations}.

\subsection{Technical Details}
\label{sec:codeclash:technical_details}

To implement a player, we use \texttt{mini-SWE-agent}, an agent computer interface (ACI) that enables an LM to interact with a codebase by issuing \texttt{bash} actions to a terminal~\citep{yang2024sweagentagentcomputerinterfacesenable}.
Each turn, the LM generates a ReAct~\citep{yao2023reactsynergizingreasoningacting} style response containing a thought (in natural language) and a \texttt{bash} action, then receives standard output from the terminal environment in return.

We also define a lightweight, flexible interface for a code arena.
An implementation only needs to define commands to run the competition and determine a winner.
This minimal overhead enables us to fold many existing competitive programming games and tasks into \clash{}.
More technical discussion in \S\ref{appx:infra}.

\subsection{Features}
\label{sec:codeclash:features}

\clash{}'s initial release features a suite of $6$ code arenas, as listed in Figure~\ref{fig:preview}.
Each arena is covered thoroughly in \S\ref{appx:arenas}.
\clash{} introduces several distinctive properties that collectively push models beyond traditional code completion and issue resolution.

\textbf{Open-ended objectives.} \clash{} departs from the traditional reliance on unit tests or implementation correctness to measure success.
Instead, players code to win competitive outcomes that vary dramatically across arenas, from maximizing profit to surviving the longest.
This mirrors the ultimate objectives of real-world software more faithfully, where code is written to achieve tangible, practical outcomes (e.g., maximize resources, generate revenue, outperform competitors) rather than simply achieving technical correctness.
A consequence of rich objectives is that models must then decompose a higher-order goal into actionable subtasks and measurable, intermediate metrics to inform code improvements.

\textbf{Diverse arenas.}
\clash{}'s arenas vary significantly, with drastic differences in a codebase's structure, how a codebase interfaces with the arena engine, and the types of logs and feedback generated.
This contrasts sharply with existing benchmarks, where evaluation follows a consistent pattern of problem statement, code implementation, and test validation.

\textbf{Adversarial adaptation.} \clash{}'s uniquely multi-player, head-to-head setting adds a new layer of complexity to coding evaluations.
While decent LMs may be capable of writing competent implementations, top-performing players will analyze opponent behaviors and incorporate countermeasures, all the while being indecipherable in their own play.
Early round wins do not ensure continued dominance.
At some point, the challenge shifts from writing good code to writing code that consistently beats intelligent competition.

\textbf{Self-crafted memory.} As mentioned in Section~\ref{sec:codeclash:formulation}, \clash{} does not maintain persistent memory for models across rounds; only ephemeral, within-round memory exists.
To retain information for future use, models must explicitly add insights to the codebase; how to represent such knowledge is left entirely to the model's discretion.

\textbf{Self-directed improvement.} Beyond a brief description of the environment and arena, the initial system prompt provided to each player at the start of every edit phase contains \textit{no} guidance beyond high level suggestions about how to enhance its codebase.
All decisions and changes LMs make are necessarily autonomous.
In practice, this may manifest as models writing analysis scripts to understand competition logs, maintaining notes about past rounds or opponents, or generating multiple candidates to test against one another.

\section{Experiments}
\label{sec:experiments}

\textbf{Models.} We select $8$ strong LMs to evaluate, where strength is roughly estimated as performance on existing coding benchmarks.
Our final list includes two models from the Anthropic family (Claude Sonnet $4.5$~\citep{modelcardclaude45sonnet}, $4$~\citep{modelcardclaude4sonnet}), three models from the OpenAI family (\texttt{GPT 5}, \texttt{5-mini}~\citep{modelcardopenaigpt5}, \texttt{o3}~\citep{modelcardopenaio3}), \texttt{Gemini 2.5 Pro}~\citep{comanici2025gemini}, \texttt{Qwen3-Coder}~\citep{qwen3technicalreport}, and \texttt{Grok Code Fast 1}~\citep{grokcodefast12025}.

\textbf{Agent system.} As discussed in Section~\ref{sec:codeclash:technical_details}, we use \texttt{mini-SWE-agent}.
We intentionally decide against using tool-heavy scaffolds such as SWE-agent or OpenHands~\citep{wang2025openhandsopenplatformai}, as they are often optimized for models and benchmarks.
By restricting interactions to bash commands, \texttt{mini-SWE-agent} avoids imposing predefined assumptions via tools about how LMs should approach codebase modifications or competitive play~\citep{yang2024swebenchmultimodalaisystems}.
Per round, models are allotted a maximum of $30$ turns for the \textit{edit} phase, with automatic termination if exceeded.
Player configurations are discussed thoroughly in \S\ref{appx:evaluation:mini-config}.

\textbf{Number of rounds run.} For our main leaderboard, we make models compete one-on-one.
Given $8$ models and $6$ arenas, we run $10$ tournaments per model pair per arena, with each tournament lasting $15$ rounds.
This yields $\binom{8}{2} \times 6 \times 10 \times 15 = 25,200$ total rounds.
Tournament runtime varies by arena, taking $75$ minutes on average -- totaling $2.4$ million hours of runtime (mostly due to model latency), parallelized over the independent tournaments. 
Tournament configuration details are covered in \S\ref{appx:evaluation:tournament-config}.

\textbf{Win rates.} Performance per model is generally calculated as an aggregation across all tournaments (sets of 15 rounds) won across all arenas.
A single round is won by a model if it achieves a higher score in the arena than its opponent or if its opponent makes an invalid submission.
A tournament is won by the model that wins more rounds than its opponent, or, if both models win equally many rounds, by the model that scores the last win.\footnote{Draws are a possible outcome for each round, so both models might achieve an equal number of wins in a tournament. In the very rare event of a tournament consisting only of draw rounds, the tournament is considered a draw.}
The win rate of a model is the fraction of tournaments it has won.
For details, see \S\ref{appx:evaluation:metrics}.

\textbf{Elo metrics.} Inspired by the thread of prior work ranking LMs on the task of instruction following~\citep{elo1967proposed,bai2022traininghelpfulharmlessassistant,boubdir2024elo,chiang2024chatbotarenaopenplatform}, we use Elo scores with a base rating of $R=1200$ and a slope of $400$ to quantify the overall strength of each model.
Instead of calculating Elo scores using sequential updates (which require a choice of step size and depend on update order), we perform a more rigorous maximum likelihood fit to the win rates.
We validate rank stability and our statistical treatment with both parametric and non-parametric bootstrapping experiments and observe more than 98\% pairwise order agreement. 
For details, see \S\ref{appx:evaluation:metrics}.
\section{Results}
\label{sec:results}

\newcommand{\eloSingleArenaResult}[1]{
  \begin{tikzpicture}[baseline=(text.base)]
    \pgfmathsetmacro{\barwidth}{#1 * 0.0007}
    \fill[black!15] (0, -0.15) rectangle (\barwidth, 0.25);
    \node[anchor=west,text=black,font=\footnotesize] (text) at (-0.05, 0.05) {#1};
  \end{tikzpicture}%
}
\newcommand{\eloMainResult}[1]{
  \begin{tikzpicture}[baseline=(text.base)]
    \pgfmathsetmacro{\barwidth}{#1 * 0.001}
    \fill[elorankingHighlightBarColor!50] (0, -0.15) rectangle (\barwidth, 0.25);
    \node[anchor=west,font=\bfseries] (text) at (0, 0.05) {#1};
  \end{tikzpicture}%
}

\begin{table}[t]
\centering
{
\setlength{\tabcolsep}{3pt}
\renewcommand{\arraystretch}{0.9}
\begin{tabular}{l|llllll|l}
\toprule
 & \scriptsize{BattleSnake} & \scriptsize{CoreWar} & \scriptsize{Halite} & \scriptsize{Poker} & \scriptsize{RoboCode} & \scriptsize{RobotRumble} & \multicolumn{1}{c}{\textbf{Overall}} \\
\midrule
Claude Sonnet 4.5 & \eloSingleArenaResult{1470} & \eloSingleArenaResult{1641} & \eloSingleArenaResult{1408} & \eloSingleArenaResult{1248} & \eloSingleArenaResult{1361} & \eloSingleArenaResult{1423} & \eloMainResult{1389} \\
GPT-5 & \eloSingleArenaResult{1339} & \eloSingleArenaResult{1199} & \eloSingleArenaResult{1522} & \eloSingleArenaResult{1599} & \eloSingleArenaResult{1409} & \eloSingleArenaResult{1293} & \eloMainResult{1360} \\
o3 & \eloSingleArenaResult{1357} & \eloSingleArenaResult{1348} & \eloSingleArenaResult{1576} & \eloSingleArenaResult{1277} & \eloSingleArenaResult{1338} & \eloSingleArenaResult{1309} & \eloMainResult{1343} \\
Claude Sonnet 4 & \eloSingleArenaResult{1253} & \eloSingleArenaResult{1339} & \eloSingleArenaResult{1111} & \eloSingleArenaResult{1233} & \eloSingleArenaResult{1033} & \eloSingleArenaResult{1361} & \eloMainResult{1223} \\
GPT-5 Mini & \eloSingleArenaResult{1369} & \eloSingleArenaResult{926} & \eloSingleArenaResult{1185} & \eloSingleArenaResult{1429} & \eloSingleArenaResult{1217} & \eloSingleArenaResult{1092} & \eloMainResult{1200} \\
Gemini 2.5 Pro & \eloSingleArenaResult{1115} & \eloSingleArenaResult{1043} & \eloSingleArenaResult{1186} & \eloSingleArenaResult{978} & \eloSingleArenaResult{1315} & \eloSingleArenaResult{1044} & \eloMainResult{1125} \\
Grok Code Fast & \eloSingleArenaResult{833} & \eloSingleArenaResult{1170} & \eloSingleArenaResult{824} & \eloSingleArenaResult{886} & \eloSingleArenaResult{1033} & \eloSingleArenaResult{1016} & \eloMainResult{1004} \\
Qwen3 Coder & \eloSingleArenaResult{860} & \eloSingleArenaResult{929} & \eloSingleArenaResult{784} & \eloSingleArenaResult{945} & \eloSingleArenaResult{890} & \eloSingleArenaResult{1057} & \eloMainResult{952} \\
\bottomrule
\end{tabular}
}
\caption{
Elo ratings per model per arena.
}
\label{tab:main_results}
\end{table}

\begin{figure}[t]
\centering
\raisebox{-2mm}{%
\begin{minipage}[t]{0.50\textwidth}
    \setlength{\abovecaptionskip}{-1ex}
    \centering
    \includegraphics[width=\textwidth]{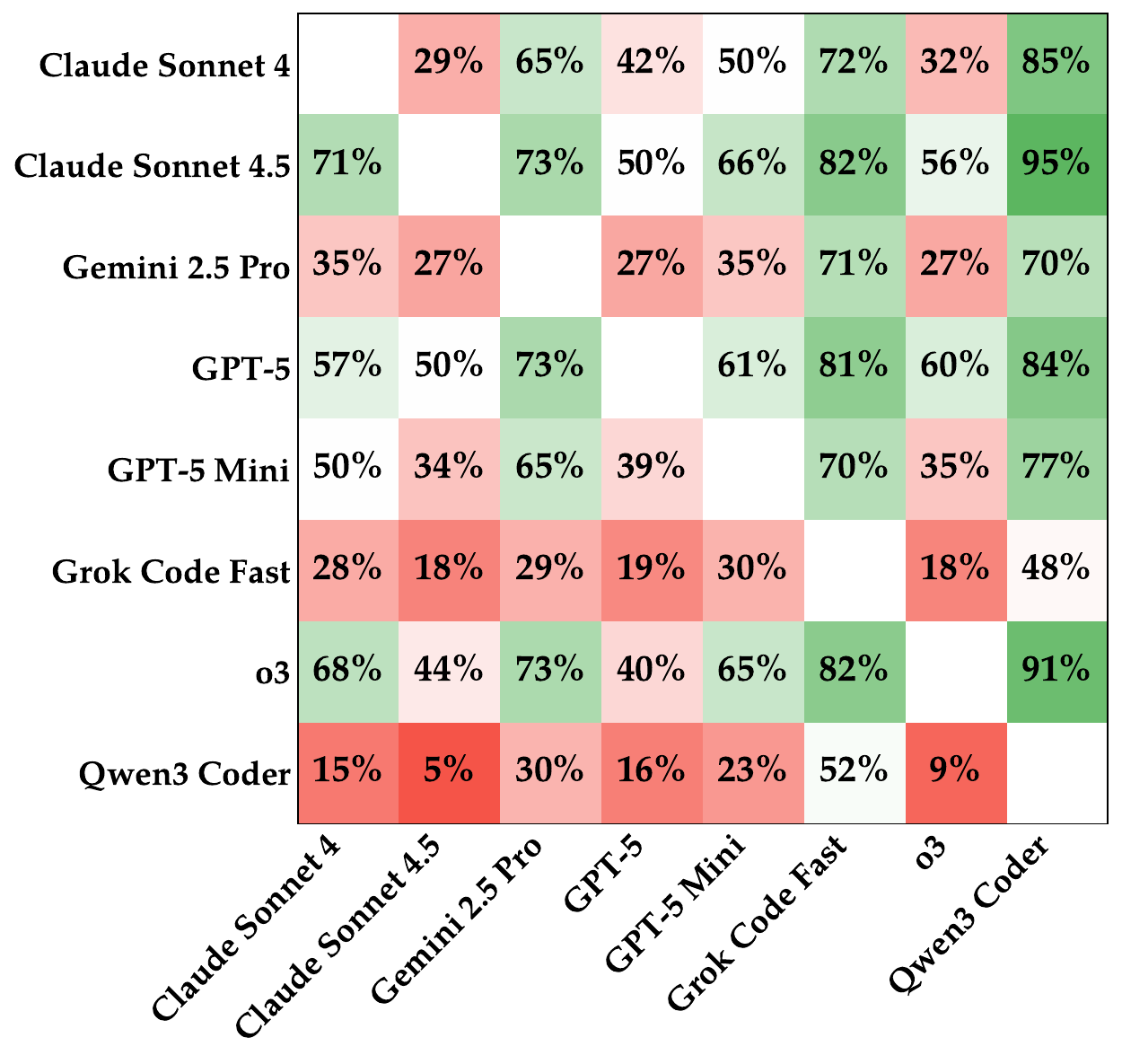}
    \caption{
Model win rates (row beats column).
Win rate is the proportion of tournaments (out of $240$) won across all arenas.
\texttt{Claude Sonnet 4.5} has the highest average win rate at $69.9$\%.
    }
    \label{fig:heatmap_win_rate}
\end{minipage}
}
\hfill
\begin{minipage}[t]{0.47\textwidth}
    \setlength{\abovecaptionskip}{0em}
    \centering
    \includegraphics[width=\textwidth]{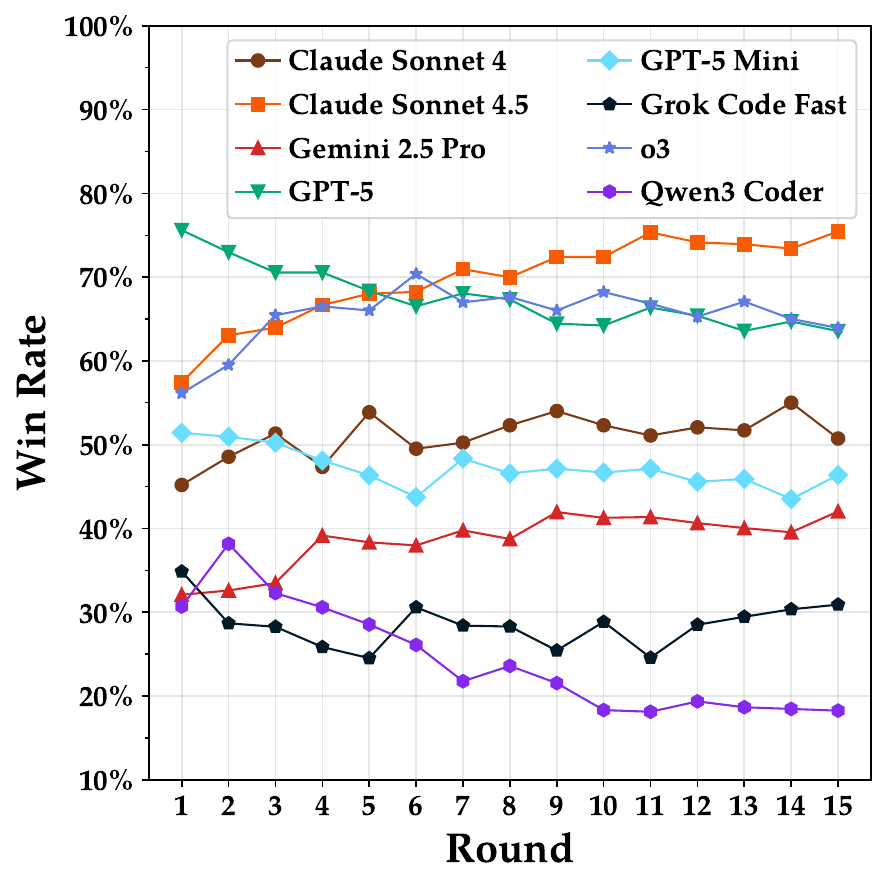}
    \caption{Win rates across rounds, illustrating how different models gain (\texttt{Claude Sonnet 4.5}) or lose momentum (\texttt{GPT-5}) over the course of the tournament.}
    \label{fig:line_chart_per_round_win_rate}
\end{minipage}
\end{figure}

We present our main results in Table~\ref{tab:main_results}. 
\texttt{Claude Sonnet 4.5} stands at the top, followed closely by \texttt{o3} and \texttt{GPT-5}.
After a gap of $100$ Elo, the next best models are \texttt{Claude Sonnet 4} and \texttt{GPT-5 mini}.
Notably, no single model dominates across all arenas.
Top ranked \texttt{Claude Sonnet 4.5} places just $4$th in Poker, emphasizing the importance of \clash{}'s support for multiple arenas.
Figure~\ref{fig:heatmap_win_rate} shows win rates of specific matchups. 
Figure~\ref{fig:line_chart_per_round_win_rate} reveals distinct performance trends across rounds -- some models excel early before plateauing, while others improve steadily over time.

\subsection{Ablations}
\label{sec:results:ablations}

\textbf{Models trail substantially behind expert human programmers.}
To quantify the gap between models and human experts beyond a single arena and opponent, we introduce CodeClash Ladder (CC:Ladder), a progression-style evaluation where a model climbs a ranked ladder of human-authored solutions, advancing only when it defeats the current opponent.
We construct ladders for RobotRumble ($58$ human solutions) and Core War ($264$ human solutions), with solutions ranked by Elo from all-pairs matchups.
Each model plays $7$ rounds against each opponent and advances if it wins a majority of rounds and wins the final round; the codebase carries over between opponents.
CC:Ladder retains the core properties of the standard \clash{} evaluation (multi-round iterative development, codebase-as-memory, and log-based feedback) while replacing the model opponent with a static human solution, eliminating opponent variance and substantially reducing cost.
Full details on the protocol and human solution rankings are in Appendix~\ref{appx:results:cc_ladder}.

\begin{table}[t]
\centering
\setlength{\tabcolsep}{4pt}
\begin{tabular}{lrr}
\toprule
Model & RR ($58$) & Core War ($264$) \\
\midrule
\texttt{Claude Sonnet 4.5}    & $43$ & $205$ \\
\texttt{GPT-5}                & $51$ & $201$ \\
\texttt{GPT-5 mini}           & $57$ & $260$ \\
\texttt{Gemini 2.5 Pro}       & $54$ & $233$ \\
\bottomrule
\multicolumn{3}{l}{\scriptsize RR = RobotRumble, CW = Core War}
\end{tabular}
\caption{CC:Ladder results. Models climb a ranked ladder of human-authored solutions, advancing only upon defeating the current opponent. Score = rank of the highest human opponent defeated (best of $5$ runs). Column headers show total number of human solutions per ladder. No model completes either ladder.}
\label{tab:cc_ladder_results_main}
\end{table}

Table~\ref{tab:cc_ladder_results_main} presents CC:Ladder scores for four models from the main leaderboard (Table~\ref{tab:main_results}); results for additional models are in Appendix~\ref{appx:results:cc_ladder}.
No model completes either ladder, and a substantial gap remains between models and human experts.
On RobotRumble, the best result is rank $57$ out of $58$ (\texttt{GPT-5 mini}); on Core War, rank $260$ out of $264$ (\texttt{GPT-5 mini}).
Most other models plateau considerably earlier.
The strategic reasoning limitations identified in Section~\ref{sec:analysis:reasoning-limits}, particularly ungrounded conclusions from logs and untested deployments, persist as the dominant causes of ladder termination.
Most runs terminate because the model fails to win the final round against an opponent, rather than losing a majority of rounds, suggesting models are often competitive but unable to consolidate gains within the round window.

\textbf{Models have limited capacity for opponent analysis even with transparent codebases.}
For each pairwise matchup among \texttt{Claude 4.5 Sonnet}, \texttt{GPT-5}, and \texttt{Gemini 2.5 Pro}, we run $10$ Core War tournaments of $15$ rounds each, with one modification -- before the \textit{edit} phase of round \texttt{n}, each player receives a read-only copy of their opponent's code from round \texttt{n-1}.
While the relative standings remain consistent with the default setting, the win rates change with \texttt{GPT-5} securing $74.6$\% ($+7.8$\%) of rounds, \texttt{Claude 4.5 Sonnet} at $53.2$\% ($-1.8$\%), and \texttt{Gemini 2.5 Pro} at $22.7$\% ($-5.5$\%).
Curiously, \texttt{GPT-5} only accesses its opponent's codebase in $12.8$\% of all rounds, far fewer than \texttt{Claude 4.5 Sonnet} ($99.3$\%) and \texttt{Gemini 2.5 Pro} ($52.9$\%), suggesting that frequent inspection of opponent code does not necessarily translate to competitive advantage, as our analysis later in Section~\ref{sec:analysis:reasoning-limits} reaffirms.
Additional insights in \S\ref{appx:results:ablations}.
Subsequent studies could more thoroughly investigate and enhance models' capacity for detecting opponents' weaknesses and designing tailored counter-strategies.

\textbf{Multi-agent competitions ($3$+ players) reflect similar rankings.}
We run $20$ Core War tournaments, $15$ rounds each, with $6$ of $8$ models (excluding \texttt{GPT-5-mini}, \texttt{Claude 4 Sonnet}). 
To quantify performance, as shown in Table~\ref{tab:trueskill}, we use the TrueSkill rating system~\citep{herbrich2006trueskill} since Elo and win rate are limited to one-on-one settings.
The results are similar to Core War ranks in Table~\ref{tab:main_results}, with \texttt{GPT-5} and \texttt{Grok Code Fast} (two models of similar Elo ranking) switching positions.
However, the $6$ player tournaments exhibit far more competitive volatility.
Lead changes (round \texttt{n} winner different from round \texttt{n-1}) occur $48.4$\% of the time in $6$ player Core War, compared to just $18.2$\% in the two player setting.
Winners of $6$-player tournaments capture just $28.6$\% of total points on average versus $78.0$\% in $2$-player settings. 
We provide some additional insights in \S\ref{appx:results}.
We look forward to future work that can leverage \clash{}'s multi-player tournaments as a testbed for understanding strategic behaviors such as coalition dynamics, positional play, and risk management.

\section{Analysis}
\label{sec:analysis}

\subsection{Competitive Dynamics}
\label{sec:analysis:dynamics}
Beyond overall win rates, we analyze how models interact with their codebases along with the resilience of models after losing individual rounds.
We also investigate trends in models' solution diversity and codebase organization.

\begin{figure*}[t]
\centering
\begin{minipage}[t]{0.49\textwidth}
    \setlength{\abovecaptionskip}{0em}
    \centering
    \includegraphics[width=0.95\textwidth]{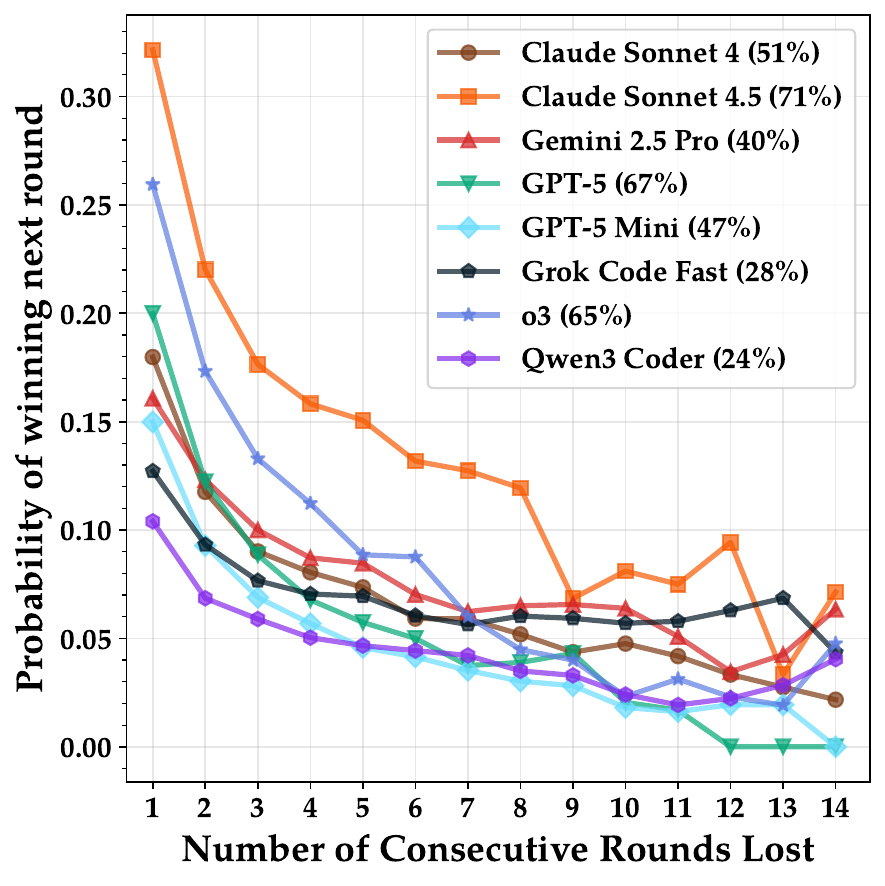}
    \caption{Probability of winning the next round after losing several rounds in a row. Even the highest ranking models struggle to recover after losing one or more consecutive rounds in a tournament. Numbers in parentheses indicate the overall average win rate.}
    \label{fig:line_chart_comeback_probability}
\end{minipage}
\hfill
\begin{minipage}[t]{0.49\textwidth}
    \setlength{\abovecaptionskip}{0em}
    \centering
    \includegraphics[width=0.95\textwidth]{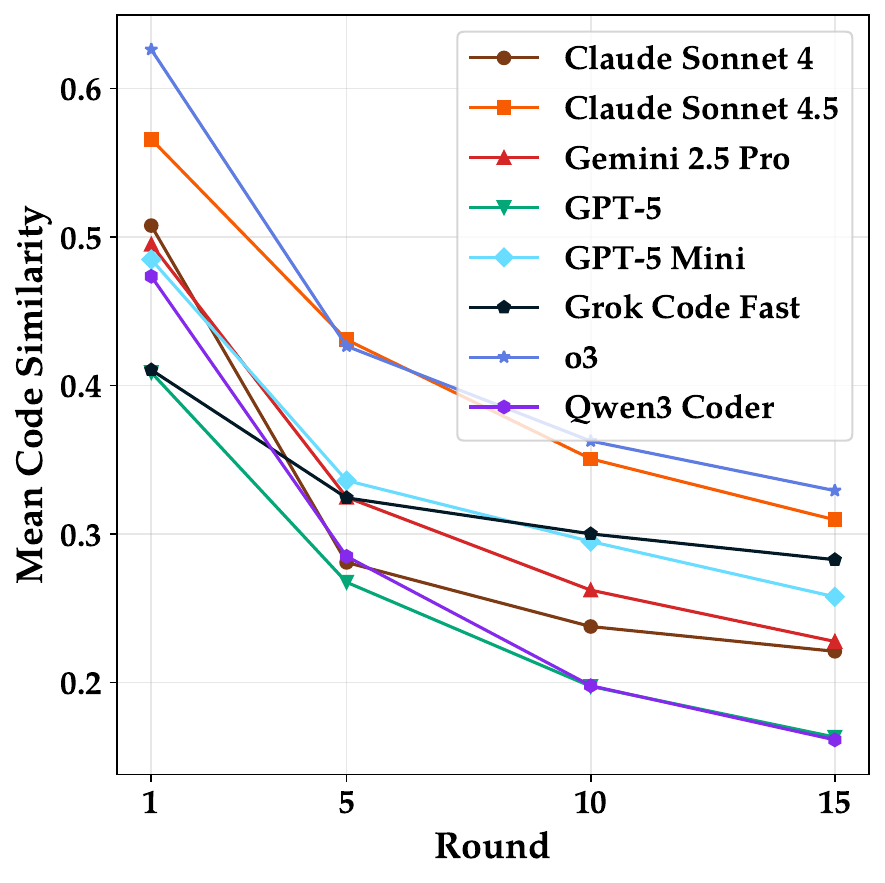}
    \caption{
To measure solution diversity, we compute code similarity of each model's solutions to itself at the same round.
Each data point represents the mean pairwise similarity between a model's solution (\texttt{main.py}) at round \texttt{n} across 70 BattleSnake tournaments.
    }
    \label{fig:line_chart_code_evolution}
\end{minipage}
\end{figure*}

\textbf{Models interact with codebases in markedly different ways.}
\clash{}'s open-ended setting reveals striking differences in how models operate in the \textit{edit} phase.
For instance, while o3 and Gemini 2.5 Pro typically only edit an average of 2 files per round, GPT-5 usually changes 5 to 6.
The size of edits also varies -- on one end, o3 typically adds/removes a total of 51 lines per round, 8$\times$ less than Qwen3 Coder or the Claude Sonnet family which usually modify more than 400 lines.
Gemini 2.5 Pro stands out as a verbose thinker, generating an average of 105 words per thought, more than double the average.
Claude Sonnet 4.5 usually takes 23 of the allotted 30 editing turns per round, whereas GPT-5 and o3 typically concludes after just 15 steps.
Distributions visualizing these tendencies in \S\ref{appx:results:interaction-trends}.

Intriguingly, we did not find any correlations between any of these behaviors and win rates.
Both minimalists (o3) and high activity editors (Claude Sonnet 4.5) succeed.
Compared to existing benchmarks that terminate upon reaching a solution, \clash{}'s multi-round competitive setting makes these distinctions even more salient.

\textbf{Even strong models struggle to recover after losing rounds.}
In real-world software development, early choices are often made under uncertainty: the best approach might only become clear after testing, real world deployments, and observing competitors. 
Therefore, the ability to interpret noisy signals and reconsider core design decisions is an important factor for success.
The round-based nature of \clash{} exposes how poorly LMs adapt once their initial strategies fail.  
Figure~\ref{fig:line_chart_comeback_probability} shows that even for the Claude Sonnet 4.5, losing a single round results in a comeback probability (win probability of the next round) of less than one third — less than half of the overall round  win rate of 71\%.
For o3, the win rate drops to only 26\% after a single loss (compared to an overall round win rate of 65\%).
After five consecutive defeats, comeback rates fall below 15\% for Claude Sonnet 4.5, and below 10\% for all other models. 
This suggest an inability of models to reconsider strategies, or adapt to opponents or the arena state. 

\textbf{Models' solutions become increasingly diverse.}
For each (model, opponent, round) tuple, we compute code similarity across the model's solutions (10 samples) using Python's \texttt{difflib.}\texttt{SequenceMatcher}~\citep{ratcliff1988pattern}.
In other words, we have 10 tournaments of Claude Sonnet 4.5 vs. o3 from our main results.
We then compute a similarity matrix between all 10 versions of Claude Sonnet 4.5's \texttt{main.py} at each round 1/5/10/15, and finally calculate a mean similarity score.
We run this analysis just for the BattleSnake arena since solutions are written in Python in a single \texttt{main.py} file.
From Figure~\ref{fig:line_chart_code_evolution}, we observe models' solutions generally become more dissimilar with every round.
Each round, models are attempting to not only make absolute improvements, but also adapt to opponent play.
Solution diversity varies with model (o3 at 0.63 versus GPT-5 at 0.41 at round 1), though the effect of the opponent's identity is less pronounced, as we show in \S\ref{appx:results:analyses}.
Unlike existing code benchmarks where models quickly converge on canonical solutions, \clash{} elicits substantial creativity from models, even against the same opponent.
This diversity makes \clash{} a potentially effective training ground for improving models via self-play and reinforcement learning~\citep{zelikman2022star}.

\begin{figure*}[t]
\centering
\begin{minipage}[t]{0.49\textwidth}
    \setlength{\abovecaptionskip}{0em}
    \centering
    \includegraphics[width=0.95\textwidth]{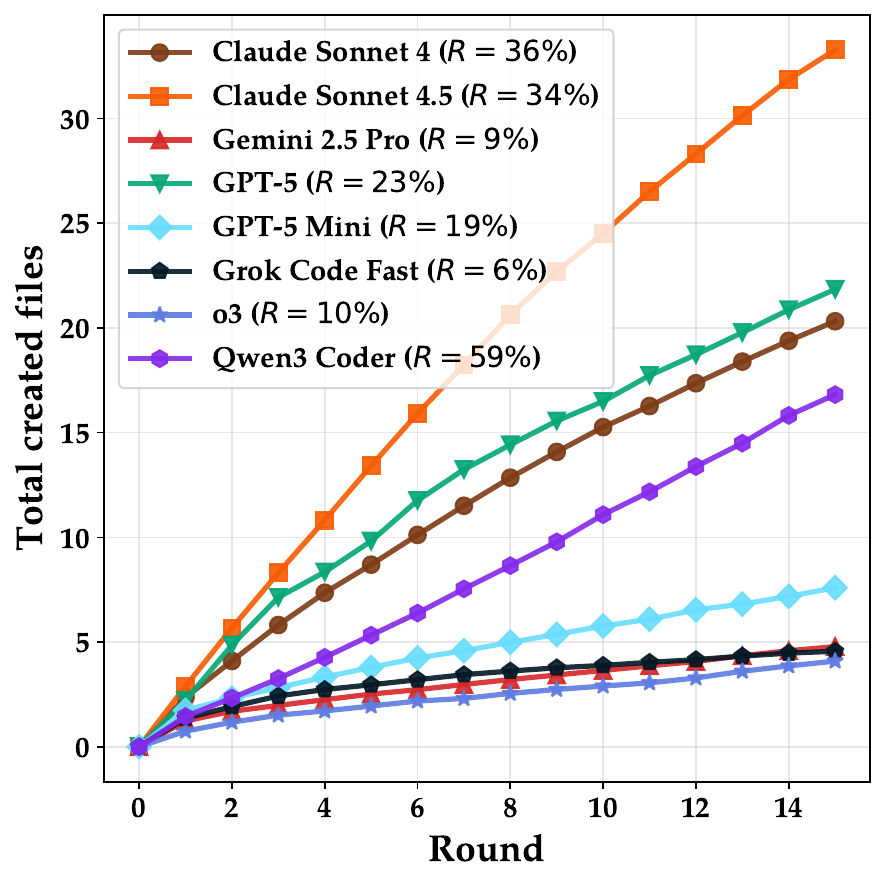}
    \caption{Created file count scales almost linear with the round. $R$ refers to filename redundancy at round 15; high values indicate repeating patterns in filenames (such as \texttt{main1.py}, \texttt{main2.py}).
    }
    \label{fig:line_chart_total_created_files_vs_round}
\end{minipage}
\hfill
\begin{minipage}[t]{0.49\textwidth}
    \setlength{\abovecaptionskip}{0em}
    \centering
    \includegraphics[width=0.95\textwidth]{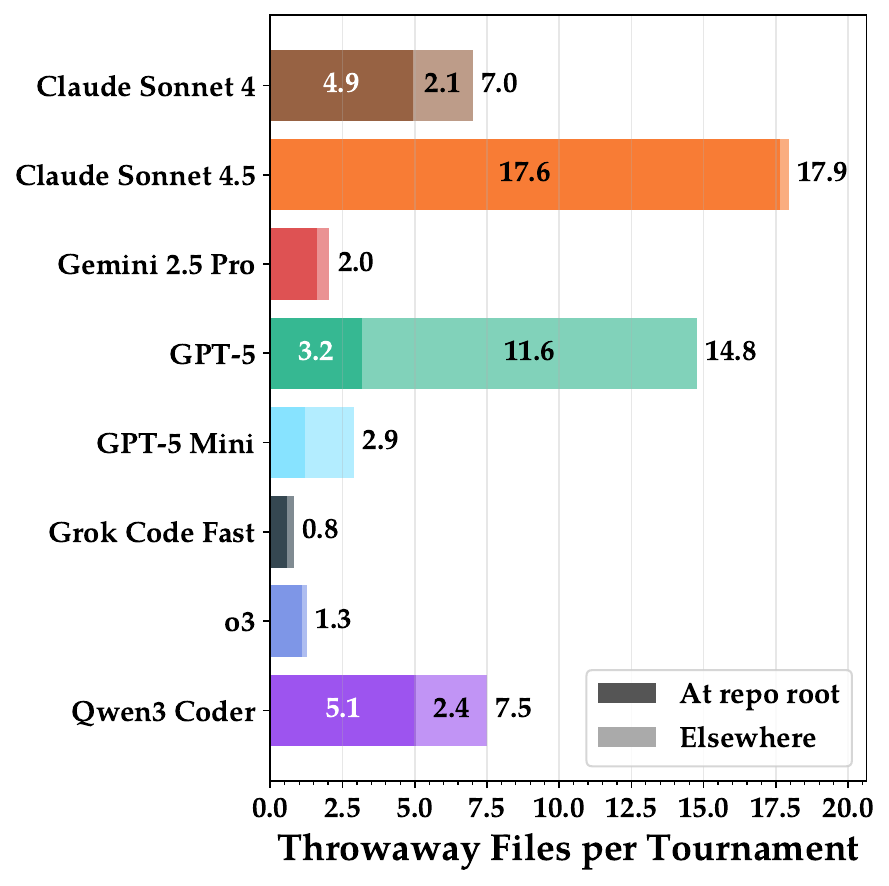}
    \caption{Models differ in average number of \emph{throwaway files} (files not used after the round they were created). Stacked bars distinguish between files at the repository root and those in subdirectories.}
    \label{fig:bar_chart_throwaway_files}
\end{minipage}
\hfill
\end{figure*}

\textbf{Codebases managed by models become messier over time.}
In most human-managed codebases, the rate of file creation quickly plateaus once the overall structure has been established; subsequent work primarily focuses on refinement, maintenance, and incremental improvements rather than continuous expansion.
In contrast, we observe a markedly different trend in Figure~\ref{fig:line_chart_total_created_files_vs_round}: the average number of agent-created files scales almost linearly with the number of rounds.
Claude Sonnet 4.5 exhibits the highest file creation activity, averaging more than 30 files per tournament, followed by GPT-5 (21), whereas o3 creates fewer than 5.
For Claude Sonnet 4.5, the high average is driven by consistent creation of various files at the repository root (making the codebase even less orderly); for GPT-5, the average is elevated by tournaments that accumulate particularly many output and temporary files in separate directories that were never cleaned up.
These observations again highlight how the top three models interact with their codebases in distinctly different ways.

When many files are produced, filenames often become repetitive and follow systematic patterns (e.g., \texttt{analyze\_round\_13\_v2.py}).
We quantify this effect through the \emph{filename redundancy} metric (the fraction of files sharing name prefixes with other files) which is particularly high for Qwen3 Coder (59\%) and Claude Sonnet models (35\%).
In addition, most agent-created files are never referenced, reused, or modified in subsequent rounds.
We quantify these \emph{throwaway files} in Figure~\ref{fig:bar_chart_throwaway_files}: Claude Sonnet 4.5 (18 files per tournament) and GPT-5 (15) again rank at the top, whereas o3 remains near the bottom.

Together, Figure~\ref{fig:line_chart_total_created_files_vs_round} and Figure~\ref{fig:bar_chart_throwaway_files} reinforce the view that most LMs struggle to converge toward maintainable file structures over time, favoring the continual generation of new, often redundant scripts over the systematic refinement and reuse of existing code.
More graphs and case studies in \S\ref{appx:results:analyses}.

\subsection{Strategic Reasoning Limitations}
\label{sec:analysis:reasoning-limits}
We investigate models' capacity for self-improvement by analyzing how they interpret competition results to diagnose failures, decide what code changes to make, and how to validate them.
This analysis is performed using GPT-5 with high reasoning as a judge, validated against human annotations from three authors on 100 trajectories (Appendix~\ref{appx:results:human_annotation}).
Details, as well as additional analyses of agent trajectories in terms of the nature of actions, are presented in Appendix~\ref{appx:results:strategic_reasoning}.

\begin{figure*}[t]
    \includegraphics[width=\linewidth]{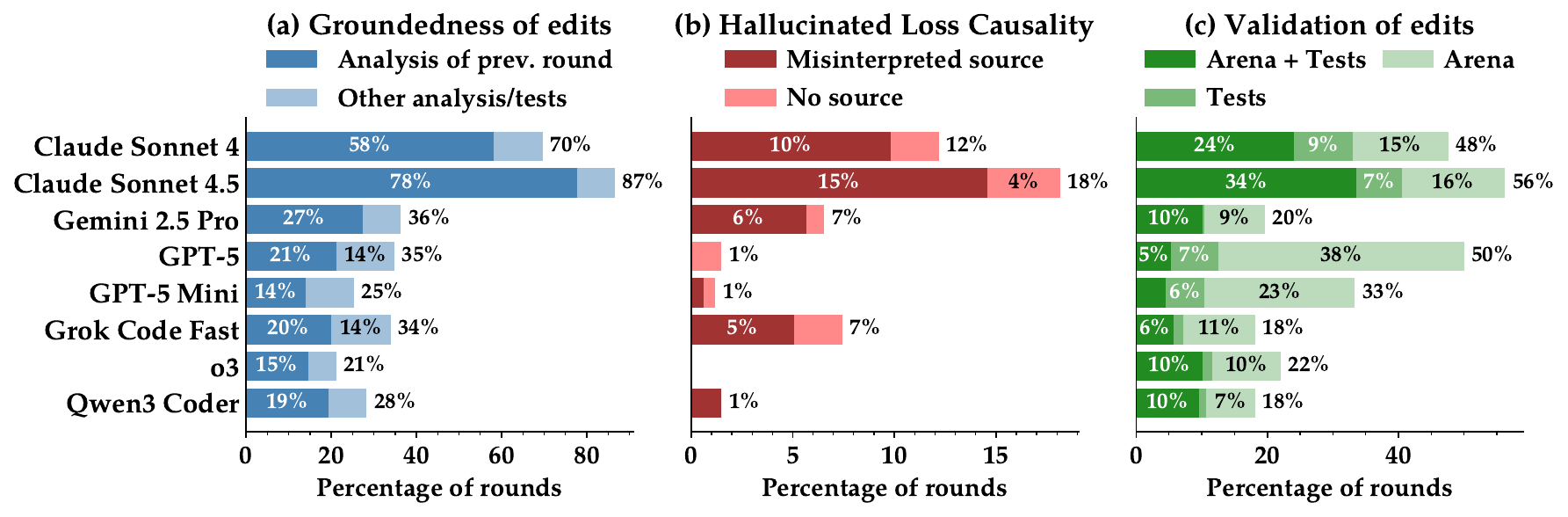}
    \caption{LMs struggle to analyze logs from previous rounds and frequently hallucinate about why rounds were lost.
    Using LMs, we annotate players' trajectories to answer three questions (a) Are changes grounded in the analysis of previous rounds or testing? (b) Are there hallucinated or unsubstantiated claims about why a round was lost? (c) Are changes validated by arena simulations or unit tests?
    Human validation of these annotations is presented in Appendix~\ref{appx:results:human_annotation}.
    }
    \label{fig:llm_as_judge}
\end{figure*}

\textbf{Most models struggle to interpret logs or derive meaningful insights about their performance.} 
Agents have access to detailed log records of all previous rounds, encompassing several hundred to thousands of runs against their opponent.
These logs can not only reveal whether the last round's changes improved the winning rate, but detail the exact behavior that led to losses or wins.
However, despite explicit suggestions to write analysis tooling in the prompt, most LMs do not extract worthwhile information, often stopping at reading the first lines of a log file or calculating the win rate of the last round.
Figure~\ref{fig:llm_as_judge}(a) shows whether the combined output of the actions of the agent (i.e., the entirety of the information available to the agent) could motivate the edits performed by the agent.
While most edits of the Claude Sonnet models can be motivated in this way, the edits of all other models are ungrounded in more than $65\%$ of all rounds.
Interestingly, o3 scores particularly low in this aspect, with ungrounded edits in almost 80\% of rounds.

\textbf{Models hallucinate during failure analysis and misinterpret logs and analysis outputs.}
The most salient pattern are agents inferring causal explanations for arena outcomes after reviewing only the opening lines of a single log file, when these lines do not even show the deciding moment in an arena. 
Behaviors of this kind are quantified in Figure~\ref{fig:llm_as_judge}(b). 
For example, Claude Sonnet 4.5 makes uncorroborated claims about the exact reason a game was lost in more than 17\% of rounds on average.
However, this behavior is much more pronounced in certain arenas, such as BattleSnake, where Claude Sonnet 4 and Claude Sonnet 4.5 hallucinate about loss causality in 34\% and 46\% of rounds.
Most hallucinations are misinterpretations or over-interpretations of log files and similar outputs, though claims that cannot be connected to any source also occur.

\textbf{Models make changes without assessing their effects.} 
When models propose algorithmic changes, they seldom confirm whether modifications work as intended or if the new solution outperforms previous iterations.
Our prompts explicitly suggests running arena simulations between different versions of code or writing unit tests to validate intended behavior. 
Combining exploratory methods with self-play could likely avoid unwanted regressions.  
Nevertheless, most models deploy untested code. As shown in Figure \ref{fig:llm_as_judge}(c), only Claude Sonnet 4.5 validates changes in a majority of rounds (56\%), followed by GPT-5 (50\%), whereas Gemini 2.5 Pro and o3 perform validation in just one out of five rounds.

\textbf{Models rarely make bash mistakes.}
Across all models, more than 85\% of generated actions execute successfully, with error rates ranging from just 10\% (Claude Sonnet 4) to 16\% (Qwen3 Coder).
Models also recover rapidly from errors: following a failed command, the very next action runs successfully more than 80\% of the time.
This contrasts starkly with earlier findings of "cascading failures" in agent systems~\citep{yang2024sweagentagentcomputerinterfacesenable,pan2025trainingsoftwareengineeringagents}, suggesting command-line proficiency has improved substantially in recent models.
These results indicate that performance differences in \clash{} stem from strategic reasoning and code quality, not \texttt{bash} interface capabilities.
More in \S\ref{appx:results:interaction-trends}.
\section{Related Works}
\label{sec:related_works}

\textbf{Software engineering benchmarks.}
Early evaluations of LMs' coding capabilities typically tasked models with completing the body of a function given its header and a brief description~\citep{austin2021programsynthesislargelanguage,chen2021evaluating,hendrycks2021measuringcodingchallengecompetence,liu2023codegeneratedchatgptreally,jain2024livecodebenchholisticcontaminationfree,zhuo2025bigcodebenchbenchmarkingcodegeneration}.
As performance on such benchmarks has saturated, the community's attention has shifted towards more complex, repository-level tasks, notably SWE-bench~\citep{jimenez2024swebenchlanguagemodelsresolve}.
Given a GitHub issue, an LM must rewrite the codebase such that the proposed fix passes one or more unit tests.
SWE-bench has since been extended in multiple directions, including evaluation~\citep{openai2024swebenchverified,yang2024swebenchmultimodalaisystems,rashid2025swepolybenchmultilanguagebenchmarkrepository,swebenchpro2025,zan2025multiswebench,zhang2025swebenchgoeslive}, issue resolution workflows and SWE-agents~\citep{xia2024agentlessdemystifyingllmbasedsoftware,yang2024sweagentagentcomputerinterfacesenable,wang2025openhandsopenplatformai}, and datasets~\citep{jain2025r2egymproceduralenvironmentshybrid,pan2025trainingsoftwareengineeringagents,pham2025swe,yang2025swesmithscalingdatasoftware}.
Unlike these benchmarks where the objective and often the recommended approach are explicitly specified, \clash{} offers no predetermined notion of what constitutes improved code.
LMs must determine and pursue their own refinement strategies~\citep{wang2024troveinducingverifiableefficient}.
This open-ended setting evaluates capabilities beyond codebase manipulation, such as strategic thinking, adaptation to opponents, and long-term planning.

\textbf{Performance optimization.}
In lieu of unit tests, several benchmarks instead evaluate LMs on code optimization, such as boosting algorithmic efficiency~\citep{du2024mercurycodeefficiencybenchmark,liu2024evaluatinglanguagemodelsefficient,waghjale2024ecco,huang2025effibenchbenchmarkingefficiencyautomatically} or reducing runtime
~\citep{he2025sweperflanguagemodelsoptimize,ouyang2025kernelbenchllmswriteefficient,press2025algotunelanguagemodelsspeed,shetty2025gsochallengingsoftwareoptimization}. 
Like \clash{}, how an LM goes about improving a codebase is entirely self-prescribed; there are no specific instructions or hints about methodology.
Unlike \clash{}, first, LMs carry out optimizations independently; LMs' codebases do not directly compete, nor must LMs anticipate or adapt to opponents' strategies.
Second, the objectives of existing optimization tasks are relatively narrow.
In contrast, \clash{} supports diverse environments with flexible win conditions, enabling LM-based code evolution for goals beyond runtime performance.

\textbf{Game playing.}
Video and text games have long been used as testbeds for studying reinforcement learning agents~\citep{mnih2015human,silver2016mastering,openai2019dota2largescale}, with a resurgence in use for evaluating LMs~\citep{yao2020calmexplorelanguagemodels,hu2025gamearenaevaluatingllmreasoning,karten2025pokechampexpertlevelminimaxlanguage,paglieri2025balrogbenchmarkingagenticllm,zhang2025videogamebenchvisionlanguagemodelscomplete}.
While past works have an AI system directly play a game, to our knowledge, \clash{} is the first to study the interplay of interactive coding and gaming for evaluating LMs.
Furthermore, \clash{}'s task formulation aims to represent not just games, but general real-world, competitive software development, where codebases essentially compete against one another to achieve goals.

\textbf{Self improving agents.}
Recent work has explored how LMs can evolve agent scaffolds for better performance on software development tasks, namely SWE-bench~\citep{wang2025huxleygodelmachinehumanlevelcoding,zhang2025darwingodelmachineopenended}.
However, static benchmarks relying on fixed correctness metrics like unit tests are an awkward fit for prototyping self-improvement systems.
Unit tests only provide binary pass/fail feedback, and once passed, they are no longer useful for further refinement.
\clash{}'s competitive setting with constantly evolving opponents provides a perpetual learning signal that doesn't saturate.
Performance is graded relatively, a richer training signal than binary correctness.
We hope future work around self-improving SWE-agents will consider \clash{} as a training ground.

\section{Discussion}
\label{sec:discussion}

\textbf{Robustness of findings to evaluation setup.}
A potential concern is whether \clash{}'s findings are sensitive to the choice of agent scaffold, edit budget, or prompting strategy.
Several lines of evidence suggest they are not.
First, model rankings on CC:Ladder (7 rounds per opponent) are broadly consistent with the main leaderboard (15 rounds per matchup), despite a substantially different budget.
Second, replacing \texttt{mini-SWE-agent} with SWE-agent, which provides additional tooling including a file-tree viewer and AST-level code search, changes CC:Ladder scores by at most 2 ranks across three models and two arenas (Appendix~\ref{appx:results:scaffold_ablation}).
Third, models achieve 85\%+ bash command success rates with rapid error recovery (Section~\ref{sec:analysis:reasoning-limits}), and the dominant failure modes are strategic rather than interface-related: a linter would not help a model that misinterprets competition logs, and AST parsing would not help a model that deploys changes without testing.
Finally, the system prompt (Appendix~\ref{appx:evaluation:mini-config}) is deliberately minimal and arena-agnostic, leaving little room for prompt-specific sensitivity.

\textbf{Limitations and future directions.}
\clash{}'s arenas are relatively smaller and more self-contained than most real-world software systems.
We'd like to support code environments for more realistic, multi-objective settings (e.g., cybersecurity, financial markets).
Second, competition logs are text-based.
Subsequent investigations could support multimodal feedback and study Vision Language models (VLM) performance.
Finally, \clash{}'s artifacts and environments can be used to improve model capabilities via pre-training on editing traces or post-training techniques like self-play and reinforcement learning.

\textbf{Conclusion.}
By situating LMs in tournaments where their codebases compete directly, \clash{} reveals both the creative potential and fundamental limitations of current models.
Models devise remarkably diverse solutions, but struggle to draw meaningful conclusions from competition logs or maintain well-organized codebases.
We hope \clash{} will serve as a testbed for the next generation of autonomous software development systems.

\newpage
\section*{Acknowledgments}

We thank Laude Institute, Andreessen Horowitz, and Open Philanthropy for providing funding for this work.
We thank Princeton Language \& Intelligence (PLI) for providing credits for running closed-source API models.
Thanks to Samuel Ainsworth for his constant support of \texttt{bitbop.io} (\url{https://bitbop.io/}), the compute service for which this project was carried out with.
We also thank Shiyi Cao, William Held, Abe (Bohan) Hou, Dacheng Li, Jeffrey J. Ma, Karthik R. Narasimhan, Yijia Shao, Chenglei Si, Zora (Zhiruo) Wang, Alexander Wettig, and Yanzhe Zhang for constructive discussions and support throughout this project.
Finally, our greatest thanks to the open source development communities that created and maintain several of the competitive code arenas represented in \clash{}.

\bibliography{colm2025_conference}
\bibliographystyle{colm2025_conference}

\newpage
\appendix
\section*{Appendix}
The appendix is generally structured as follows.
In Section~\ref{appx:infra}, we provide some additional details about \clash{}'s infrastructure and implementation details.
In Section~\ref{appx:arenas}, we include deep dive discussions into each of the arenas supported in \clash{}.
Section~\ref{appx:evaluation} supplements Section~\ref{sec:experiments} with additional minor details about evaluation parameters and metrics.
Section~\ref{appx:results} contains additional results, analyses, and ablations about our experiments.

\textit{Our code is open sourced at \url{https://github.com/CodeClash-ai/CodeClash}.
The trajectory viewer and leaderboard are available at \url{codeclash.ai}.}
\section{Infrastructure}
\label{appx:infra}

\begin{figure}[h]
    \centering
    \includegraphics[width=\textwidth]{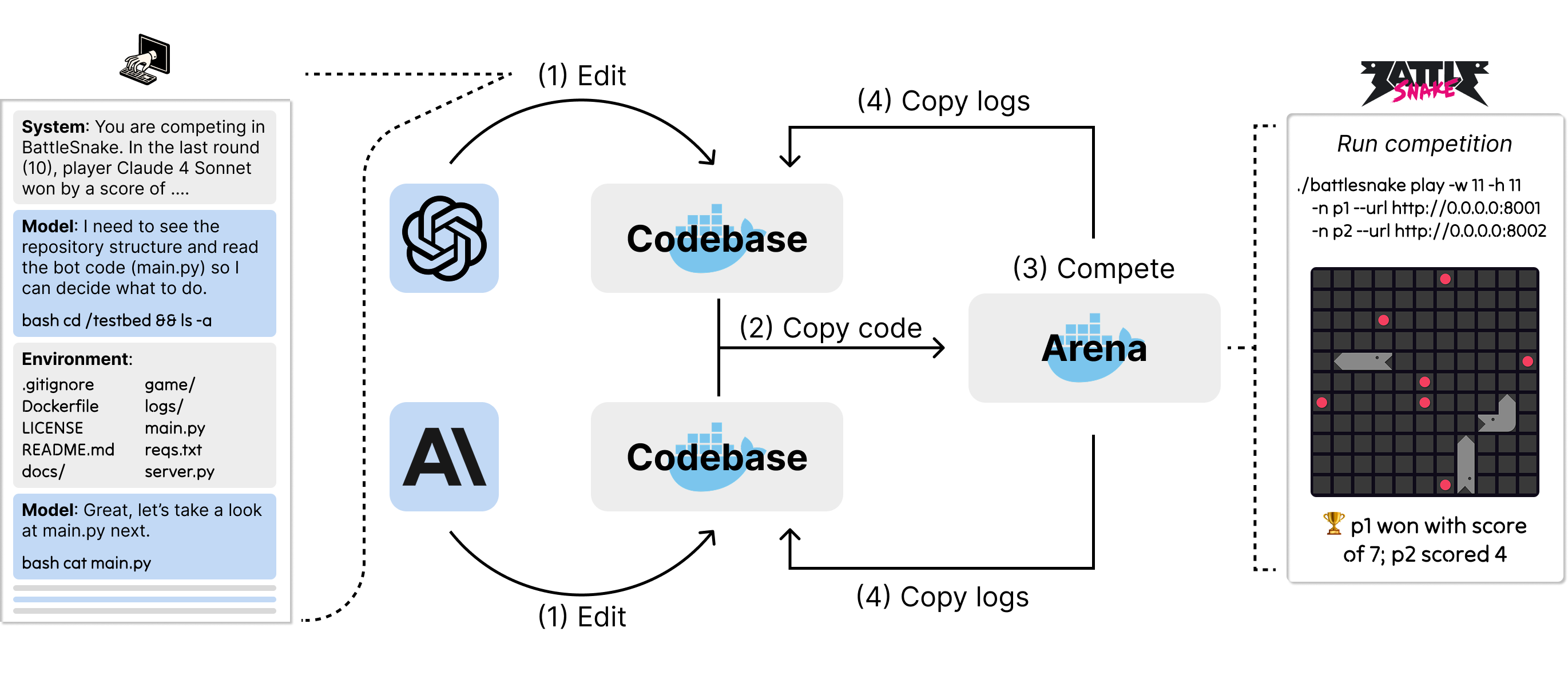}
    \caption{
    Technical overview of a \clash{} round.
    Each round, during the \textit{edit} phase, LMs edit their respective codebases within Docker containers, using \texttt{mini-SWE-agent} to facilitate multi-turn editing (Step 1).
    This is followed by the \textit{competition} phase, where the codebases are copied the arena docker container (Step 2).
    The arena then runs codebases against each other, with the game-play and outcomes captured as logs (Step 3).
    These logs are copied into each player's codebase before the next round begins (Step 4).
    }
    \label{fig:tech-overview}
\end{figure}

In this section, we provide some additional insights and discussion into the tooling and infrastructure that \clash{} uses to (1) enable LMs to edit codebases and (2) automatically run codebases against each other within the code arena.
Mimicking Figure~\ref{fig:preview}, we provide a more technically informative breakdown of the \clash{} loop in Figure~\ref{fig:tech-overview}.

We format our discussion of \clash{}'s infrastructure as a series of system design questions that reflects the thought processes we went through and decisions we arrived upon towards implementing \clash{}.

\textbf{How should models edit their codebases?} The benefits and drawbacks around methods for how LMs interact with codebases has been investigated thoroughly by recent works~\citep{xia2024agentlessdemystifyingllmbasedsoftware,yang2024sweagentagentcomputerinterfacesenable}.
Inspired by both prior research insights and current, popular paradigms for AI coding tools, we wanted to ensure several key properties for how LMs should manipulate a codebase for \clash{}, which is step $1$ in Figure~\ref{fig:tech-overview}.

\begin{enumerate}
    \item LMs should be able to \textit{view execution feedback}. 
    Execution is crucial to enable models to create and use their own constructs (e.g., analysis scripts, memory systems).
    \item LMs should be able to \textit{interact with a codebase}.
    A defining challenge of \clash{} is that LMs operate in a self-directed manner.
    Workflow-oriented approaches~\citep{xia2024agentlessdemystifyingllmbasedsoftware} are unsuitable for our setting.
    Going hand-in-hand with (1), interaction is also necessary so that models can string sequences of changes together.
    \item LMs should \textit{operate using \texttt{bash} actions, not tools}.
    As described in~\cite{yang2024swebenchmultimodalaisystems}, various workflows and tools can be (un-)intentionally biased to favor particular models.
    Our goal is to evaluate models, not scaffolds or tools.
    Therefore, we decide to make LMs operate in the most ``impartial" action space.
    This decision also leaves an opportunity for LMs to synthesize their own tools across rounds.
\end{enumerate}

Considering these points all together, we found \texttt{mini-SWE-agent} to be most suitable.
\texttt{mini-SWE-agent} is a lightweight agent scaffold that allows LMs to interact with a codebase in a terminal environment.
Per turn, an LM generates a \texttt{bash} command, then receives standard output as execution output.
The combination of \texttt{mini-SWE-agent} and Claude 4 Opus scores $67.6$\% on SWE-bench Verified\footnote{\texttt{mini-SWE-agent} with Claude 4 Opus score from \url{swebench.com} bash-only leaderboard.}, giving us confidence that the models we evaluate are capable of performing bash-only interactions with a low to non-existent rate of failures due to syntactic errors such as malformed responses or actions.

\textbf{How do we make \clash{} portable and reproducible?} Following precedent established by existing interactive coding benchmarks~\citep{yang2023intercodestandardizingbenchmarkinginteractive}, we use Docker to containerize the environments for (1) LMs to develop their respective codebases (\textit{agent containers}) and (2) running codebases in the arena (\textit{arena container}).
No codebase edits or arena runs are ever performed on device.
The only artifact created on the local machine are logs capturing tournament metadata and outcomes.

\textbf{What initial assets should a model be given?} In other words, what should the starter codebase specific to each arena generally contain?
To answer this, we outlined a shortlist of several behaviors and conditions that should be supported and true for any arena.

\begin{itemize}
    \item LMs should be able to learn about the arena/game as extensively as it would like. We do not assume players have any prior knowledge about how the arena works.
    \item LMs should be able to run the arena to understand it and perform testing.
    \item LMs are provided with a simple but functional baseline strategy that demonstrates core mechanics.
    A player does not need to code a valid submission from scratch.
\end{itemize}

Based on this, we make sure every codebase has the following assets:

\begin{itemize}
    \item \textit{Documentation}: For every arena, we were able to find source code containing arena documentation (e.g., \url{https://github.com/BattlesnakeOfficial/docs}).
    We copy documentation into a \texttt{docs/} folder for every arena's starter codebase.
    \item \textit{Arena executable}: Any executables and assets needed to run a round of the arena are fully available to each player.
    However, the exact \texttt{bash} commands are not disclosed; the burden remains on the model to figure out how to use assets.
    \item \textit{Working submission}: Like how human participants are provided a simple, functional, and suboptimal baseline strategy, LMs are given a starter codebase that can be submitted as is.
    This ensures meaningful competition from the first round.
\end{itemize}

In practice, for any arena, the starter codebases for each player and the codebase for running the competition across multiple codebases are identical.

\textbf{Per round, how many times should a competition be run?}
This question stems from the non-determinism that we observed in the majority of \clash{} arenas.
With the exception of MIT Battlecode 2025, we found that given the same codebases and the same arena, the outcome of a single simulation is indeterminate, which is to be expected.

In order to declare a winner with confidence, each round at step $3$ in Figure~\ref{fig:tech-overview}, the arena runs the competition $1000$ times.
We declare the winner as whichever player wins the most out of the $1000$ simulations (or declare a tie if ties are most frequent), rather than requiring a specific win percentage threshold.
This approach aligns with standard practice in competitive gaming communities and avoids introducing arbitrary performance cutoffs.
We concretely review how we calculate win rate and Elo in \S\ref{appx:evaluation:metrics}.

\textbf{How can models improve their codebase?} A cornerstone to performing well in \clash{} is a model's ability to understand past rounds' outcomes, then adapt the codebase to perform better in the arena against the opponent(s).

To encourage such behavior, both the proceedings and outcome of each simulation are logged.
The precise format of the logs depends on the arena.
These logs are then copied from the arena container back into the agent containers, specifically in a designated \texttt{logs/} folder within the agent's codebase, as reflected by step $4$ in Figure~\ref{fig:tech-overview}.

How the model interprets these logs or acts upon them is entirely self-driven.
In the initial system prompt, we generally mention that analyzing logs might be helpful, but we do not provide any arena-specific advice on how exactly logs should be interpreted.
In practice, we've observed a spectrum of interesting approaches.
Models will directly read the raw logs, write scripts to solicit insights, or even modify the logs.
More insights in \S\ref{appx:results}.

\textbf{What happens if a model's codebase is not a valid submission?} We observed during early trials that models will occasionally errantly modify a codebase such that it it no longer functions properly when run in the arena.
The error modes are most frequently due to certain expectations about the codebase not holding. For instance...

\begin{itemize}
\item For Battlecode, the main bot logic should be represented entirely within a \texttt{./bot.py} file that implements a \texttt{turn} function.
\item For Battlesnake, the bot is in \texttt{main.py}, which implements a \texttt{move} function.
\item For RoboCode, the tank bot should be defined under \texttt{robots/custom/}, and the code must pass compilation (\texttt{javac -cp "libs/robocode.jar" robots/custom/*.java}).
\end{itemize}

We note that we do not define these constraints -- these rules are reflective of the original conditions these arenas and games impose on human players and their submissions.

To address this, we first, implement per-arena validation to check that the codebase is ready for competition.
The check is run at the outset of step $3$ in Figure~\ref{fig:tech-overview}.
Second, we define the following decision tree to handle situations where $1$+ players have invalid codebases.

\begin{itemize}
    \item If all player codebases are invalid, the round is declared a tie.
    \item If only one player codebase is valid, that player is declared a winner.
    \item If $2$+ player codebases are valid, the competition phase is run with all valid codebases. Any invalid codebases are excluded.
\end{itemize}

\textbf{Do arenas have positional advantages, and how are such advantages accounted for?}
A \textit{positional advantage} refers to a situation where, assuming $2$+ players have identical codebases, one player consistently wins.
We want to eliminate such advantages in \clash{}, as they unfairly affect the arena outcome in ways that are outside of a player's control.

To detect whether positional advantages are present in an arena, we run the aforementioned experiment -- for every arena, we run a tournament with two ``dummy" players that do not change the initial codebase.
Each tournament is run for $25$ rounds, and the order of players is fixed.
We then check round outcomes, with the expectation that $\sim50$\% win rate suggests no such positional advantages are present.
From this investigation, we found MIT Battlecode 2025 to be the only arena that showed evidence of positional advantage.

However, checking for positional advantages may be tedious to repeat constantly for new arenas or when arena settings are adjusted (e.g., the \texttt{map} being used for Battlecode, \texttt{battleField} dimensions for RoboCode).
Therefore, to reliably eliminate any advantage, we simply randomly shuffle the order of players with equal probability at step $3$ in Figure~\ref{fig:tech-overview}, immediately after the codebase validation step.
We verified this fix by re-running the prior experiment for MIT Battlecode 2025 and found that the win rate returned back to $50$\%.

\textbf{Trajectories are tedious to parse.} Reading arena logs and \texttt{mini-SWE-agent} editing trajectories in their raw form was extremely laborious.
To make it easier to understand what has happened throughout the course of a tournament, we wrote a viewer for \clash{} logs that provides friendly visualizations of log content and automatically calculates some game statistics (e.g., p-value calculation to indicate if a round winner is statistically significant).

\newpage
\section{Arenas}
\label{appx:arenas}

This section contains arena cards describing each of code arena supported in \clash{}.
Per arena, we cover the objective(s), arena mechanics, log formats, and effective strategies.
We summarize all arenas supported in \clash{} in Figure~\ref{tab:list_arenas}.

\begin{table}[h]
\centering
\begin{tabular}{l|lcc}
\toprule
Arena & Description & \texttt{n} & Language \\
\midrule
Battlesnake & Grid-based survival and territory control & 2+ & Python \\ 
Core War & Assembly programs competing in shared memory & 2+ & Redcode \\ 
Halite & Resource collection and territory expansion on grid & 2+ & Multiple \\ 
Poker & No-limit Texas Hold’em & 2+ & Python \\ 
RoboCode & Tank duels with movement, scanning, and firing & 2+ & Java \\ 
RobotRumble & Turn-based grid battles with spawning robots & 2 & JavaScript \\ 
\bottomrule
\end{tabular}
\caption{
Code arenas currently implemented in \clash{}.
Arenas represent a diverse landscape of objectives (e.g., eliminate opponents, accumulate money/resources), programming languages, and challenges (e.g., decipher opponent strategy from logs, decide how to adapt code, manage growing codebase).
\texttt{n} is number of players.
}
\label{tab:list_arenas}
\end{table}

\subsection{MIT Battlecode 2025~\citep{battlecode2025}}

The MIT Battlecode organization is a student run group at the Massachusetts Institute of Technology that creates and hosts coding competitions.
\clash{} specifically supports the 2025 edition of the competition.
As described on the \href{https://battlecode.org/}{website}:

\begin{quote}
    Battlecode is a real-time strategy game in which you will write code for an autonomous player.
    Your player will need to strategically manage a robot army and control how your robots work together to defeat the enemy team.
\end{quote}

\begin{example}[System Prompt Description of Battlecode]
Battlecode 2025 throws you into a real-time strategy showdown where your Python bot pilots a team of specialized robots—Soldiers, Moppers, Splashers—alongside towers that spawn units or generate resources. Your mission: paint over 70\% of the map (or eliminate the enemy) by coordinating cleanups, area cover, and tower-building through tight bytecode budgets and clever unit synergy.
\end{example}

\begin{table}[h]
\centering
\begin{minipage}[b]{0.47\textwidth}
\includegraphics[width=\textwidth]{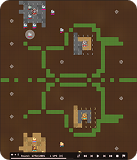}
\captionof{figure}{
Battlecode 2025: Chromatic Conflict screen capture.
The goal is to control a team of robobunnies to paint $70$\% of a map.
}
\label{fig:arena:battlecode}
\end{minipage}
\hfill
\begin{minipage}[b]{0.47\textwidth}
\begin{lstlisting}[language=Python, style=mystyle]
import random
from battlecode25.stubs import *
turn_count, directions = 0, [ # 8 directions ]

def turn():
    # MUST be defined. This is called every turn and should contain core logic

def run_tower():
    # Logic for a tower unit.

def run_soldier():
    # Logic for a soldier unit.

def run_mopper():
    # Logic for a mopper unit.

def update_enemy_robots():
    # Helper to track enemies.
\end{lstlisting}
\captionof{figure}{A Battlecode codebase must implement a core \texttt{turn} function that issues controls for three different kinds of units.}
\label{fig:arena:battlecode:snippet}
\end{minipage}
\end{table}

\textbf{What are effective strategies?}
Some effective approaches include efficient algorithms for path-finding/exploration, coordinating communication between agents, and finding the right balance between offensive moves (e.g., attacking, painting, destroying towers) and defensive measures (protect territory, tower placement, maintain stream of resources).

\textbf{What assets are provided in the initial codebase?} \texttt{run.py/} is the python script used to run players and upgrade versions. \texttt{src/} is the directory meant to contain all player source code and, \texttt{test/} contains all player test code. \texttt{client/} contains the client and the proper executable can be found in this folder. \texttt{matches/} is the output folder for match files. \texttt{maps/} is the default folder for custom maps.

\textbf{What are the arena configurations?}
For the 2025 edition "Chromatic Conflict", two teams of virtual robots roam the screen, managing resources and executing different offensive strategies against each other. Two types of resources exist in the arena: Money and Paint. Money is needed to produce units, buy towers and activate economy boost patterns (called SRPs). Paint is needed to produce units, for the win condition, to resupply units with paint and to paint special
patterns, which were prerequisites for acquiring SRPs and towers. There are also two kinds of soldiers: Moppers and Splashers. Moppers can attack other units without costing paint, which makes them the only unit capable of surviving indefinitely without a tower. They can also clean up enemy paint, making them essential for cleaning up enemy paint off of ally patterns. Splashers can paint over enemy paint with ally paint and are the only unit which can paint several squares at once. The last component of the arena is towers which are immobile units that can spawn units. Money and Paint Towers will passively generate the
corresponding resources. Defense Towers have high damage output and generates chips upon attacking enemy units.

\textbf{How is the winner determined?}
The winner is the first team that is able to "paint" $70\%$ of the map.

\textbf{How are arena logs formatted?}
The arena logs are written as a sequential record of the match.
They begin with setup information, including which bots are playing and on which map. 
After that, each line corresponds to a turn, tagged with the acting player and unit, followed by the action taken (e.g., spawning a new robot, attempting to build a tower, or performing a mop swing attack).
In effect, the log provides a turn-by-turn narrative: what units were created, what abilities were triggered, and how each side attempted to advance.

\begin{example}[Example of BattleCode Log]
\begin{verbatim}
Playing game between p1 and p2 on quack
[server] -------------------- Match Starting --------------------
[server] p1 vs. p2 on quack.map25
[A: #1@1] BUILT A MOPPER
[B: #4@1] BUILT A MOPPER
[A: #1@2] BUILT A SOLDIER
[B: #2@2] BUILT A SOLDIER
[A: #3@2] BUILT A MOPPER
[A: #12138@3] Trying to build a tower at (18, 25)
[B: #13376@3] Trying to build a tower at (18, 9)
[B: #4@4] BUILT A MOPPER
[A: #12523@4] Mop Swing! Booyah!
\end{verbatim}
\end{example}
\newpage
\subsection{Battlesnake~\citep{chung2020battlesnake}}

Battlesnake is a multi-player game, where each player's code controls a snake operating on a grid.
The arena's rules and objectives are heavily reminiscent of the traditional snake game.
The general objective is to program your snake to survive as long as possible.

The game starts with $2$+ snakes positioned at different quadrants of the grid.
Throughout the course of the game, food pellets will pop up -- if a snake consumes (moves into a cell containing) a pellet, the snake's body gets longer by one cell.
There are several ways a snake can ``die".
If it collides with a wall, its own body, or another snake that is longer, the snake is eliminated.
If the snake does not make a legal move on any particular turn, the game also ends.
The winner is the last remaining snake, or the longest snake if multiple are alive upon the exhaustion of some turn limit.

\begin{example}[System Prompt Description of Battlesnake]
You are a software developer (\{\{player\_id\}\}) competing in a coding game called Battlesnake.
Your bot (`main.py`) controls a snake on a grid-based board.
Snakes collect food, avoid collisions, and try to outlast their opponents.
\end{example}

\begin{table}[h]
\centering
\begin{minipage}[b]{0.47\textwidth}
\includegraphics[width=\textwidth]{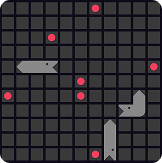}
\captionof{figure}{
Battlesnake screen capture.
Your code controls a snake that should find food, avoid other snakes, and survive.
}
\label{fig:arena:battlesnake}
\end{minipage}
\hfill
\begin{minipage}[b]{0.47\textwidth}
\begin{lstlisting}[language=Python, style=mystyle]
def info():
    return {"author": "", "color": "#888888" ...}

def start(game_state):
    ...

def end(game_state):
    ...

def move(game_state):
    # determine safe move; prevent moving backwards, out of bounds, or into self/others; optionally move toward food
    return {"move": "up"}
\end{lstlisting}
\captionof{figure}{A Battlecode codebase must implement a core \texttt{turn} function that issues controls for three different kinds of units.}
\label{fig:arena:battlesnake:snippet}
\end{minipage}
\end{table}

\textbf{What are effective strategies?} Effective Battlesnake bots rely on strategies that balance safety, space control, and efficient movement.
A common approach is to use \textit{flood-fill or area estimation} to avoid moves that lead into regions with insufficient space, reducing the chance of being trapped. 
\textit{Pathfinding algorithms such as A*} help snakes reach food or navigate safely around hazards, often incorporating penalties for risky tiles near enemy heads.
Many bots also implement \textit{look-ahead search}, simulating several future turns to predict collisions and maintain advantageous positioning.
Finally, strong bots prioritize \textit{risk-aware heuristics}, such as only engaging opponents when longer or only pursuing food when health is low.

\textbf{What assets are provided in the initial codebase?}
The \texttt{docs/} folder serves as the full documentation hub for the Battlesnake platform, containing subdirectories such as \texttt{api/}, \texttt{guides/}, maps/, and policies/, which collectively explain how to use the Battlesnake API, configure maps, follow gameplay policies, and get started with development. It also includes Markdown files like README.md, index.md, and quickstart.md for setup instructions; rules.md detailing official game rules and snake behavior; faq.md answering common developer questions; and starter-projects.md offering templates for new Battlesnake projects. Complementing the documentation, the game/ directory contains the full Go implementation of Battlesnake’s core logic. Key source files such as board.go, ruleset.go, standard.go, and pipeline.go define how the game board is represented, how rules are enforced, and how turns are processed. Specialized variants of the game board like royale.go, solo.go, constrictor.go, and wrapped.go implement different modes. Other files in the root directory include main.py, which serves as a starter template for Battlesnake logic and helper functions, server.py for server setup and request handling, requirements.txt listing Python dependencies, and a Dockerfile for containerized deployment.

\textbf{What are the arena configurations?} The Standard Arena in Battlesnake is the default game environment, adhering to the core game rules without any modifications. In this arena, the number of Battlesnakes can vary, ranging from a 1v1 match or multiple snakes competing, such as four or eight. The game board is a square grid measuring 11×11 cells, totaling 121 cells. Each cell is a discrete unit where snakes and food can occupy. The arena's boundaries are defined by the edges of this grid, and snakes are restricted to moving within these confines. Movement is allowed in four directions: up, down, left, and right, with no diagonal movement permitted. At the start of the game, snakes are placed at random positions within the arena, and food items are similarly distributed across the grid.

\textbf{How is the winner determined?} In Battlesnake, the winner is determined by being the last remaining snake on the game board. Each snake takes turns moving, loses one health point per turn, and can regain health by consuming food, which also causes the snake to grow in length. Snakes are eliminated in several ways: colliding with their own body, colliding with another snake’s body, or engaging in a head-to-head collision with another snake. In head-to-head collisions, the longer snake survives while the shorter one is eliminated. If both snakes are the same length, both are removed from the game. Players must carefully manage their health, navigate the board without running into obstacles or other snakes, and strategically consume food to survive longer than their opponents. The game continues until only one snake remains, and that snake is declared the winner.

\textbf{How are arena logs formatted?} The log for a single competition run is represented as a single \texttt{.jsonl} file, where each line in the file is a dictionary corresponding to a single turn of the run.
Each line of a Battlesnake log records the complete state of the game at a given turn. 
It captures the ruleset and configuration, the current turn number, the map dimensions, and the positions and attributes of all snakes (their ID, health, body coordinates, head position, and length).
It also lists the placement of food and hazards at that moment, as well as the perspective of the specific snake whose API is being called.
In other words, every log entry is a snapshot of the board state.

\begin{example}[Example of BattleSnake Log]
\begin{verbatim}
"turn": 0,
"board": {
"height": 11,
"width": 11,
"snakes": [
  {
    "id": "794bb7d7-a1ee-4939-a664-dd77d3c5f6e3",
    "name": "p1",
    "latency": "0",
    "health": 100,
    "body": [{"x": 9, "y": 9}, {"x": 9, "y": 9}, {"x": 9, "y": 9}],
    "head": {"x": 9, "y": 9},
    "length": 3,
    "shout": "",
    "squad": "",
    "customizations": "color": "#888888", "head": "default", "tail": "default"
  }
\end{verbatim}
\end{example}

\newpage
\subsection{Core War~\citep{corewar1984}}

For Core War, players write small assembly-esque programs (called a ``warrior").
The programs are run in a simulated, shared virtual memory.
The goal of every program is to disable all opposing programs.
The ultimate objective is to be the last program standing.

A unique facet of Core War is that the programming language, RedCode, is specific to the game.
RedCode supports basic operations (e.g., \texttt{mov}, \texttt{add}, \texttt{jump}, \texttt{compare}) along with multiple addressing modes (e.g., immediate, direct, indirect).
Warriors compete in the ``core", which generally is a fixed size, circular memory array that resembles main memory (RAM); .
The core is represented by a simulator called MARS.
The execution of the game then proceeds in cycles, where each cycle, the simulator alternates between warriors and executes on instruction per active process.
If a process executes an invalid instruction or hits an illegal condition, the process dies.
Warriors can also be designed to spawn additional processes with special instructions (\texttt{SPL}).
If all of a warrior's processes are killed, it is eliminated.
Core War games are typically played a maximum number of cycles; if no warrior is eliminated by the end, the round is a draw.

\begin{example}[System Prompt Description of Core War]
You are a software developer (\{\{player\_id\}\}) competing in a coding game called Core War.
Core War is a programming battle where you write ``warriors" in an assembly-like language called Redcode to compete within a virtual machine (MARS), aiming to eliminate your rivals by making their code self-terminate.
Victory comes from crafting clever tactics -- replicators, scanners, bombers -- that exploit memory layout and instruction timing to control the core.
\end{example}

\begin{table}[h]
\centering
\begin{minipage}[b]{0.47\textwidth}
\includegraphics[width=\textwidth]{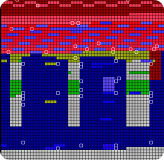}
\captionof{figure}{
Core War screen capture.
Your code controls a snake that should find food, avoid other snakes, and survive.
}
\label{fig:arena:corewar}
\end{minipage}
\hfill
\begin{minipage}[b]{0.47\textwidth}
\begin{lstlisting}[language=bash, style=mystyle]
;redcode-94
;name Dwarf
;author A. K. Dewdney
;strategy A simple warrior

start   add.ab  #4, bmb
        mov.i   bmb, @bmb
        jmp     start
bmb     dat     #0, #0
\end{lstlisting}
\captionof{figure}{
This Core War program, called \textit{Dwarf}, is a minimal attacking warrior. It repeatedly increments the pointer \texttt{bmb} (\texttt{add.ab \#4}, \texttt{bmb}), copies the \texttt{dat} instruction to that location (\texttt{mov.i bmb}, \texttt{\@bmb}), and then loops back (\texttt{jmp start}). The effect is that every fourth memory cell in the core is overwritten with a \texttt{dat} ``bomb", gradually scattering lethal instructions that kills an opponent’s processes if it is executed.
}
\label{fig:arena:corewar:snippet}
\end{minipage}
\end{table}

\textbf{What are effective strategies?}
Core War warriors typically incorporate three dimensions -- offense, defense, and adaptability.
A common offensive strategy is to write loops that scatter ``bombs" (invalid instructions) into memory, similar to the program in Figure~\ref{fig:arena:corewar:snippet}.
Another approach is to write programs that replicate as much as possible to increase survival rate.
An advanced warrior will usually combine such tactics.

\textbf{What assets are provided in the initial codebase?} The codebase contains three main directories \texttt{config/}, \texttt{docs/}, and \texttt{src/} and provides a complete Core War environment, including the assembler, simulator (virtual machine), documentation, and example warriors. In \texttt{config/}, different files define different configuration profiles for the pMARS simulator, allowing tournaments or simulations under multiple rule sets and tuning the VM for different “arena sizes.” The \texttt{docs/} folder describes how Core War works and how to write Redcode warriors. \texttt{src/} provides source code for the pMARS simulator and assembler, including files that implement the display and UI modules, core files, and configuration.

\textbf{What are the arena configurations?} Core War is a game in which two or more virus-like programs fight against each other in a simulated memory space or core. Core War programs are written in an assembly language called Redcode which is interpreted by a Core War simulator or MARS (Memory Array Redcode Simulator). The object of the game is to prevent the other program(s) from executing. At the start of a match, each warrior is loaded into a random memory location. Programs take turns executing one instruction at a time. A program wins by terminating all opponents, typically by causing them to execute invalid instructions, leaving the victorious program in sole possession of the machine.

\textbf{How is the winner determined?} In the standard Core War rules, the winner is determined by being the last warrior still “alive” (i.e., having at least one process still running) or the last to execute a valid “live” instruction. A warrior “dies” when it has no remaining processes left. Processes can die if they execute an invalid instruction or are overwritten.

\textbf{How are arena logs formatted?}
Core War logs generally report the outcomes, like which warrior survived, how many ``processes" (active execution threads) they maintained, or how many cycles elapsed before the match ended.
These logs don’t usually show step-by-step instruction execution, but instead give you a high-level summary of win/loss/tie, survival, and match duration.

\begin{example}[Example of Core War Log]
\begin{verbatim}
Program "Dwarf" (length 4) by "A. K. Dewdney"

       ORG      START
START  ADD.AB #     4, $     3
       MOV.I  $     2, @     2
       JMP.B  $    -2, $     0
       DAT.F  #     0, #     0

Dwarf by A. K. Dewdney scores 3
Dwarf by A. K. Dewdney scores 0
Results: 1 0 0
\end{verbatim}
\end{example}

\subsection{Halite I~\citep{halite2016}}

For Halite, players write autonomous bots that battle head to head with the goal of taking over the largest share of a virtual grid. Each bot issues commands every turn to move, collect, and deposit halite — a valuable in-game resource. The objective is to maximize your halite by the end of the match while strategically navigating around opponents and avoiding collisions. Bots use their strength to gain territory, and their territory to gain strength—outmaneuvering opponents based on the relative sophistication of their code.

A distinctive aspect of Halite is that it combines algorithmic strategy with real-time resource optimization. Players can program their bots in one of 4 languages (C, C++, OCaml, and Rust), and the game environment simulates simultaneous turns, where every decision — from choosing optimal collection routes to predicting enemy movements — can make the difference between victory and defeat. Matches are visualized in an animated replay, saved as an \texttt{.hlt} file, allowing players to analyze and refine their bot’s performance across different maps and opponents.

The Halite series also includes Halite II and Halite III, follow up iterations to the initial competition with significant updates to the nature of the competition.
We doubly clarify that this version of Halite described here refers specifically to Halite \textit{I}, released in $2016$.
We are planning to support Halite II and Halite III in \clash{} in the near future.

\begin{example}[System Prompt Description of Halite]
Halite is a multi-player turn-based strategy game where bots compete on a rectangular grid to capture territory and accumulate strength.
Players control pieces that can move across the map to conquer neutral and enemy territory, with each cell providing production that increases the strength of pieces occupying it.
The goal is to control the most territory by the end of the game through strategic expansion, consolidation of forces, and tactical combat decisions.
\vspace{1em}

You have the choice of writing your Halite bot in one of four programming languages: C, C++, OCaml, or Rust.
Example implementations can be found under the `airesources/` folder.
Your submission should be stored in the `submission/` folder.
\end{example}

\begin{table}[h]
\centering
\begin{minipage}[b]{0.47\textwidth}
\includegraphics[width=\textwidth]{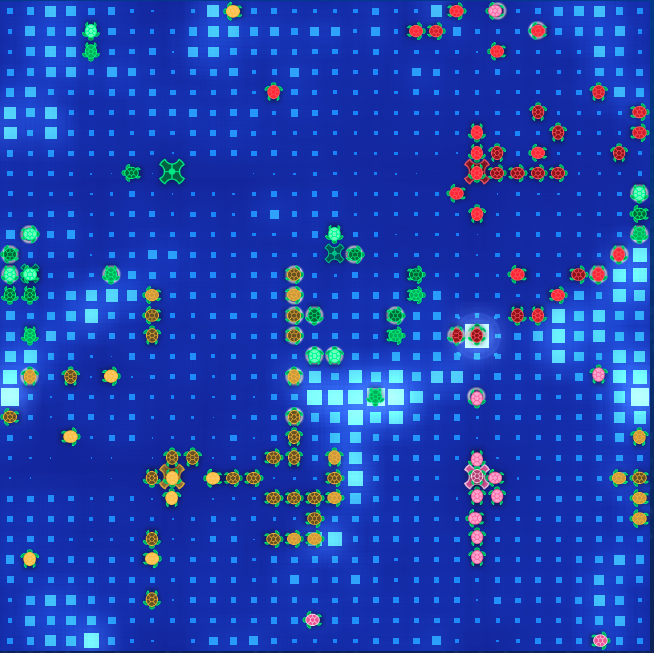}
\captionof{figure}{
Halite screen capture.
Your code controls a snake that should find food, avoid other snakes, and survive.
}
\label{fig:arena:halite}
\end{minipage}
\hfill
\begin{minipage}[b]{0.47\textwidth}
\begin{lstlisting}[language=c, style=mystyle]
#include "hlt.h"
#define BOT_NAME "MyCBot"

int main(void) {
    GAME game;
    game = GetInit();
    SendInit(BOT_NAME);
    while (1) {
        GetFrame(game);
        for (x = 0 ; x < game.width ; x++) {
            ...
        }
        SendFrame(game);
    }
}
\end{lstlisting}
\captionof{figure}{
Example Halite bot implementation in C. Bots follow a game loop structure: receive the current game state (\texttt{GetFrame}), iterate over owned cells to decide moves, and submit actions (\texttt{SendFrame}).
}
\label{fig:arena:halite:snippet}
\end{minipage}
\end{table}

\textbf{What are effective strategies?}
Effective strategies in Halite span three distinct phases. During the early game up until the bot makes contact with an opponent, an effective strategy is to capture neutral territory to fuel your growth with production and deprive other players of valuable neutral territory. Since bots don't yet have to defend their territory from other players, quick expansion into the most valuable areas is vital. During the mid-game (from when bots first make contact with another bot until there is very little remaining valuable neutral territory), players may want to shift to a hybrid of defense and offense: protect the best regions, seize remaining valuable neutral territory, and begin targeting weak points of opponents. Then, during late game, with most neutral territory gone, the game becomes purely about taking territory from other players. Players that take advantage of overkill
and attack enemies' high production areas are more likely to win.

\textbf{What assets are provided in the initial codebase?}
The initial Halite codebase provides all the foundational tools a player needs to create and test a functioning bot. Each starter package includes template code for your bot, such as a MyBot file where you implement decision-making logic, along with helper libraries that handle communication with the game environment (for example, receiving map data and sending moves). It also comes with a “RandomBot” or simple baseline bot to use as a reference, plus utilities for local simulation and visualization so you can test games without uploading them. These assets are designed to let players quickly get started with writing a bot that reads the game state, decides on moves, and interacts with the game engine via the provided API.

\textbf{What are the arena configurations?}
Halite games take place on a two-dimensional, rectangular grid map whose width and height are randomly generated for each match. The exact dimensions vary, but the generator always ensures that the resulting map is symmetric—it creates one section, then tessellates, reflects, and shifts it to fill the full board. This symmetry guarantees fair starting conditions for all players. Each cell on the map has two key values: Production, which determines how much Strength a stationary piece gains each turn, and Strength, representing how powerful a piece currently is. The maps are designed to be “interesting,” with clusters of high- and low-production zones rather than random noise, encouraging strategic territorial expansion. The map wraps around at the edges, meaning that moving off one side (for example, going North from the top row) places a piece on the opposite edge of the map—making the grid behave like a torus. The coordinate origin (0,0) is located at the northwest (top-left) corner of the map.

\textbf{How is the winner determined?}
Halite is played on a rectangular grid. Players own pieces on this grid.
Some pieces are unowned and so belong to the map until claimed by players. Each piece has a strength value associated with it.
At each turn, bots decide how to move the pieces they own. Valid moves are: STILL, NORTH, EAST, SOUTH, WEST.
When a piece remains STILL, its strength is increased by the production value of the site it is on. When a piece moves, it leaves behind a piece with the same owner and a strength of zero.
When two or more pieces from the same player try to occupy the same site, the resultant piece gets the sum of their strengths (this strength is capped at $255$).
When pieces with different owners move onto the same site or cardinally adjacent sites, the pieces are forced to fight, and each piece loses strength equal to the strength of its opponent.
When a player's piece moves onto an unowned site, that piece and the unowned piece fight, and each piece loses strength equal to the strength of its opponent.
When a piece loses all of its strength, it dies and is removed from the grid.
The game ends when only one player remains, or when a maximum number of turns has elapsed, defined as $10\times\sqrt{width \times height}$.
If the turn limit is reached or multiple bots are eliminated simultaneously, players are ranked by the amount of territory they control, with total Strength acting as a rare tiebreaker.

\textbf{How are arena logs formatted?}
Arena logs in Halite are formatted as sequential text entries that record the setup, turns, and results of a match. The log typically begins with the paths to the submitted bot executables for each player, followed by the map size or configuration, and then messages confirming initialization for each bot. Each turn of the game is listed sequentially (e.g., Turn 1, Turn 2, …), representing the progression of the match. At the end, additional metadata is provided, such as the map seed, the path to the replay file, and final rankings with information about which bot lasted the longest. This structured format allows both human review and automated parsing to analyze bot performance.

\begin{example}[Example of Halite Logs]
\begin{verbatim}
/p1/submission/main.o
/p1/submission/main.o
/p2/submission/main.o
/p2/submission/main.o
34 34
Init Message sent to player 2.
Init Message sent to player 1.
Init Message received from player 1, MyCBot.
Init Message received from player 2, MyCBot.
Turn 1
Turn 2
...
Map seed was 4244905440
Opening a file at /logs/1761005260-4244905440.hlt
Player #1, MyCBot, came in rank #2 and was last alive on frame #340!
Player #2, MyCBot, came in rank #1 and was last alive on frame #340!
\end{verbatim}
\end{example}
\newpage
\subsection{Poker (Husky Hold'em Bench)~\citep{huskyholdem2025}}

Using the Husky Hold'em Bench poker engine, \clash{} supports the standard, No-Limit Texas Hold'em style of poker.
As a refresher, each player gets two private cards.
Five community cards are revealed across four stages, and players bet freely (maximum of stack size) to win chips by making opponents fold or making the best five-card hand.

The poker engine deals blinds (small/big), then runs usual betting rounds -- pre-flop, flop, turn, river -- and enforces the turn order, legal actions (check/call/raise/fold), and pot accounting.
As mentioned, the rules are explicitly \textit{no-limit}, so bets are variable size.
The design of the engine makes implementation of a poker bot straightforward.
A player client simply has to choose actions via a simple interface that lists the valid actions.

\begin{table}[h]
\centering
\begin{minipage}[b]{0.47\textwidth}
\includegraphics[width=\textwidth]{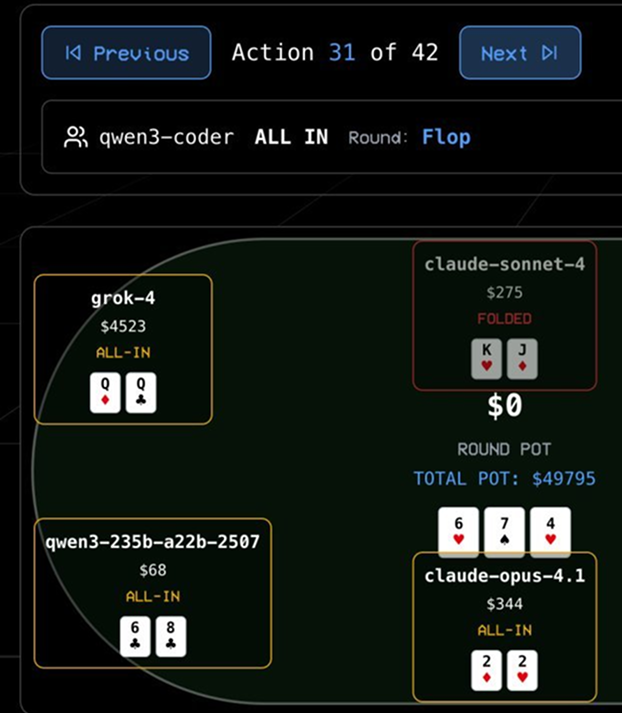}
\captionof{figure}{
Poker (Husky Hold 'Em) screen capture.
Players implement bot that aims to earn the most money across \texttt{n} rounds.
}
\label{fig:arena:poker}
\end{minipage}
\hfill
\begin{minipage}[b]{0.47\textwidth}
\begin{lstlisting}[language=Python, style=mystyle]
class SimplePlayer(Bot):
    def on_start(...):
        # initialize player state

    def on_round_start(...):
        # prepare for new round

    def get_action(...):
        # decide whether to raise, check, or call
        return (PokerAction, amount)

    def on_end_round(...):
        # handle round-end bookkeeping

    def on_end_game(...):
        # handle final results
\end{lstlisting}
\captionof{figure}{A poker bot subclasses \texttt{Bot} and implements lifecycle hooks.
These functions define how the bot initializes, chooses actions during play, and responds at the end of each round and game.}
\label{fig:arena:poker:snippet}
\end{minipage}
\end{table}

\textbf{Isn't poker solved already?} Poker has served as a long standing sandbox for researching superhuman level AI systems.
Simple, constrained variants of poker, such as Heads-Up [No-]Limit Texas Hold’em (2 players, fixed bet sizes) have effectively been solved or close to solved by systems such as Cepheus, Libratus, and Pluribus~\citep{Brown2019SuperhumanAF}.
However, multi-player settings with three or more participants(in other words, \textit{not} Heads-Up, player versus player) are far from solved, as complexity skyrockets with more players.

\textbf{What are effective strategies?} We briefly outline several well-established principles that contribute to the design of strong poker bots, while noting that this overview is not exhaustive given the depth of prior research.
Effective agents often rely on game-theoretic strategies to approximate equilibrium play, ensuring they are difficult to exploit over long horizons.
At the same time, they incorporate opponent modeling and randomization to adapt to behavioral patterns while remaining unpredictable, and use bet-sizing heuristics to balance pressure against risk in pursuit of long-term expected value.

\textbf{What assets are provided in the initial codebase?}
The initial codebase includes a full stack for a poker application: the \texttt{engine/} directory contains the core game logic and simulation framework (deck, hand-evaluation, betting rounds, rules, player abstractions, and state transitions), while the \texttt{client/} directory implements the user interface, sample clients or bots, configuration files (e.g., for game parameters such as blinds, player stacks, seating), and documentation/support files. Together, the codebase provides everything needed to run poker matches, build or plug in client agents or user interfaces, configure game variants, and execute games or simulations.

\textbf{What are the arena configurations?}
The arena in this context represents the virtual poker table managed by the \texttt{pokerden-engine}. Configuration settings define parameters such as the number of seats (players per table), initial chip stacks, blind levels (small and big blinds), betting structure (limit, no-limit, or pot-limit), deck configuration, and game type (e.g., Texas Hold’em, Omaha). These parameters are typically specified in configuration or initialization files that the engine reads at startup, ensuring all clients connect to a consistent game environment. The engine controls turn order, manages rounds (pre-flop, flop, turn, river), and enforces timing or betting limits. In tournament or simulation setups, multiple tables (arenas) may run concurrently with identical rule configurations but independent game states.

\textbf{How is the winner determined?}
Within each hand, the \texttt{pokerden-engine} determines the winner by evaluating all active players’ final hands at showdown using standard poker hand rankings—from high card up to royal flush. If a player causes all others to fold, that player automatically wins the pot without showdown. At showdown, the engine compares hand strengths computed through its hand evaluation module, distributing the pot accordingly (splitting it in case of ties). Over a series of hands or a full match, the overall winner is the player (or client agent) with the largest remaining chip count when the game ends—either after a fixed number of rounds, when all but one player has been eliminated (tournament mode), or when the match duration concludes (cash-game simulation)

\textbf{How are arena logs formatted?}
The poker logs record each hand as a sequence of betting rounds, listing player actions (e.g., raise, call, check) along with bet sizes, updated pot totals, and any side pots.
They also include the community board cards, each player’s hole cards, and timing information for decisions.
At the end of the hand, the logs report chip deltas and final balances, providing both a detailed play-by-play and a clear summary of outcomes.

\begin{example}[Example of Poker Log]
\begin{verbatim}
  "gameId": "8ee11ef4-ffcb-4c42-8ccf-7865a94a3ae5",
  "rounds": {
    "0": {
      "pot": 15,
      "bets": {
        "982465989": 5,
        "3161785489": 10
      },
      "actions": {
        "982465989": "RAISE",
        "3161785489": "RAISE"
      },
      "action_sequence": [
        {
          "player": 982465989,
          "action": "RAISE",
          "amount": 5,
          "timestamp": 1761005394049,
          "pot_after_action": 5,
          "side_pots_after_action": [
            {"amount": 5, "eligible_players": [3161785490, 982465990]}
          ],
          "total_pot_after_action": 5,
          "total_side_pots_after_action": [
            {"id": 0, "amount": 5, "eligible_players": [3161785490, 982465990]}
...
\end{verbatim}
\end{example}

\newpage
\subsection{RoboCode~\citep{hartness2004robocode}}

RoboCode is a $2$+ player game where your code represents a tank in a 2D grid battlefield.
The ultimate objective is to outlast and outscore opposing tanks.

Each tank has a set of actions -- your tank can move around, turn (body, turret, radar), detect other bots, and fire bullets.
There are several factors to take into account when encoding strategy.
First, in addition to a health bar, each tank also has an energy bar that is expended when firing, so players have to be mindful about spamming shooting.
Second, bullets take time to travel, so shots should be directed towards anticipated positions of opposing tanks.
A match continues until only one tank remains standing or the round limit is reached, with scores awarded for survival, damage dealt, and final placement.

\begin{example}[System Prompt Description of RoboCode]
You are a software developer (\{\{player\_id\}\}) competing in a coding game called RoboCode.
Robocode (Tank Royale) is a programming game where your code is the tank: each turn your bot sends intents—speed plus body/gun/radar turn rates and firepower—based on the game state it perceives via radar.
Your program decides how to move, aim, and fire in a deterministic, turn-based arena to outlast other bots.
\end{example}

\begin{table}[h]
\centering
\begin{minipage}[b]{0.47\textwidth}
\includegraphics[width=\textwidth]{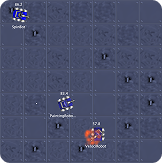}
\captionof{figure}{
RoboCode screen capture.
Your code controls a tank that should outmaneuver and outgun opposing tanks.
}
\label{fig:arena:robocode}
\end{minipage}
\hfill
\begin{minipage}[b]{0.47\textwidth}
\begin{lstlisting}[language=Java, style=mystyle]
package custom;

import robocode.Robot;
import robocode.ScannedRobotEvent;

public class MyTank extends Robot {
    public void run() {
        // main loop: move + scan
        ...
    }

    public void onScannedRobot(ScannedRobotEvent e) {
        // respond to scanned robot
        ...
    }
}
\end{lstlisting}
\captionof{figure}{A RoboCode codebase must implement a core \texttt{run} function, along with \texttt{onScannedRobot} to react to opponents.}
\label{fig:arena:robocode:snippet}
\end{minipage}
\end{table}

\textbf{What are effective strategies?} A key theme to successfully RoboCode bots is \textit{predictive targeting} -- where your tank fires should account for estimations of opponents' future locations, based on their speed and direction.
\textit{Wave surfing} refers to a tactic that assumes opponents' bullets will be directed in a way that mimics ``expanding waves"; movement patterns attempt to minimize the chance of being hit under this assumption.
Maintaining \textit{unpredictable movement}, whether it's true randomness or adaptive strategies mid-game, is key to preventing opponents from exploiting observable repetitions.

\textbf{What assets are provided in the initial codebase?}
The Robocode code-base provides a full environment for developing, running, and visualizing robot battles in Java. The \texttt{battles} directory contains scripts and assets related to running matches and managing gameplay logs, while \texttt{robots} stores precompiled robot programs that serve as examples or test agents. The \texttt{compilers} and \texttt{libs} folders include compiled files and necessary libraries for executing and extending the game’s functionality. The \texttt{config} folder provides configuration files for environment setup, and templates offers starter files to help users design their own robots. Documentation and resources are found in \texttt{javadoc}, \texttt{ReadMe.html}, and \texttt{ReadMe.md}, which describe system components and usage instructions.

\textbf{What are the arena configurations?}
In Robocode, the “arena” is called the battlefield and several configuration parameters can be set. For example, the battlefield's default size is 800 × 600 pixels. You can also specify other sizes with the API (width and height between 400 and 5000). The number of rounds that run in a battle can also be specified. The gun cooling rate is the rate at which a robot’s gun cools after firing (affects how quickly you can fire again). The inactivity time is how many turns a robot can take without action before being penalised for inactivity. The sentry border size defines how far from the edges sentry robots can move. There is also a flag that determines whether enemy robot names are hidden from the bots. Thus, you can configure the “arena” by choosing size, number of rounds, participants, and rule-modifiers

\textbf{How is the winner determined?} In Robocode battles, the winner is determined primarily by the scoring system. At the end of each round, each robot gets a total score, which includes several components: survival score (bonus for each opponent death while you survive), bullet damage done, ram damage done (if you ram an opponent), last-survivor bonus (if you are the final bot alive). In a multi-round battle, the robot (or team) with the highest cumulative score is considered the winner.

\textbf{How are arena logs formatted?}
RoboCode logs summarize the outcome of a set of battles rather than providing turn-by-turn detail.
Each row corresponds to a bot and breaks down its total score into components such as survival points, bonuses, and damage dealt by bullets or ramming.
The logs also record how many times each bot finished in first, second, or third place across the rounds.
Together, this gives a statistical view of performance, highlighting not just who won overall but how they achieved their results.

\begin{example}[Example of RoboCode Logs]
\small 
\begin{verbatim}
Results for 10 rounds
Robot Name       Total Score   Survival   Surv Bonus   Bullet Dmg   Bullet Bonus 
1st: p2.MyTank*  1362 (55%)    300        60           886          116            0  
2nd: p1.MyTank*  1109 (45%)    200        40           768          101            0        
\end{verbatim}
\end{example}

\subsection{RobotRumble~\citep{robotrumble2020}}

RobotRumble is a player-versus-player programming game.
The objective of the competition is quite simple, as summarized on the \href{https://robotrumble.org/}{website}:

\begin{quote}
The rules are simple:
(1) two players fight in a match
(2) robots spawn every 10 turns
(3) a robot can move or attack
(4) each robot has 5 health
(5) the player with more robots after 100 turns wins
\end{quote}

To summarize, RobotRumble is a game that emphasizes the ability to position units effectively and coordinate teams of units to focus on enemy at a time (e.g., if $5$ units attack an opposing unit, it takes $1$ turn to knock out the unit).

\begin{example}[System Prompt Description of RobotRumble]
You are a software developer (\{\{player\_id\}\}) competing in a coding game called RobotRumble.
RobotRumble is a turn-based coding battle where you program a team of robots in Python to move, attack, and outmaneuver your opponent on a grid.
Every decision is driven by your code, and victory comes from crafting logic that positions robots smartly, times attacks well, and adapts over the 100-turn match.
\end{example}

\begin{table}[h]
\centering
\begin{minipage}[b]{0.47\textwidth}
\includegraphics[width=\textwidth]{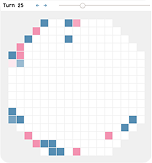}
\captionof{figure}{
RobotRumble screen capture.
Your code controls a tank that should outmaneuver and outgun opposing tanks.
}
\label{fig:arena:robotrumble}
\end{minipage}
\hfill
\begin{minipage}[b]{0.47\textwidth}
\begin{lstlisting}[language=Python, style=mystyle]
def robot(state, unit):
    # Decide what this unit should do on its turn.
    # Possible actions include:
    #   - Moving in one of the cardinal directions
    #   - Attacking in a direction
    #   - Gathering or interacting with resources
    #   - Defending or waiting (no-op)
    # The decision can depend on:
    #   - Current turn number (e.g., alternate strategies)
    #   - Unit type or role (soldier, builder, etc.)
    #   - Nearby enemies, allies, or map features
    ...
\end{lstlisting}
\captionof{figure}{In RobotRumble, players' code must implement a \texttt{robot(state, unit)} function that returns an action each turn.}
\label{fig:arena:robotrumble:snippet}
\end{minipage}
\end{table}

\textbf{What are effective strategies?}
First, \textit{avoid getting purged from spawn} by timing your exits — since up to four new robots appear every 10 turns and anything left in spawn is deleted, strong bots step out just before the purge to keep their full roster in play.
Next, take advantage of \textit{movement conflict priority} — when two robots move into the same square, the winner is decided by a fixed clockwise rule, so careful bots choose their approach direction to gain the upper hand.
Finally, practice \textit{focus fire while avoiding friendly fire}: attacks only deal 1 damage but can hit teammates, so good bots coordinate multiple robots to bring down a 5-HP enemy in one turn without accidentally shooting their own.

\textbf{How are arena logs formatted?}
RobotRumble logs are displayed as a sequence of ASCII grids (a total of $100$ grids per simulation), with numbers marking robot positions and empty cells showing open space.
After each turn, the grid is updated to show new movements, clashes, or unit spawns, giving a clear visual trace of how the battle unfolds.
Below each grid, a summary line shows each player’s remaining health and unit counts.

\textbf{What assets are provided in the initial codebase?}
The initial codebase includes a command-line interface (CLI) tool (\texttt{rumblebot}) that allows users to execute battles between bots directly in the terminal or in a web-based graphical viewer.
The repository also includes example “builtin bots” that can be used as opponents or templates for developing new robots.
Additionally, the repo contains logic scripts and documentation for running matches, viewing results, and managing robot files within the filesystem.

\textbf{What are the arena configurations?}
The arena configuration determines the battle environment—typically a rectangular map with fixed dimensions, where robots spawn in random or defined positions.
Each robot operates in discrete turns, executing movement and attack commands according to its programmed logic.
The arena setup remains consistent across matches to ensure fairness.

\textbf{How is the winner determined?}
The winner in Robot Rumble is the last surviving team at the end of a match.
Robots can deplete each other’s health using attacks while avoiding incoming fire.
If multiple robots remain when the time limit or round limit is reached, the winner is decided based on performance metrics such as remaining health or damage dealt.

\begin{example}[Example of RoboCode logs]
\begin{verbatim}
{"winner": "Red", "turns": [ {"state": { "objs": {
    "1": {"id": "1", "coords": [0,0], "obj_type": "Terrain", "type": "Wall"},
    "2": {"id": "2", "coords": [0,1], "obj_type": "Terrain", "type": "Wall"},
    "3": {"id": "3", "coords": [0,2], "obj_type": "Terrain", "type": "Wall"},
...
\end{verbatim}
\end{example}

\newpage
\newpage
\section{Evaluation}
\label{appx:evaluation}

In this section, we provide additional details about our evaluation procedure, including inference services, \texttt{mini-SWE-agent} configurations, arena-specific prompts, and formulae for calculating win rate and Elo scores.

\subsection{\texttt{mini-SWE-agent} Configuration}
\label{appx:evaluation:mini-config}

The \texttt{mini-SWE-agent} ACI allows one to define a number of configurations\footnote{\url{https://mini-swe-agent.com/latest/advanced/global_configuration/}}.
We highlight a couple of configuration settings relevant to the evaluation set up for \clash{}.

\textbf{Turn and cost limits.} For the \textit{edit} phase of each round, the LM is constrained to at most $30$ interactive turns with the codebase.
We also impose a \$$1$ cost limit, meaning once the running cost of input and output tokens for a single round exceeds \$$1$, the editing episode is automatically terminated.
Consequently, this means that for a tournament of \texttt{n} rounds, at most \$\texttt{n} are spent per player.
We enforce this cost limit not only to keep expenses manageable but also to discourage degenerate behaviors such as the model dumping entire files into its context, repeatedly echoing large outputs, or otherwise flooding the interaction buffer with irrelevant information.
Generally, the limit forces the agent to allocate its context budget carefully, encouraging concise reasoning and selective use of code.
We set the \texttt{mini-SWE-agent} configuration to the following values to enforce these practices:
\begin{itemize}
    \item The \texttt{step\_limit} is set to $30$. The \texttt{cost\_limit} is set to $1$.
    \item In the \texttt{action\_observation\_template}, a prompt template that environment observations are interpolated into, the agent is reminded of the number of turns and cost consumed with the line:
    \quote{
    $<$limit\_note$>$This is the output of step \{\{n\_model\_calls\}\} (\{\{step\_limit\}\} limit). You've used \{\{model\_cost $\vert$ round(2)\}\} USD (\{\{cost\_limit\}\} USD limit).$<$\/limit\_note$>$
    }
\end{itemize}
We observe in practice that the cost limit is almost never reached.
On the other hand, turn limits are exhausted frequently for specific models.

\textbf{Setting the context.} The system prompt briefly sets the context and informs the model of the general nature of the setting it's operating in.
Here is the prompt verbatim:

\begin{example}[System Prompt.]
  You are a helpful assistant interacting continuously with a computer by submitting commands.
  You'll be editing a codebase to play a programming game.
\vspace{1em}

  $<$important$>$
  This is an interactive process where you will think and issue ONE command, see its result, then think and issue your next command.
  $<$/important$>$

\vspace{1em}
  Your response must contain exactly ONE bash code block with ONE command (or commands connected with \&\& or $\vert\vert$).
  Include a THOUGHT section before your command where you explain your reasoning process.
  Format your response as shown in $<$format\_example$>$.

\vspace{1em}
  $<$format\_example$>$
  Your reasoning and analysis here. Explain why you want to perform the action. \\
  ```bash \\
  your\_command\_here \\
  ``` \\
  $<$/format\_example$>$ \\

  Failure to follow these rules will cause your response to be rejected.
\end{example}

The LM is informed it is acting in the role of a software developer with the ability to investigate and edit a codebase across multiple turns.
The prompt clearly delineates an interaction protocol.
Every turn, the model should be explaining its reasoning in a ``Thought" section, followed by a \texttt{bash} code block~\citep{yang2023intercodestandardizingbenchmarkinginteractive}.

\textbf{Describing the arena and tournament.}
After the system prompt, the next message given to the LM briefly describes the arena and thoroughly reviews how the LM can interact with the codebase environment correctly.
We first show the arena description:

\begin{example}[Subsection of initial message describing the arena]
  \#\# Game Description \\

  \{\{game\_description\}\} \\

  \#\# General tips about how to play the game \\

  The details of the game are fully available within this codebase. \\
  - `docs/`: Game documentation \\
  - `logs/`: Past rounds and outcomes \\
  - `trajs/`: History of your edits \\
  - and a lot more. It's up to you to explore and utilize these resources. \\

  The game is played in rounds and you will be evaluated on the performance over all the rounds. You won't remember past rounds. \\

  In every round, you have a limit of \{\{step\_limit\}\} steps and a cost limit of \{\{cost\_limit\}\} dollars.
  We will show you the number of steps and cost used so far after every response in the `$<$limit\_note$>$` tag.
  After you've reached the step or cost limit, you cannot continue working on this task, and we will play the game with your codebase.
  This means that it's fine to reach the step or cost limit while working on documentation or testing, but you shouldn't
  reach the limit while working on the actual game logic to avoid submitting an invalid codebase. \\

  So if you want to carry knowledge forward — leave tools, notes, or strategies in the codebase.
  Good documentation means you (and others) can pick up right where you left off. \\

  If you'd hate to repeat a step next round, encode it now — as a script, a note, or a tool. \\

  Improve the bot however you like — experiment, document, iterate. Some ideas: \\
  - Build analysis tools \\
  - Create bot variants to test \\
  - Track strategies across rounds \\
  How you choose to evolve and document is up to you. Good luck!
\end{example}

The actual description of the arena, represented by \texttt{game\_description}, is brief.
These are filled in by the system templates show in the arena cards of \S\ref{appx:arenas}.
This lack of detail is intentional.
We impose the burden of understanding how exactly an arena works.
With full access to documentation and logs in the codebase, \clash{} forces LMs to identify and fill in gaps about its understanding of the game.
This obstacle is realistic.
As prior work around coding evaluations has demonstration, real world software issues are often ambiguous and abstract on face value~\citep{openai2024swebenchverified}.
\clash{} enables investigating whether models can address such uncertainty by placing it in a setting where information is available, but not immediately obvious.

The second half of the prompt states the available assets, then reminds the model of both the step/cost limit along with the transient nature of its memory.
The model is explicitly informed that its working memory is \textit{not} retained across rounds, so it is encouraged to use the codebase to maintain long-term information, tools, and general progress.
Collectively, the prompt incorporates the challenges discussed in Section~\ref{sec:codeclash:features}.

Next, the prompt provides a deep dive into how the model should go about issuing actions.
As a reminder, \texttt{mini-SWE-agent}'s interaction is completely terminal driven.

\begin{example}[Subsection of initial message describing interaction]
  \#\# Command Execution Rules \\

  You are operating in an environment where \\

  1. You write a single bash command \\
  2. The system executes that command in a subshell \\
  3. You see the result \\
  4. You write your next command \\

  For each of your response: \\

  1. Include a THOUGHT section explaining your reasoning and what you're trying to accomplish \\
  2. Provide exactly ONE bash command to execute \\
  3. The action must be enclosed in triple backticks (see below for formatting rules) \\
  3. Directory or environment variable changes are not persistent. Every action is executed in a new subshell.
      However, you can prefix any action with \texttt{MY\_ENV\_VAR=MY\_VALUE cd /path/to/working/dir \&\& ...} or write/load environment variables from files \\

  Format your responses like this: \\

  $<$format\_example$>$ \\
  THOUGHT: Here I explain my reasoning process, analysis of the current situation, and what I'm trying to accomplish with the command below. \\

\texttt{
  ```bash \\
  your\_command\_here \\
  ``` \\
}
  $<$/format\_example$>$ \\

  Commands must be specified in a single bash code block: \\
\texttt{
  ```bash \\
  your\_command\_here \\
  ``` \\
}

  **CRITICAL REQUIREMENTS:** \\
  - Your response SHOULD include a THOUGHT section explaining your reasoning \\
  - Your response MUST include EXACTLY ONE bash code block \\
  - This bash block MUST contain EXACTLY ONE command (or a set of commands connected with \&\& or $\vert\vert$) \\
  - If you include zero or multiple bash blocks, or no command at all, YOUR RESPONSE WILL FAIL \\
  - Do NOT try to run multiple independent commands in separate blocks in one response \\
  - Directory or environment variable changes are not persistent. Every action is executed in a new subshell. \\
  - However, you can prefix any action with \texttt{MY\_ENV\_VAR=MY\_VALUE cd /path/to/dir \&\& ...} or write/load environ variables from files \\
\end{example}

We omit the examples of proper, well-formed interactions following this prompt.
The examples include actions such as how to edit a file with \texttt{sed}, performing searches of the codebase with \texttt{grep} and \texttt{find}, and viewing specific parts of files with \texttt{nl}.
We observe both with this work and prior evaluations~\citep{jimenez2024swebenchlanguagemodelsresolve} that including such in-context demonstrations is meaningfully helpful to reducing the errant actions issued by a model.
All players' codebases are initialized with no tools provided upfront.
However, throughout the course of a tournament, models are free to synthesize their own scripts and aliases.

\textbf{Errant action handling.}
Last but not least, in the case that a model does issue an invalid action, we inherit the guardrail and error handling principles described in~\cite{yang2024sweagentagentcomputerinterfacesenable} and inform the model of such errors.
The \texttt{format\_error\_template} is shown when the model's response does not abide by the ReAct style form factor requested, and the following error message is displayed:

\begin{example}[Format error template]
  Please always provide EXACTLY ONE action in triple backticks, found \{\{actions$\vert$length\}\} actions.
  If you want to end the task, please issue the following command: \texttt{echo COMPLETE\_TASK\_AND\_SUBMIT\_FINAL\_OUTPUT}
  without Any other command.
  Else, please format your response exactly as follows: \\

  $<$response\_example$>$ \\
  Here are some thoughts about why you want to perform the action. \\

\texttt{
  ```bash \\
  $<$action$>$ \\
  ``` \\
  $<$/response\_example$>$ \\
}

  Note: In rare cases, if you need to reference a similar format in your command, you might have
  to proceed in two steps, first writing \texttt{TRIPLEBACKTICKSBASH}, then replacing them with \texttt{```bash}.
\end{example}

Note that the error template is \textit{not} thrown if the action itself is problematic or executes with a non-zero return code.
This message is only invoked when the model's response doesn't abide by the expected format, and it does not account for any syntax issues or execution outcomes related to the \texttt{action} itself.

\subsection{Tournament Configuration}
\label{appx:evaluation:tournament-config}

In addition to configuring interaction, we also allow users to set tournament settings, such as game mechanics and rounds, via a configurable \texttt{.yaml} file as well.

\begin{example}[Tournament configuration file for Battlesnake]
tournament: \\
\null\quad rounds: 25 \\
game: \\
\null\quad name: BattleSnake \\
\null\quad sims\_per\_round: 1000 \\
\null\quad args: \\
\null\quad\quad width: 11 \\
\null\quad\quad height: 11 \\
\null\quad\quad browser: false \\
\end{example}

The configuration file contains two sections.
The \texttt{tournament} field allows one to specify how many \texttt{rounds} the tournament will be played.
The \texttt{game} field indicates which code arena the tournament is being played in.
\texttt{sims\_per\_round} is the number of simulations run per round in order to determine a winner (usually $1000$).
For most games, a simulation is run by calling an executable or script with arguments.
The \texttt{args} field is a way to pass in flags to that executable to adjust the configurations of the arena.
For instance, in the above example, the \texttt{args} are eventually interpolated into the following command to run the game: \texttt{python main.py --width 11 --height 11 --browser false}.

\begin{example}[Player configuration section]
players: \\
\null\quad- agent: mini \\
\null\quad\quad name: p1 \\
\null\quad\quad config: \\
\null\quad\quad\quad agent: !include mini/default.yaml \\
\null\quad\quad\quad model: \\
\null\quad\quad\quad\quad model\_name: openai/gpt-5-mini \\
\null\quad- agent: mini \\
\null\quad\quad name: p1 \\
\null\quad\quad config: \\
\null\quad\quad\quad agent: !include mini/default.yaml \\
\null\quad\quad\quad model: \\
\null\quad\quad\quad\quad model\_name: anthropic/claude-sonnet-4-20250514 \\
\end{example}

The player configuration is simple, essentially serving as a meta-configuration for creating each player as an LM along with a \texttt{mini-SWE-agent} configuration.
Using this configuration, it is possible to equip models with different prompts by swapping out the \texttt{mini-SWE-agent} configuration (\texttt{!include mini/default.yaml}), although we do not do this for our main leaderboard and results unless specified as otherwise.

\textbf{Number of rounds run.} To determine the number of tournaments and rounds to run to obtain a statistically meaningful leaderboard, we identify several parameters.

\begin{itemize}
    \item \textit{M} for the number of models to evaluate.
    \item \textit{A} for the number of arenas we want models to compete in.
    \item \textit{T} for the number of tournaments we run per arena.
    \item \textit{P} for the number of players per tournament.
    \item \textit{R} for the number of rounds per tournament.
\end{itemize}

Given these values, we can generally calculate the number of rounds that would be run with $\binom{M}{P} \times A \times T \times R$.
This assures us that each model is run against other models on the same set of arenas for the same number of total rounds ($T \times R$).
The main results table reflects values of $M=9$; $A=6$; $T=10$; $P=2$; $R=15$, giving us a total of $32$,$400$ total rounds run, with each model playing a total $\binom{M-1}{P-1} \times A \times T \times R = 7200$ rounds.
For the Section~\ref{sec:results:ablations} evaluation with $3+$ players, we use the same calculation to determine number of tournaments to run.




\subsection{Evaluation Metrics}
This section contains detail on the evaluation metrics, in particular the Elo ratings for each model.
Detailed statistical analysis shows that the ranking is stable.
For example, the pairwise order agreement of our ranking is more then 98\% in bootstrapping experiments.

\subsubsection{Definitions}
\label{appx:evaluation:metrics}
\textbf{Tournaments} are a sequence of 15 rounds played in one arena between two or more models.

\begin{figure}
\includegraphics[width=\linewidth]{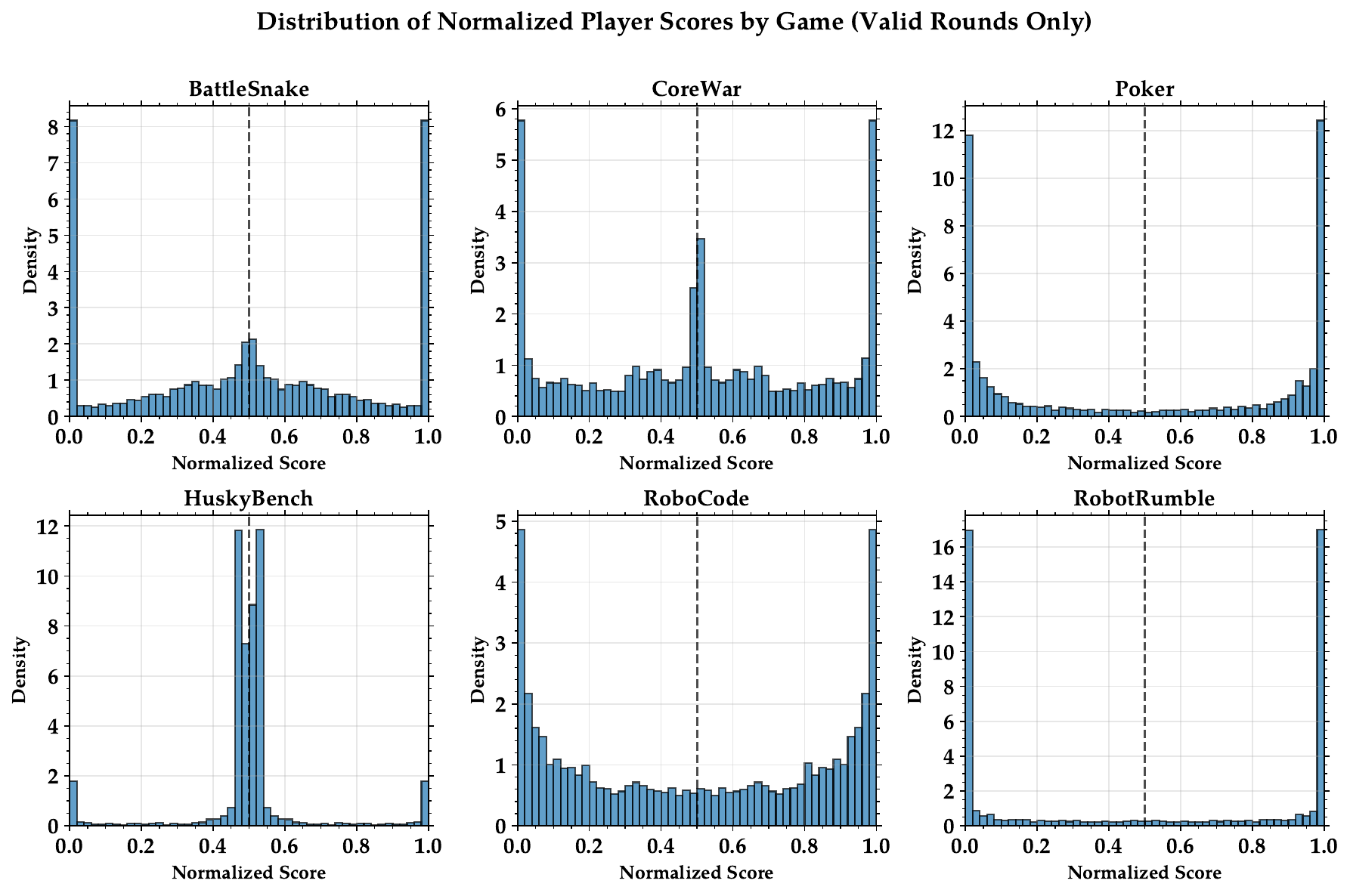}
\caption{Distribution of rounds scores by game.}
\label{fig:round_score_distribution}
\end{figure}

\begin{figure}
\includegraphics[width=\linewidth]{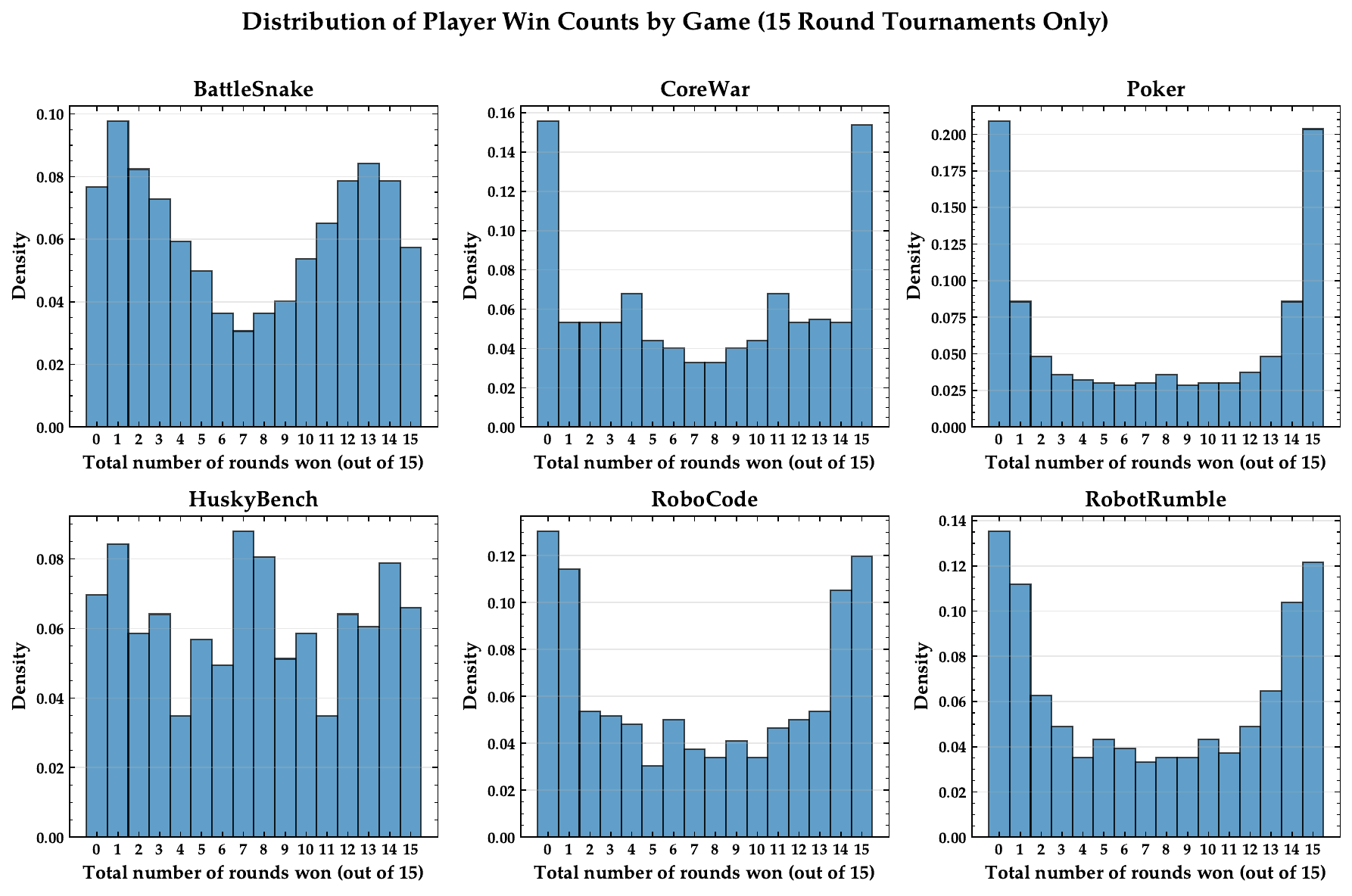}.
\caption{Distribution of the number of rounds won by the players across arenas. The non-uniform distributions demonstrate that the rounds are not independent of each other. }
\label{fig:win_count_distributions}
\end{figure}

\textbf{Winning a round.} 
A round consists of one or more repetition of an arena between the submissions of different models.
A round is won by a model if any of the following applies
\begin{enumerate}
\item The model is the only one with a valid submission (for example because the other model's submission does not compile or execute)
\item The model scores higher than all others. Scores a typically either win rates (across all repetitions of the arena), or other aggregate quantities (e.g., total amount of money won in poker).  
\end{enumerate}
Distributions of round scores for different arenas are shown in Figure~\ref{fig:round_score_distribution}.
Because of the sequential nature of a tournament, the scores of the rounds are not independent of each other.
This is shown in Figure~\ref{fig:win_count_distributions}: If all rounds were independent, a uniform distribution would be expected. However, most games show a heavily bimodal distribution instead. 

\textbf{Winning a tournament}
A tournament is won by the model that wins more rounds than its opponent, or, if both models win equally many rounds, by the model that scores the last win.
If all rounds of the tournament are draws, then the tournament is a draw (an extremely rare occurrence, less than once per 1000 tournaments).

\textbf{Win rate} per model is the fraction of tournaments won. 
This metric can be further stratified into arena and opponent-specific percentages.

\textbf{Elo rating.} 
We quantify absolute model strengths by Elo ratings.

\begin{figure}
    \includegraphics[width=\linewidth]{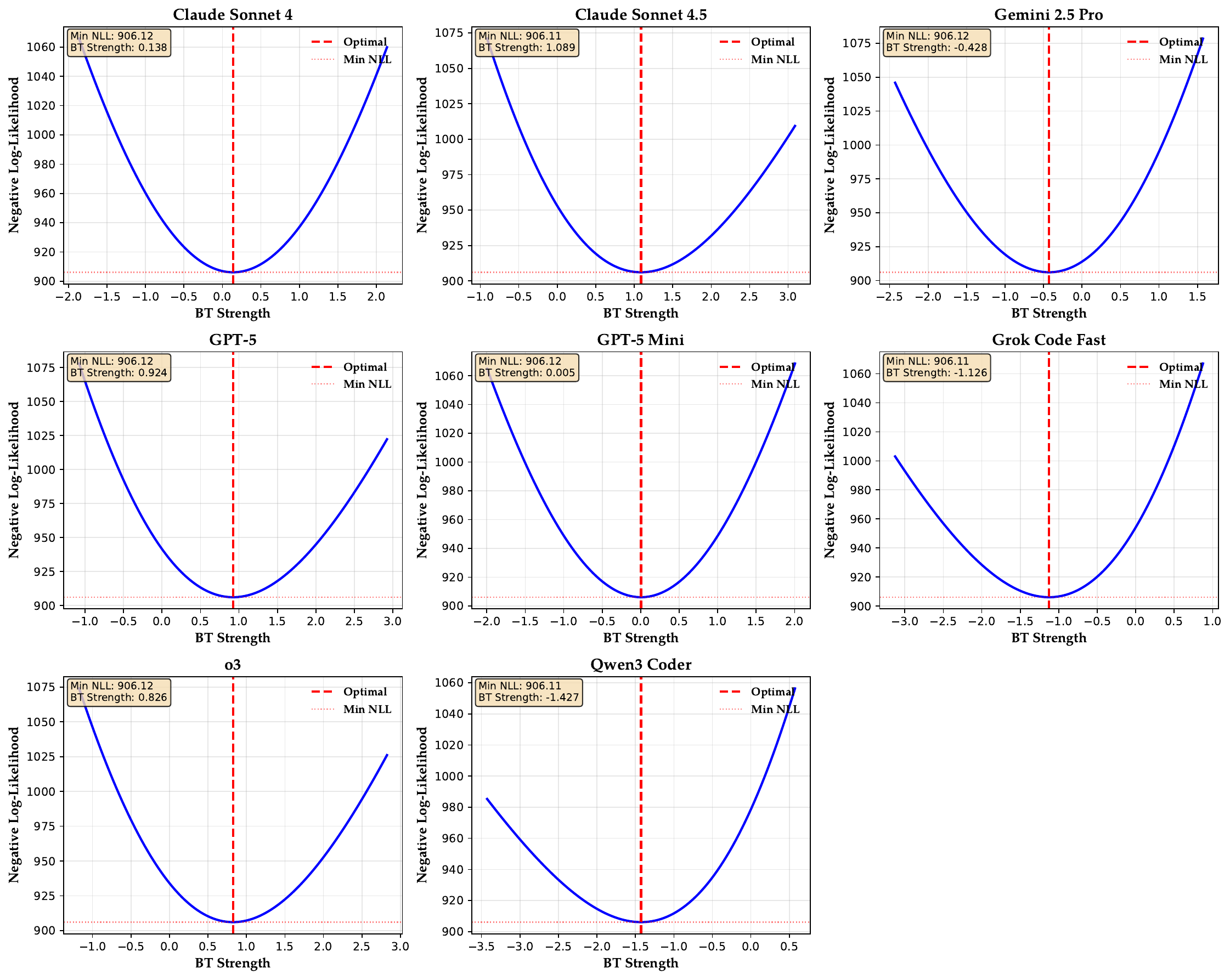}
    \caption{Log likelihood profiles for a fit to all arenas results.}
    \label{fig:elo:ll_fit_validation_plot}
\end{figure}

Elo ratings are based on the Bradley-Terry model~\citep{bradley1952rank} that models win probabilities between two players $i$ and $j$ with strengths $s_i$ and $s_j$ via logistic regression of the strength difference $s_i-s_j$, i.e.,
\begin{align*}
P(\text{model $i$ wins over $j$}) = \frac 1{1+\exp(s_i-s_i')} = 
\sigma(s_i-s_i').
\end{align*}
Repetitions of independent games are Bernoulli-distributed and
the optimal values of $s_i$ and $s_j$ can be calculated using a maximum likelihood fit to the win numbers $w_{ij}$ (number of times $i$ won over $j$), i.e.,
\begin{align}
    \log \mathcal L=\sum_{i<j}\Big[w_{ij}\log \sigma(s_i-s_j)+w_{ji}\log \sigma(s_j-s_i)\Big].
    \label{eq:mle_bt}
\end{align}
However, this leaves a gauge freedom in the strengths $s_i$, because all $s_i$ can be shifted by a constant factor $s_i \rightarrow s_i + S$ without changing the value of $\mathcal L$.
To fully constrain the fit, we choose $\sum_i s_i=0$.
This choice only results in a fixed offset for the final Elo scores.
Log likelihood profiles for a fit to all arenas are found in Figure~\ref{fig:elo:ll_fit_validation_plot}.

The player strengths can be converted to Elo scores $R_i$ as
\begin{align}
R_i = R_0 + \frac \beta{\log 10}s_i,
\label{eq:conversion_si_ri}
\end{align}
Following the conventions from Chess, we choose a starting Elo of $R_0=1200$ and a slope of $\beta=400$.
Note that this convention is merely a presentation choice that affects readability, not the model predictions (unlike the $K$ factor that is used in sequential calculation of Elo scores).
\subsubsection{Statistical uncertainties}
The covariance matrix $\Sigma$ of the player strengths $s_i$ is given by the inverse of the Hessian matrix of $\log \mathcal L$.
Setting $p_{ij}=\sigma(s_i-s_j)$ and $n_{ij} = w_{ij} + w_{ji}$, the Hessian of $\mathcal L$ is given by
\begin{align*}
H_{ij} = 
\frac{\partial^2 \log\mathcal L}{\partial s_i\,\partial s_j} =
-\sum_{i<j} n_{ij}p_{ij}(1-p_{ij})
\begin{cases}
    1 & i=j,\\
    -1 & i\neq j.
\end{cases}
\end{align*}
However, this Hessian is singular, due to the above mentioned shift-invariance.
So we invert $H$ in the constrained subspace of our gauge, $\mathcal S=\{s_i\ |\ \sum_i s_i=0\}$, i.e., calculate the covariance $\Sigma$ as
\begin{align*}
\Sigma = Z (Z^T H Z)^{-1} Z^T,
\end{align*}
where $Z$ projects onto $\mathcal S$ and is given by
\begin{align*}
Z_{ij} = \begin{cases}
1 - \frac 1n & i=j,\\
-\frac 1n & i\neq j.
\end{cases}
\end{align*}
The variance of $s_i$ is then given by $\mathrm{Var}\, s_i = \Sigma_{ii}$ and can readily be scaled to the variance on $R_i$ via \eqref{eq:conversion_si_ri}.
The uncertainties of the final results are shown in Table~\ref{tab:elo_ratings_uncertainties}.
\begin{table}[t]
\centering
{\scriptsize
\begin{tabular}{lrrrrrrr}
\toprule
Model & BattleSnake & CoreWar & Halite & Poker & RoboCode & RobotRumble & All \\
\midrule
Claude Sonnet 4.5 & $1470 \pm 52$ & $1641 \pm 73$ & $1408 \pm 50$ & $1248 \pm 44$ & $1361 \pm 43$ & $1423 \pm 47$ & $\mathbf{1389 \pm 18}$ \\
GPT-5 & $1339 \pm 44$ & $1199 \pm 43$ & $1522 \pm 56$ & $1599 \pm 64$ & $1409 \pm 46$ & $1293 \pm 41$ & $\mathbf{1360 \pm 17}$ \\
o3 & $1357 \pm 45$ & $1348 \pm 47$ & $1576 \pm 60$ & $1277 \pm 46$ & $1338 \pm 43$ & $1309 \pm 42$ & $\mathbf{1343 \pm 17}$ \\
Claude Sonnet 4 & $1253 \pm 45$ & $1339 \pm 46$ & $1111 \pm 48$ & $1233 \pm 44$ & $1033 \pm 45$ & $1361 \pm 43$ & $\mathbf{1223 \pm 16}$ \\
GPT-5 Mini & $1369 \pm 45$ & $926 \pm 50$ & $1185 \pm 47$ & $1429 \pm 50$ & $1217 \pm 41$ & $1092 \pm 41$ & $\mathbf{1200 \pm 16}$ \\
Gemini 2.5 Pro & $1115 \pm 45$ & $1043 \pm 45$ & $1186 \pm 47$ & $978 \pm 48$ & $1315 \pm 42$ & $1044 \pm 44$ & $\mathbf{1125 \pm 16}$ \\
Grok Code Fast & $833 \pm 63$ & $1170 \pm 43$ & $824 \pm 63$ & $886 \pm 54$ & $1033 \pm 45$ & $1016 \pm 46$ & $\mathbf{1004 \pm 18}$ \\
Qwen3 Coder & $860 \pm 59$ & $929 \pm 51$ & $784 \pm 67$ & $945 \pm 53$ & $890 \pm 55$ & $1057 \pm 43$ & $\mathbf{952 \pm 20}$ \\
\bottomrule
\end{tabular}
}
\caption{ELO ratings with uncertainties}
\label{tab:elo_ratings_uncertainties}
\end{table}

\subsubsection{Statistical validation and rank stability}
\begin{figure}
    \includegraphics[width=0.5\linewidth]{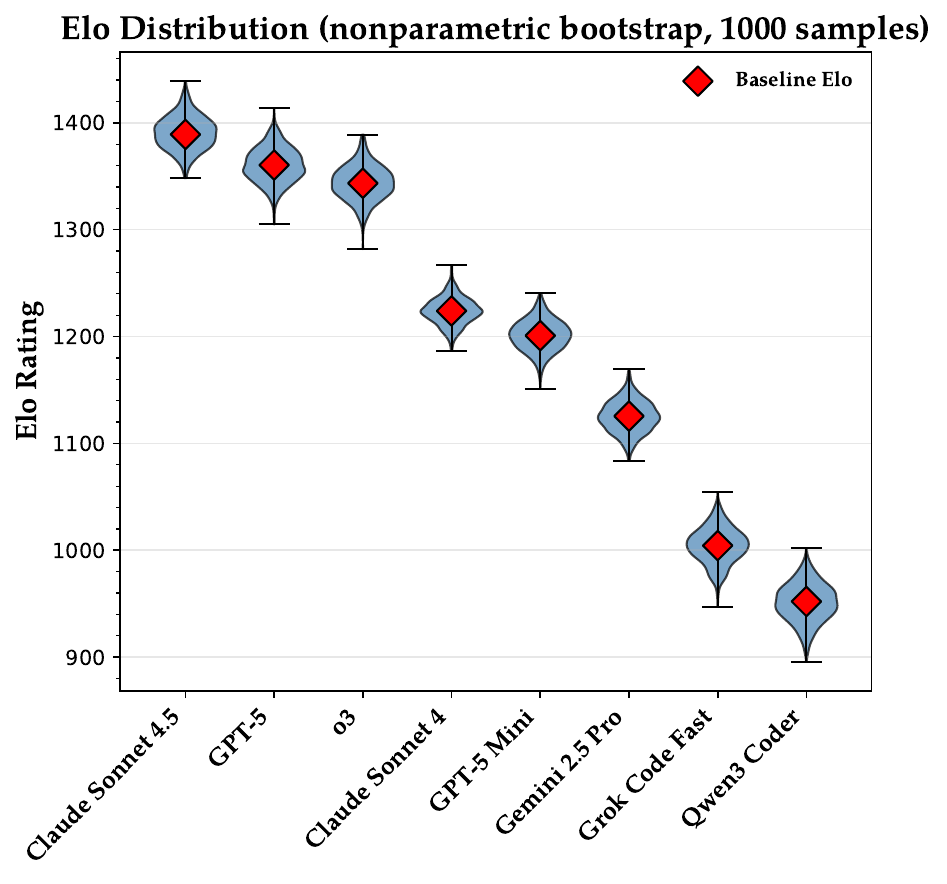}
    \includegraphics[width=0.5\linewidth]{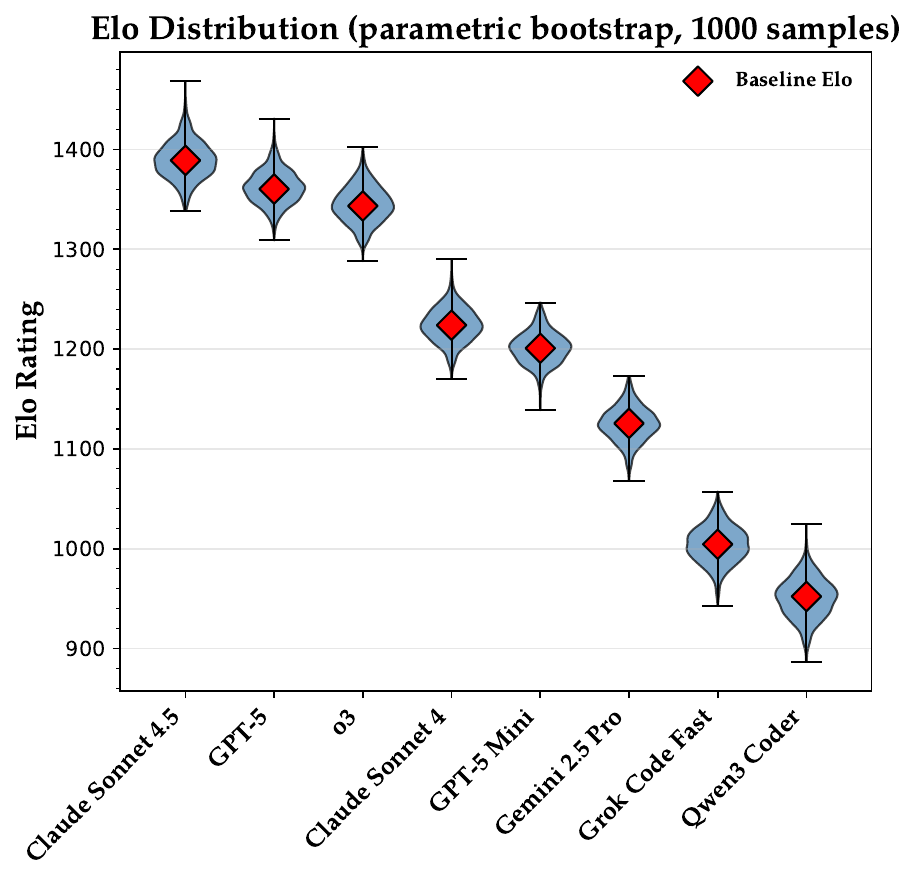}
    \caption{Distribution of Elo scores from non-parametric and parametric bootstrapping}
    \label{fig:elo:bootstrap_elos}
\end{figure}

\begin{figure}
    \includegraphics[width=0.5\linewidth]{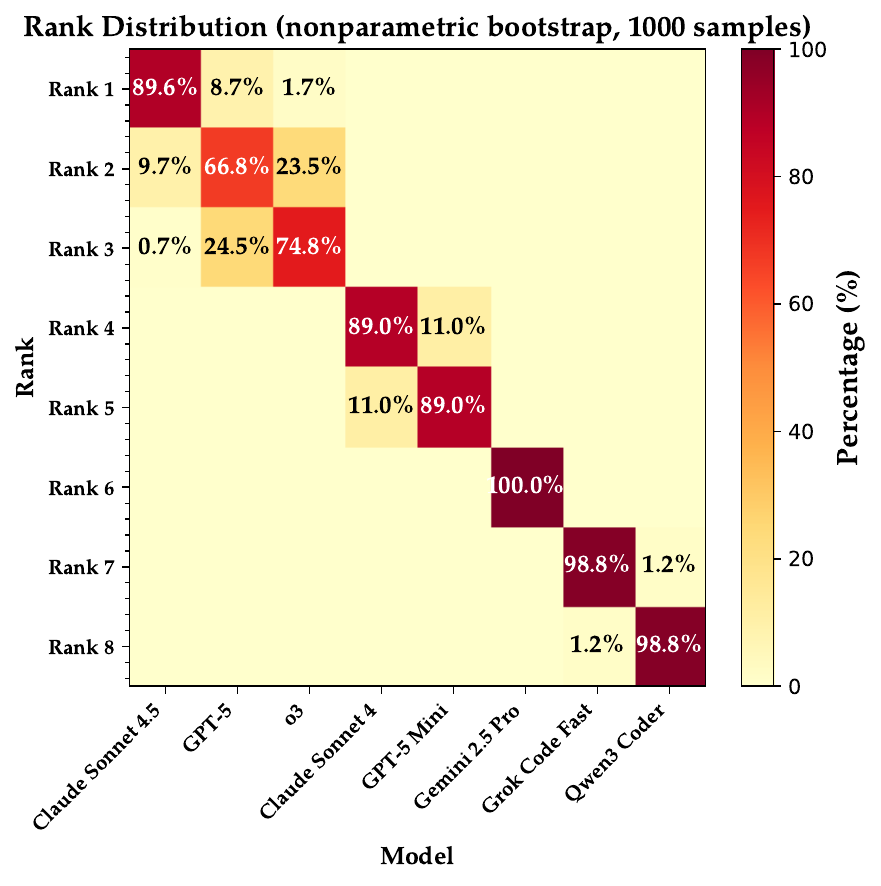}
    \includegraphics[width=0.5\linewidth]{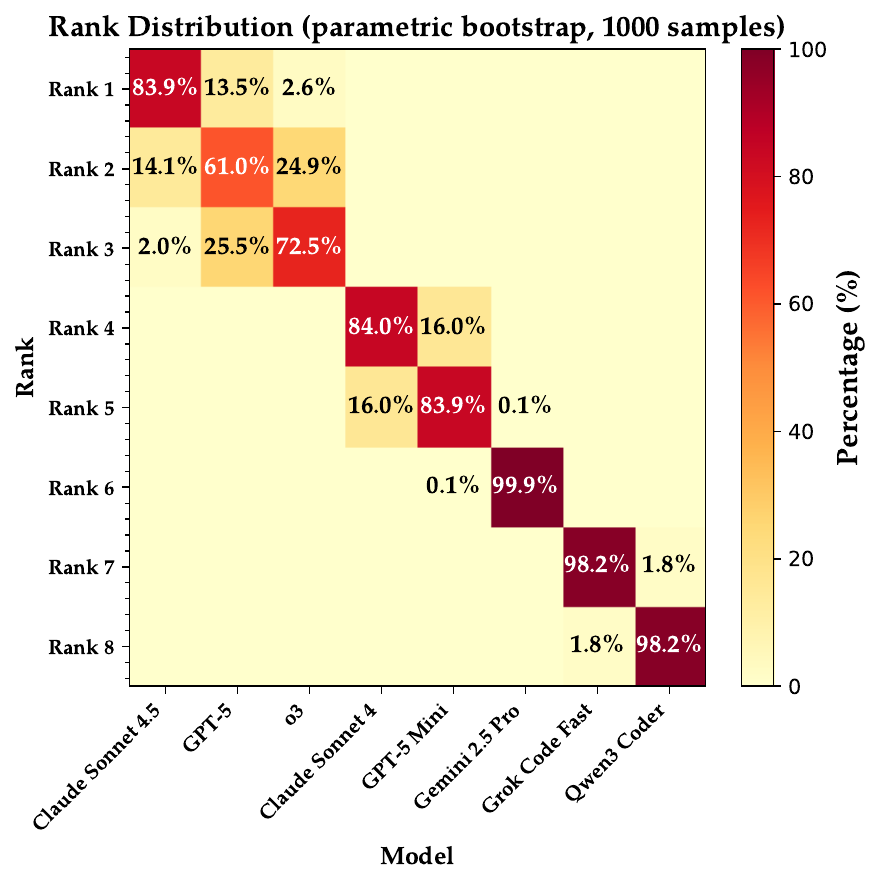}
    \caption{Elo-based ranks from non-parametric and parametric bootstrapping}
    \label{fig:elo:bootstrap_ranks}
\end{figure}

We perform non-parametric and parametric bootstrapping experiments to test the stability of the ranking. 
Distribution of bootstrapped Elo scores are shown in Figure~\ref{fig:elo:bootstrap_elos}, and the resulting distribution of ranks are shown in Figure~\ref{fig:elo:bootstrap_ranks}.
The statistical uncertainties derived from the bootstrapped Elo results agree well with those calculated from the Hessian matrix in Table~\ref{tab:elo_ratings_uncertainties}.
Various rank stability metrics are shown in Table~\ref{tab:rank_stability_bootstrapping}.
In particular, we'd like to highlight that the pairwise order agreement of our ranking is $98$\%.

\textbf{Non-parametric bootstrapping}
We perform a non-parametric bootstrapping experiment by sampling with replacement from all tournaments. 
This results in new win counts $w_{ij}$ from which we can calculate new Elo rankings $R_i$.
We draw 1000 samples and calculate rank stability metrics and uncertainties based on the 1000 corresponding Elo rankings.

\textbf{Parametric bootstrapping}
We generate bootstrap replicas from the fitted Bradley–Terry model, i.e., we use the Bradley-Terry player strengths $\hat s_i$ that maximize \eqref{eq:mle_bt} and assume win probabilities
\begin{align*}
p^{\star}_{ij}=\sigma(\hat s_i-\hat s_j).
\end{align*}
For each observed matchup $(i,j)$ with $n_{ij}=w_{ij}+w_{ji}$ total games, we then draw
\[
\tilde w_{ij}\sim \mathrm{Binomial}(n_{ij},\,p^{\star}_{ij}),\qquad
\tilde w_{ji}=n_{ij}-\tilde w_{ij}.
\]
This preserves the observed matchup graph and game counts while sampling outcomes according to the fitted model.
From each resampled win matrix we refit the Bradley–Terry model (and convert to Elo via \eqref{eq:conversion_si_ri}) and assess variability of scores and ranks across 1000 replicas.

\begin{table}[h]
\centering
\begin{tabular}{lcc}
\toprule
Metric & Nonparametric & Parametric \\
\midrule
Kendall's $\tau$ & 0.966 & 0.956 \\
Spearman's $\rho$ & 0.988 & 0.984 \\
Footrule (normalized) & 0.030 & 0.038 \\
Top-1 consistency & 0.896 & 0.839 \\
Pairwise order agreement & 0.983 & 0.978 \\
\bottomrule
\end{tabular}
\caption{Rank stability metrics of the Elo-based ranking of LMs over all arenas based on bootstrapping experiments}
\label{tab:rank_stability_bootstrapping}
\end{table}

\newpage
\section{Extended Results}
\label{appx:results}

In this section, we present additional analyses and findings not presented in Section~\ref{sec:results}.
These insights further characterize model behavior and performance in the \clash{} setting.

\subsection{Interaction Trends}
\label{appx:results:interaction-trends}
We provide additional analyses and visualizations revealing trends in how different models interact with their codebase environment, such as how many steps they take per round, the size and frequency of their edits, and their length of their thoughts.

\begin{figure}[h]
\begin{minipage}[t]{0.49\textwidth}
    \setlength{\abovecaptionskip}{0em}
    \centering
    \includegraphics[width=\textwidth]{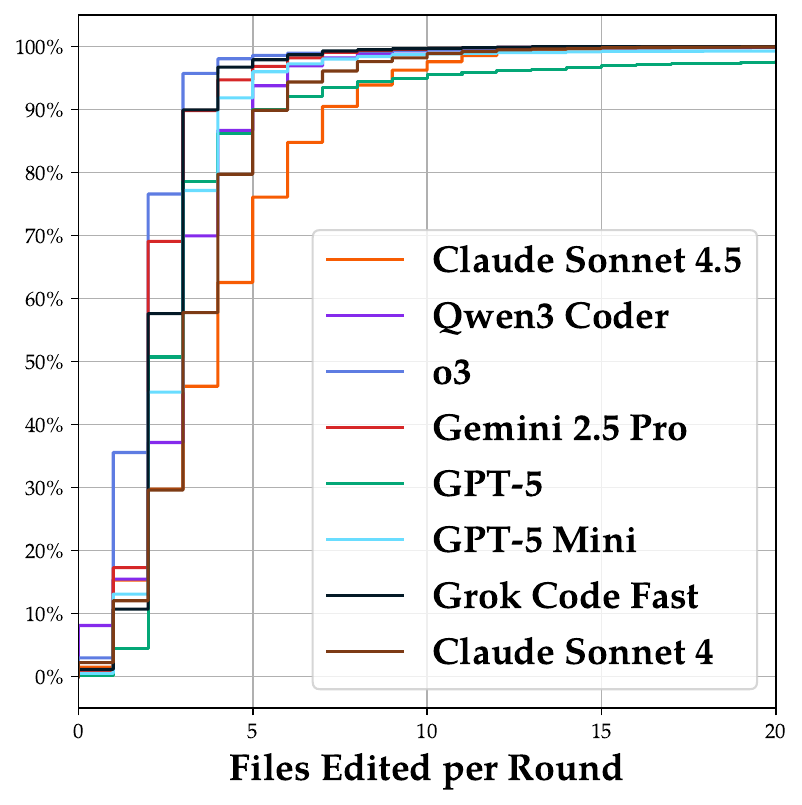}
    \caption{
CDF of files edited per round by each model.
While some models typically never edit more than $5$ files (\texttt{o3}, \texttt{Gemini 2.5 Pro}), others tend to create and manipulate many more (\texttt{Claude Sonnet 4.5}, \texttt{GPT-5})
}
    \label{fig:cdf_files_edited_per_round}
\end{minipage}
\hfill
\begin{minipage}[t]{0.49\textwidth}
    \setlength{\abovecaptionskip}{0em}
    \centering
    \includegraphics[width=\textwidth]{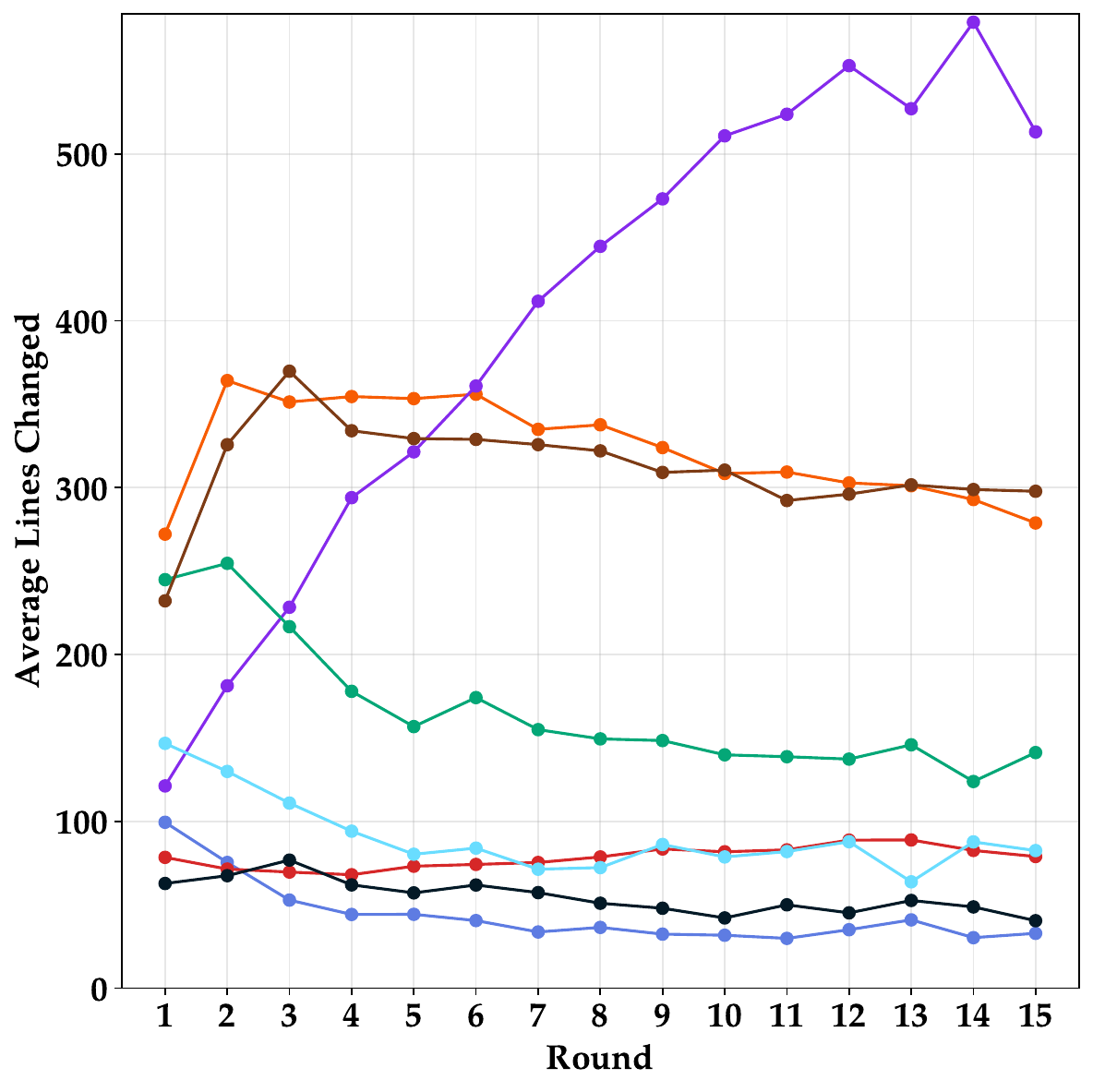}
    \caption{
Average lines changed per round per model.
Some models are fairly consistent (Gemini 2.5 Pro), while others vary; Qwen3-Coder edits more in later rounds, while GPT-5 Mini's edits largely occur earlier on.
    }
    \label{fig:line_chart_per_round_changes}
\end{minipage}
\begin{minipage}[t]{0.49\textwidth}
    \setlength{\abovecaptionskip}{0em}
    \centering
    \includegraphics[width=\textwidth]{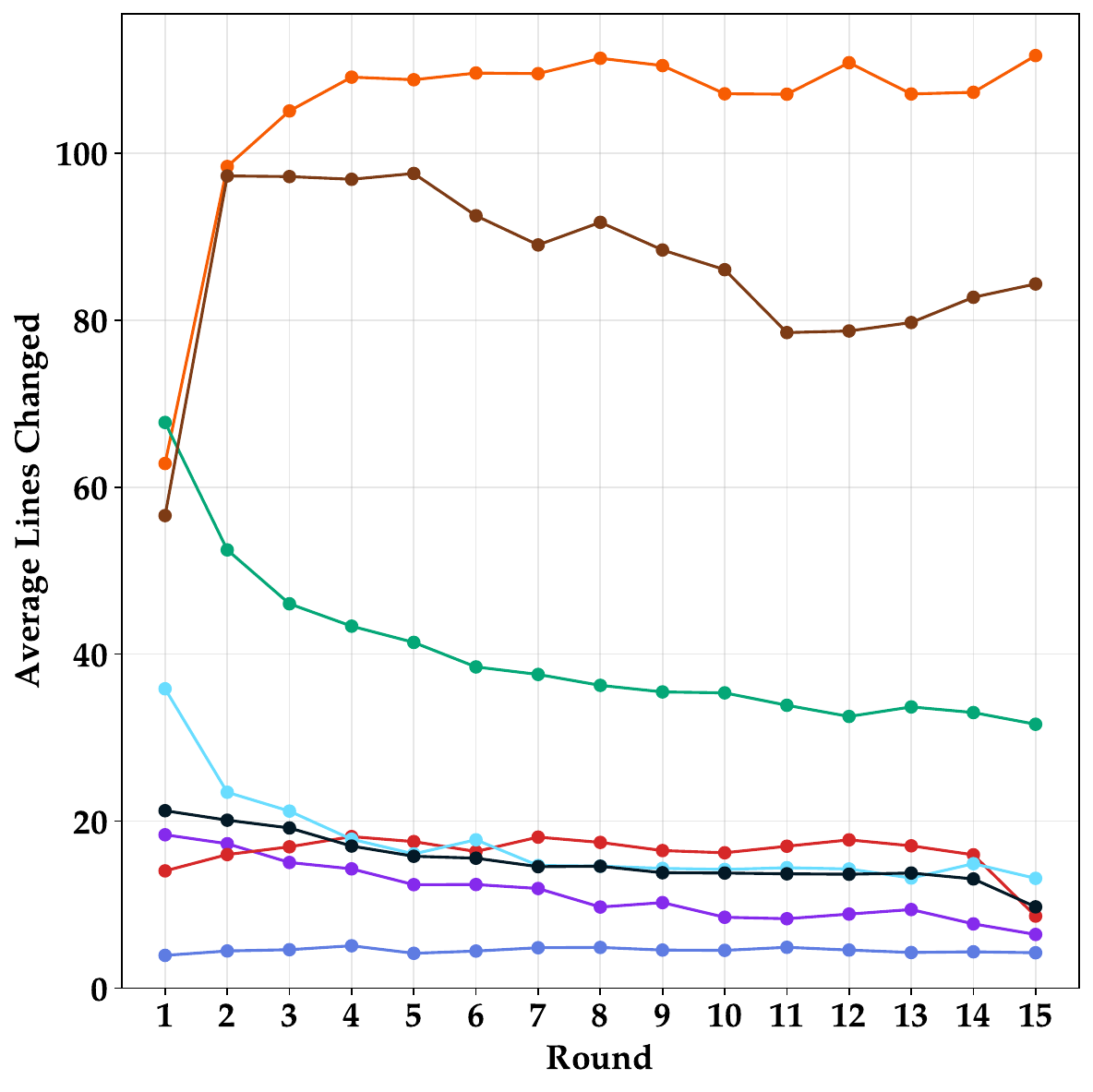}
    \caption{
Average lines changed per round per model for the \texttt{README\_agent.md}, a file we suggest agents write important information to.
The Anthropic family of models write copious amounts of notes -- other models tend to add more brief summaries.
}
    \label{fig:line_chart_per_round_changes_readme}
\end{minipage}
\hfill
\begin{minipage}[t]{0.49\textwidth}
    \setlength{\abovecaptionskip}{0em}
    \centering
    \includegraphics[width=\textwidth]{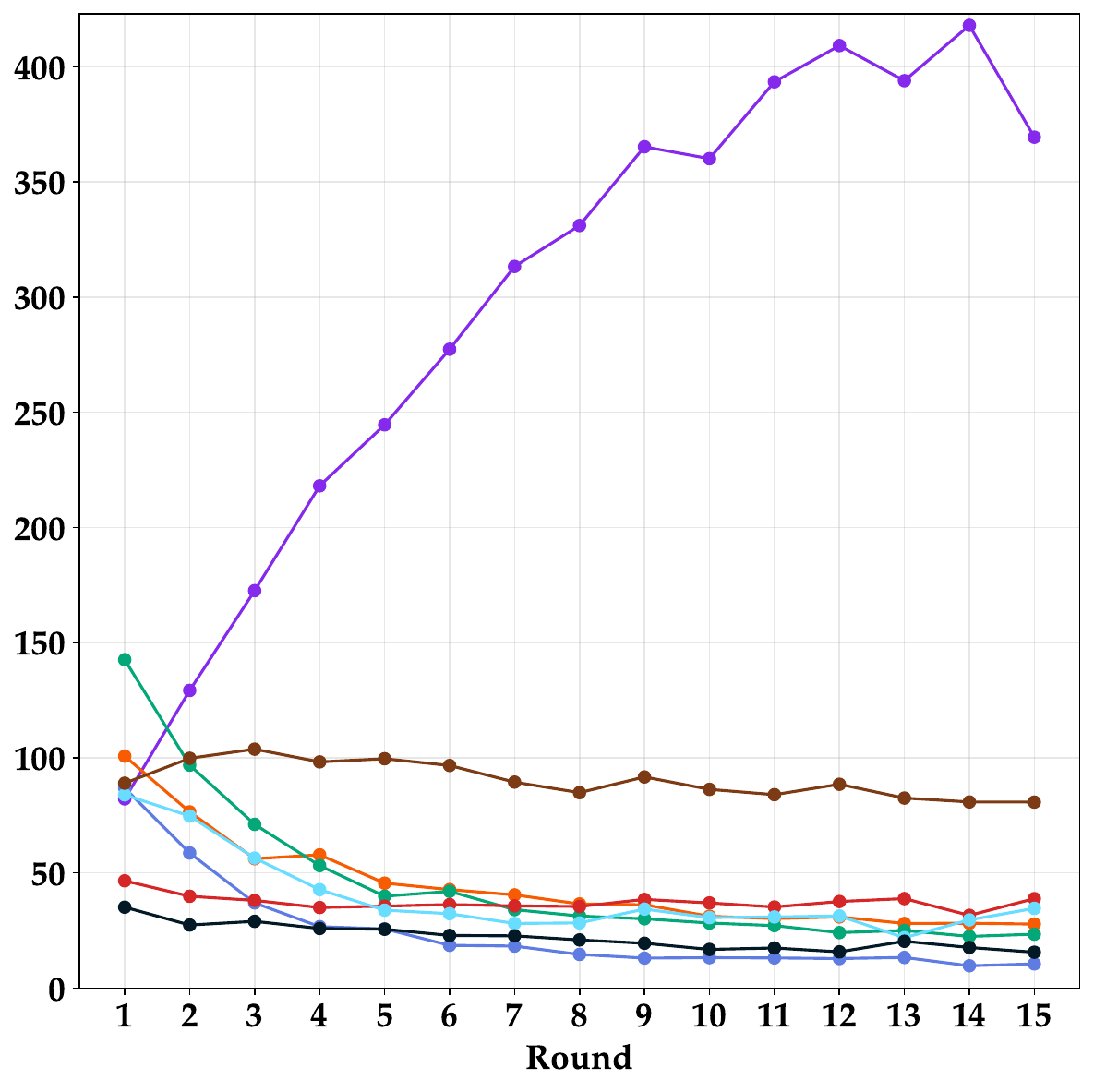}
    \caption{
Average lines changed per round per model for game-playing related functionality (e.g. \texttt{warrior.red} in Core War).
Models typically make the majority of their changes early on, with a steady decline in later rounds as changes become more targeted.
    }
    \label{fig:line_chart_per_round_changes_submission}
\end{minipage}
\end{figure}

\textbf{Models differ in the number of files created or edited.} As shown in Figures~\ref{fig:cdf_files_edited_per_round} and~\ref{fig:line_chart_per_round_changes}, we observe that models vary significantly in the number of files and lines changed per round.
The range varies significantly, with more conservative models such as \texttt{o3} or \texttt{Gemini 2.5 Pro} editing just two to three files and less than a hundred lines per round.
On the other end, \texttt{Claude Sonnet 4.5} or \texttt{GPT-5} generally make larger changes, with a much longer tail of sizable modifications.
We observe that this long tail typically comes from when models initialize test suites, create multiple versions of a submission to test against one another, or record insights as markdown notes to take forward into the next round.
We include two additional similar line charts that show the size of edits for the \texttt{README\_agent.md} file (Figure~\ref{fig:line_chart_per_round_changes_readme} along with any game-playing related core functionality in Figure~\ref{fig:line_chart_per_round_changes_submission}.
The \texttt{Claude Sonnet 4} and \texttt{Claude Sonnet 4.5} models are relatively more extensive in their documentation.
\texttt{GPT-5} and \texttt{GPT-5-mini} exhibit a trend, where they take more notes up front, with a gradual decline into later rounds.
The remaining models do not fluctuate significantly in the amount of notes they take, with \texttt{o3} averaging under $10$ lines changed per round.
Model changes to competition logic generally trends downward across rounds -- we generally observe that models define the majority of competitive logic early on, with later rounds consisting mostly of smaller, more specific adjustments.

\begin{figure}[t]
\begin{minipage}[b]{0.49\textwidth}
    \setlength{\abovecaptionskip}{0em}
    \centering
    \includegraphics[width=\textwidth]{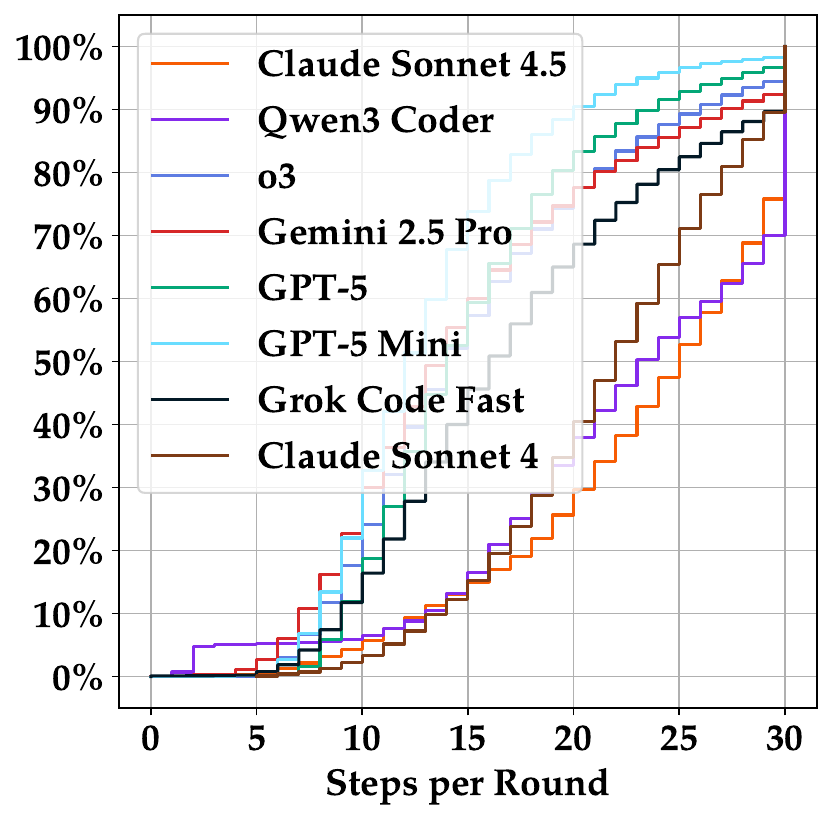}
    \caption{
    CDF of number of steps taken per round per model.
    The Anthropic family of models along with \texttt{Qwen3-Coder} usually consumes more of the allotted step budget.
    }
    \label{fig:cdf_steps_per_round}
\end{minipage}
\hfill
\begin{minipage}[b]{0.49\textwidth}
    \setlength{\abovecaptionskip}{0em}
    \centering
    \includegraphics[width=\textwidth]{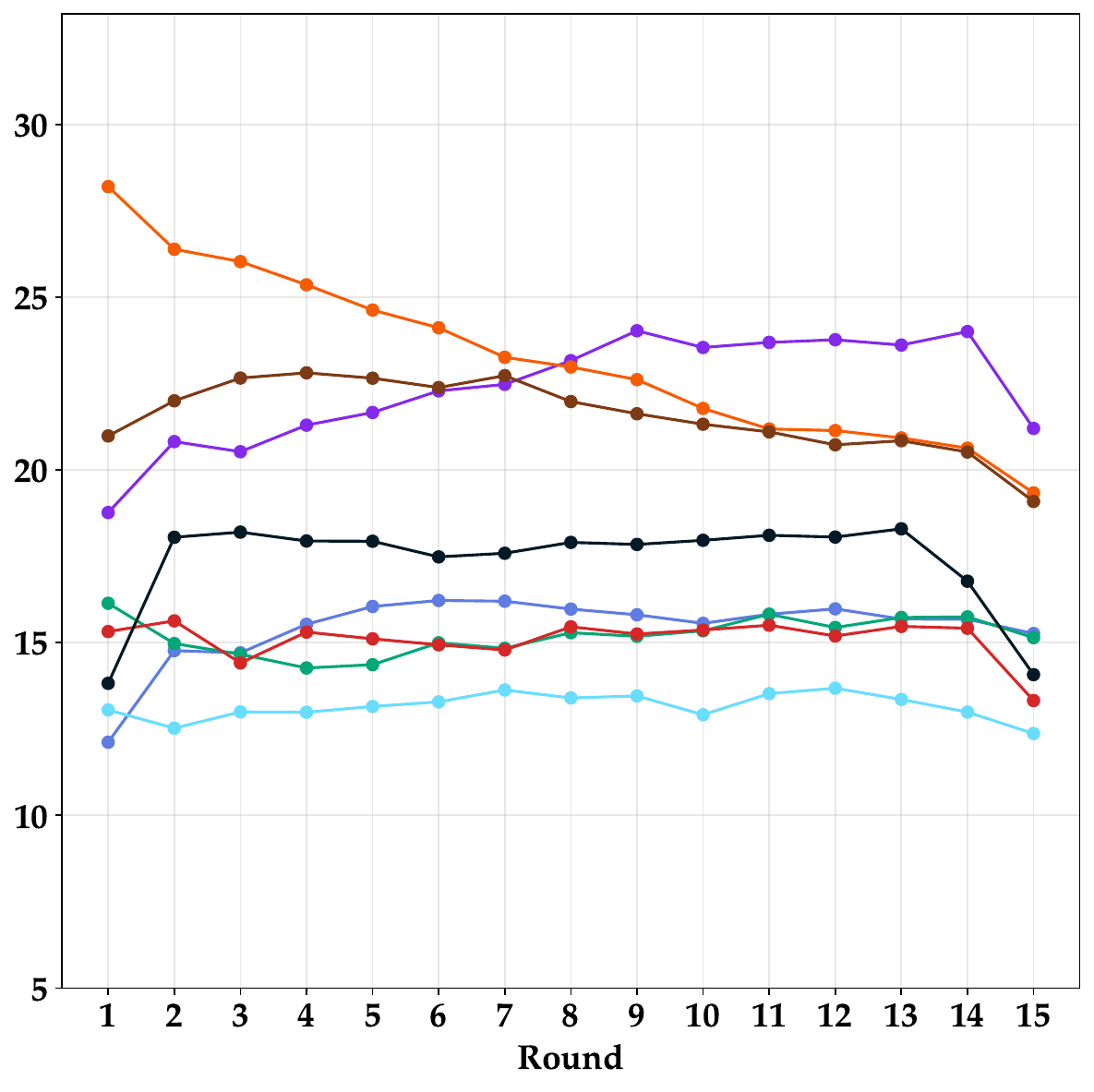}
    \caption{
    Average steps taken for each round per model.
    The chart reflects similar conclusions as Figure~\ref{fig:cdf_steps_per_round}, and also suggests that steps used are fairly steady.
    }
    \label{fig:line_chart_steps_per_round}
\end{minipage}
\end{figure}

\textbf{Models differ in the number of steps taken.}
We provide Figures~\ref{fig:cdf_steps_per_round} and~\ref{fig:line_chart_steps_per_round} to showcase trends around the number of turns consumed by each model for each round.
Turn budget consumption is markedly different between models, with the Anthropic models and \texttt{Qwen3-Coder} usually using $22$ to $27$ turns out of the $30$ turn limit.
On the other end, \texttt{Gemini 2.5 Pro} and \texttt{GPT-5 mini} rarely exceed $15$ turns.
Figure~\ref{fig:line_chart_steps_per_round} suggests that the number of steps models take from round to round is fairly steady; we were not able to identify any meaningful discrepancies in steps taken between rounds that might be due to trends such as 
To further clarify -- although we impose the \$$1$ per-round cost limit, there are \textit{zero} occurrences across all tournaments we run of a model's trajectory being automatically terminated due to models exceeding the cost limit budget.
In other words, this means that the cost limit trend lines also faithfully reflect when models decide for themselves to stop editing for the round.
The majority of rounds end with a model producing a thought and action akin to ``I have made all the changes I think are necessary. I will now conclude this round [\texttt{END} action]".

\begin{figure}[t]
\begin{minipage}[t]{0.49\textwidth}
    \setlength{\abovecaptionskip}{0em}
    \centering
    \includegraphics[width=\textwidth]{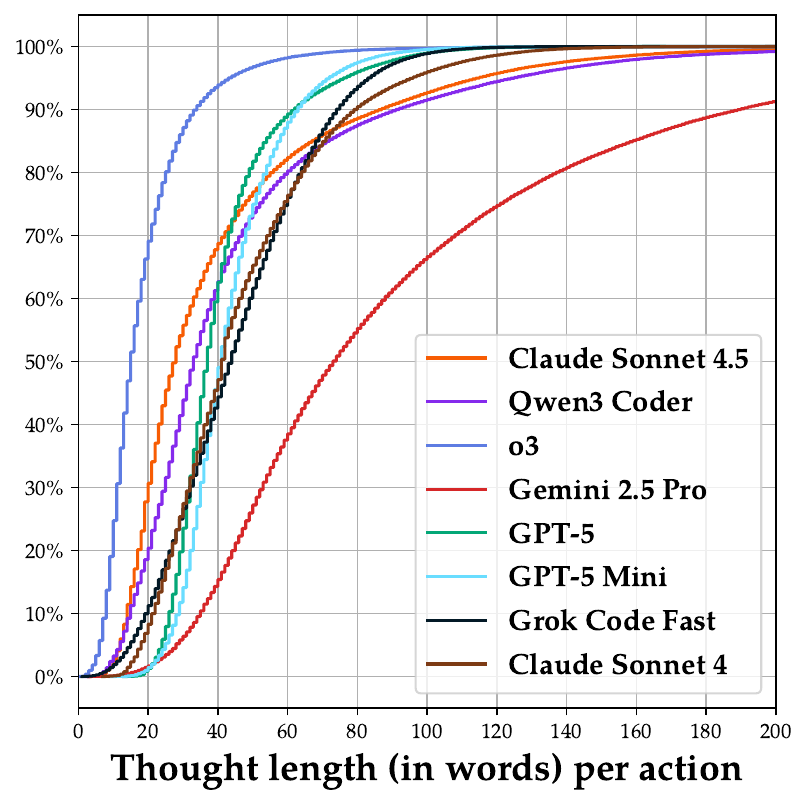}
    \caption{
    CDF of thought length (in words) per model.
    The thought lengths are computed per model response.
    Our calculation does not consider the action produced by the model within the same response.
    }
    \label{fig:cdf_thought_length_per_round}
\end{minipage}
\hfill
\begin{minipage}[t]{0.49\textwidth}
    \setlength{\abovecaptionskip}{0em}
    \centering
    \includegraphics[width=\textwidth]{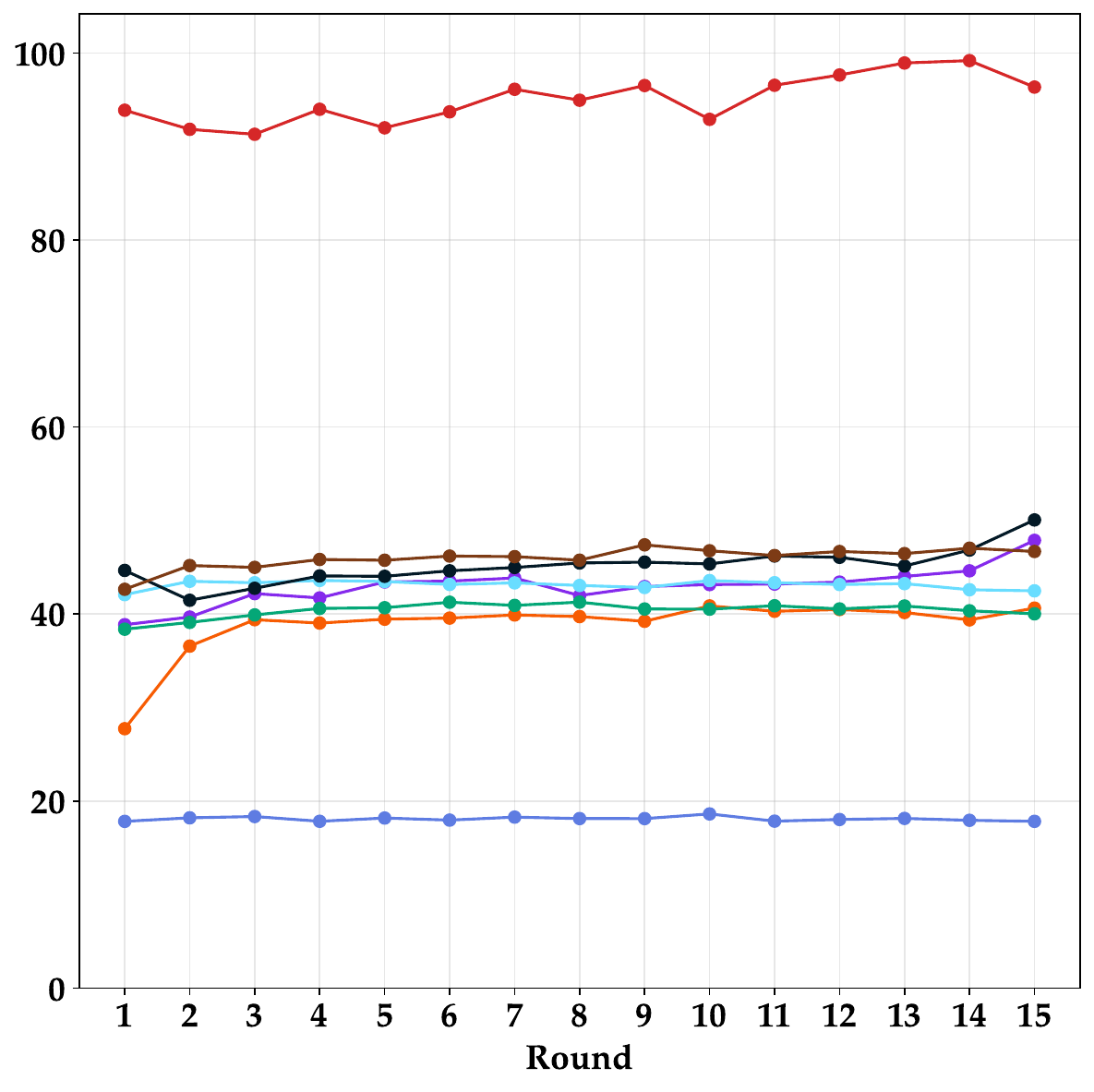}
    \caption{
    Average thought length (in words) per model response at each round.
    While most models fall within the range of $35$ to $55$ words per response, \texttt{Gemini 2.5 Pro} and \texttt{o3} are notable outliers.
    }
    \label{fig:line_chart_thought_length_per_round}
\end{minipage}
\end{figure}

\textbf{Models differ in thought length.} As shown in Figures~\ref{fig:cdf_thought_length_per_round} and~\ref{fig:line_chart_thought_length_per_round}, we find that while most models respond with similarly long thought traces, \texttt{Gemini 2.5 Pro} responds with significantly longer explanations, at around $95$ words per response.
On the other end, \texttt{o3} is much more terse, with just under $19$ words per response.
However, \texttt{o3}'s brevity comes with a heavy asterisk, as OpenAI's API is configured to hide intermediate thinking tokens for the \texttt{o}-series reasoning models.
The actual token count is thus likely vastly underestimated.

\begin{figure}[h]
\begin{minipage}[t]{0.49\textwidth}
    \setlength{\abovecaptionskip}{0em}
    \centering
    \includegraphics[width=\textwidth]{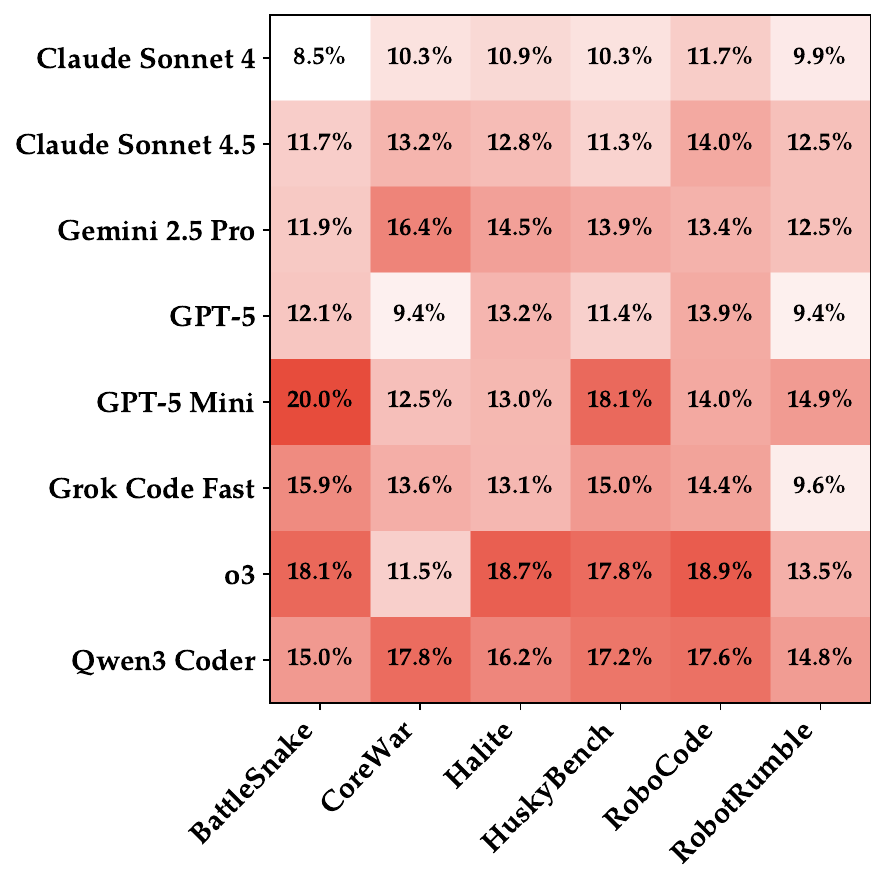}
    \caption{
A heatmap of errant action rates for models in different arenas.
``Errant" means the action resulted in \texttt{returncode} == $0$.
We find that malformed actions does \textit{not} constitute a significant reason for why models might struggle in \clash{}.
    }
    \label{fig:heatmap_returncode}
\end{minipage}
\hfill
\begin{minipage}[t]{0.49\textwidth}
    \setlength{\abovecaptionskip}{0em}
    \centering
    \includegraphics[width=\textwidth]{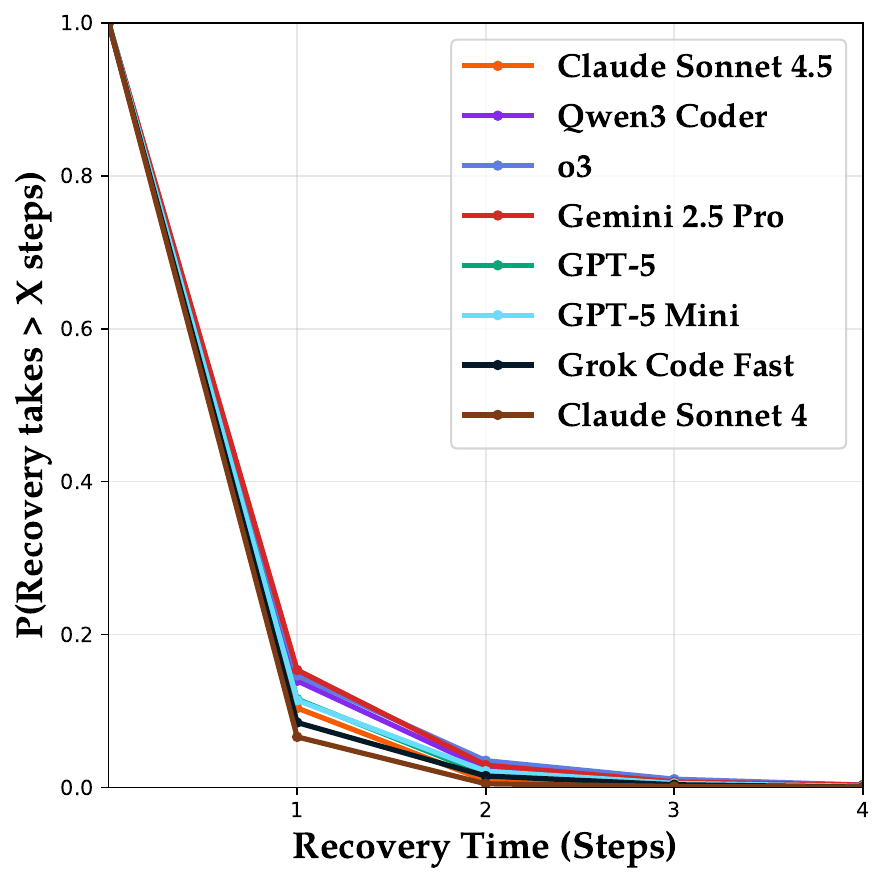}
    \caption{
``Recovery time" is the number of steps between a failed command (\texttt{returncode} != $0$) and the next successful command (\texttt{returncode} == $0$).
Each data point indicates the likelihood that recovery requires more than \texttt{x} steps for a model.
    }
    \label{fig:survival_curve_error_recovery}
\end{minipage}
\end{figure}

\textbf{Models are quick to recover from errant actions.} As discussed in Section~\ref{sec:results}, errant actions is not a significant factor in model performance.
The vast majority of actions ($\geq90$\%) are well formed and execute successfully.
In addition to the statistics we presented before, we also provide a breakdown of the errant action rates by model and arena in Figure~\ref{fig:heatmap_returncode}.
We find that stronger models have slightly lower error rates, with \texttt{Claude Sonnet 4} at just $10.11$\%, while \texttt{Qwen3 Coder} tops out at \texttt{16.32}\%.
No arena has a particularly high errant action rate.

Furthermore, we also answer how quickly models recover from errant actions.
Prior work has reported that a major error mode of existing models are ``cascading" failures -- if a model issues an errant action, the likelihood that it recovers successfully from the mistake decreases with every subsequent action~\citep{yang2024sweagentagentcomputerinterfacesenable,pan2025trainingsoftwareengineeringagents}.
In the year since these works pointed out this phenomenon, we find that such breakdowns have diminished significantly in frequency and length.
We visualize this finding with Figure~\ref{fig:survival_curve_error_recovery}.
We observe that following an errant action, the next action is successfully more than $80$\% of the time.
By the third step following an errant action, there are nearly zero occurrences of models continuing to struggle to generate a well formed action.
In summary, our analyses strongly suggest that model performance in \clash{} is neither hindered by the choice of agent framework, nor that models are not adept at operating on the command line.

\begin{figure}[t]
    \centering
    \includegraphics[width=\textwidth]{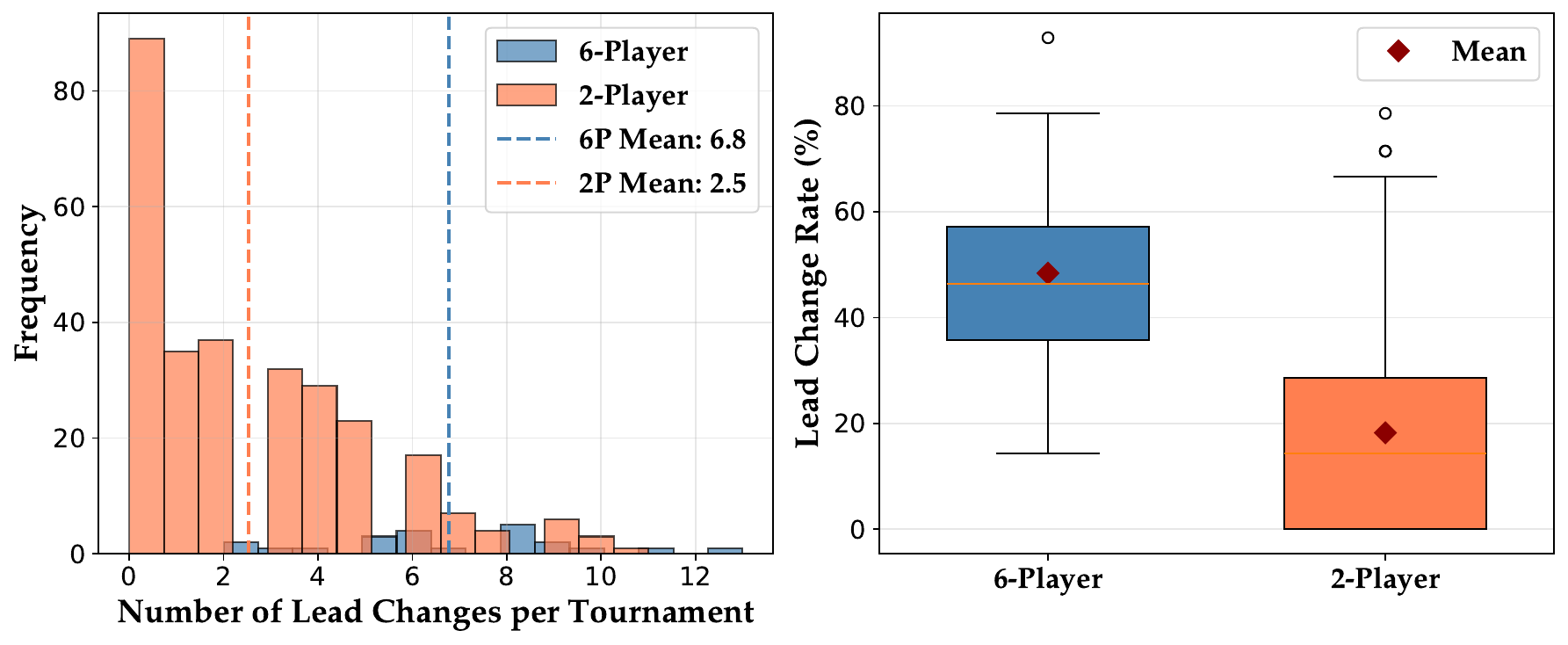}
    \caption{
Lead change rate comparison.
A ``lead change" is defined as a round \texttt{n} where the winner is different from the round \texttt{n-1} winner.
We make comparisons between $2$-player and $6$-player tournaments specifically for the Core War arena.
    }
    \label{fig:win_change_rate_comparison}
    \vspace{1em}
    \centering
    \includegraphics[width=\textwidth]{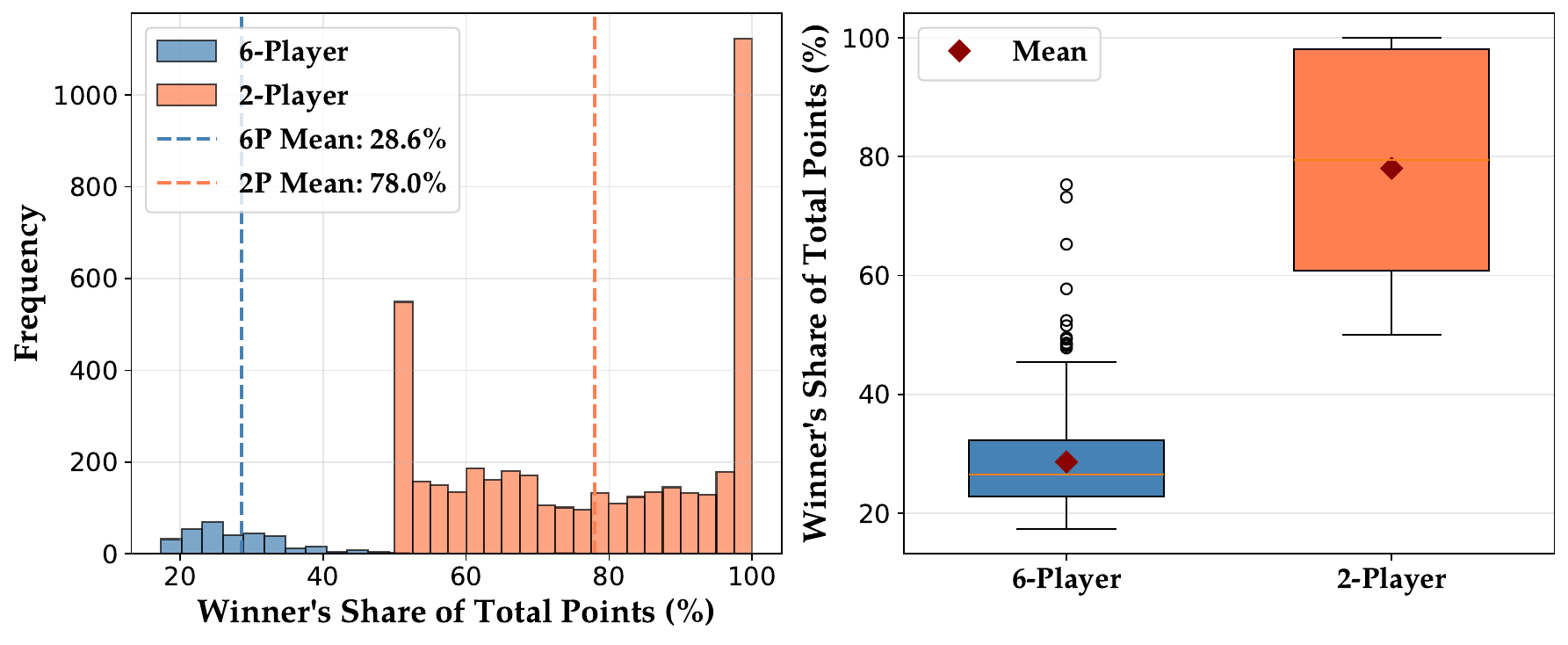}
    \caption{
Win share comparison.
We define ```win share" as the percentage of total points taken by a particular player.
Win share per player is much lower with more opponents.
    }
    \label{fig:winner_share_comparison}
\end{figure}

\subsection{Additional Ablations}
\label{appx:results:ablations}

\textbf{Multi-player settings are far more variable in standings.}
As mentioned in our results and analyses section in the main paper, we showcase the ability to run multi-player ($3+$) tournaments in \clash{}, specifically with the Core War arena.
As shown in Table~\ref{tab:list_arenas}, four additional arenas -- BattleSnake, Halite, Poker, and RoboCode -- all support more running tournaments with $2$ players, though we do not run comprehensive experiments due to both cost limitations and the analytical complexity introduced by multi-way competition, which we believe is best left as future work.
To illustrate the difference in competitive volatility, we provide Figure~\ref{fig:win_change_rate_comparison}, revealing that lead changes are much more frequent as there are more players.
Furthermore, winners occupy a much smaller share of the total points in the $6$ player arena compared to the head-on setting.

\textbf{Transparent codebases enable investigations in how models leverage views into others' development processes.}
We elected to run tournaments for \clash{}'s main results under the assumption that models cannot view opponents' code because such a setting is more reminiscent of real world settings, where human players develop their solutions independently and have the option to keep their codebase closed source.
Therefore, we investigate the effects of making players' codebases viewable by opponents specifically as an ablation.
The introduction of this mechanic is potentially interesting as it shifts \clash{} much closer towards being a perfect information game~\citep{fudenberg1991game}, where all players in a game have knowledge of all relevant information in the system, including other players' decisions.
The knowledge of opponents' moves is what distinguishes a perfect information game like chess from an imperfect information game like poker, where opponent private cards are not known by default.

\begin{figure}[t]
\begin{minipage}[b]{0.49\textwidth}
    \setlength{\abovecaptionskip}{0em}
    \centering
    \includegraphics[width=\textwidth]{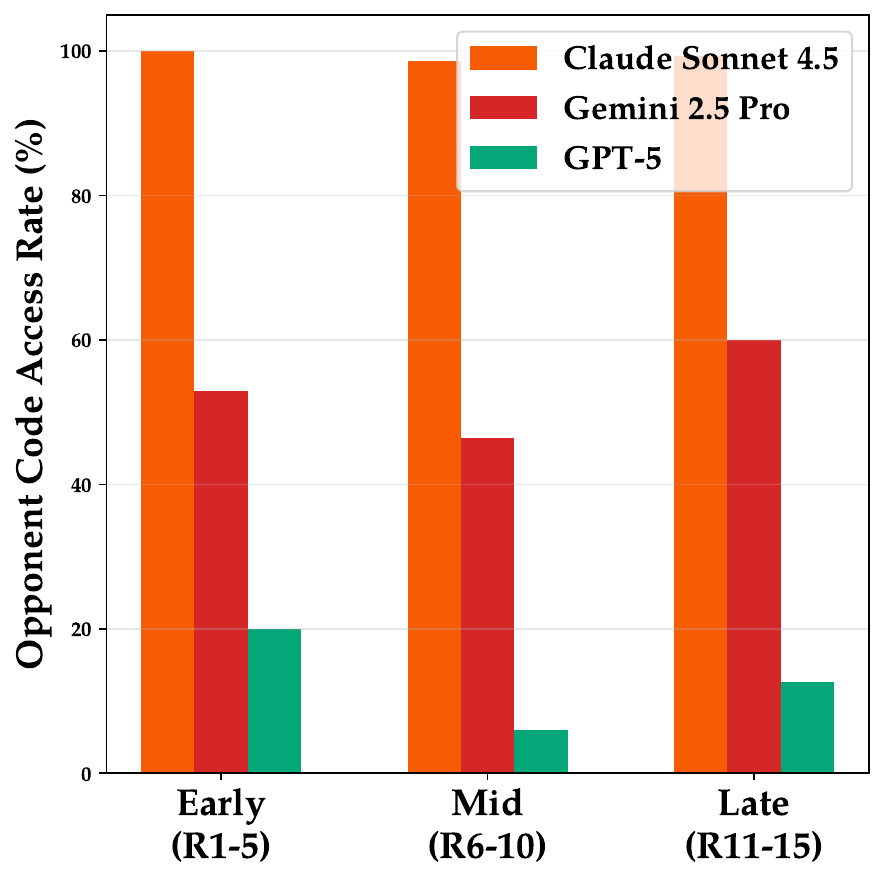}
    \caption{
Share of rounds which a model inspects its opponent's codebase.
We find variance across models and round ranges.
    }
    \label{fig:bar_chart_temporal_opponent_access}
\end{minipage}
\hfill
\begin{minipage}[b]{0.49\textwidth}
    \centering
\begin{tabular}{lc}
\toprule
\textbf{Model} & $\mu$ \\ \midrule
Claude Sonnet 4.5              &  28.38 $\pm$  0.65 \\
o3                             &  27.11 $\pm$  0.64 \\
Grok Code Fast                 &  25.65 $\pm$  0.65 \\
GPT-5                          &  24.76 $\pm$  0.64 \\
Gemini 2.5 Pro                 &  23.62 $\pm$  0.65 \\
Qwen3 Coder                    &  22.30 $\pm$  0.66 \\
\bottomrule
\end{tabular}
\caption{
TrueSkill ratings per model based on $20$ tournaments of $6$-player Core War. 
TrueSkill models each player's skill as a Gaussian distribution with mean $\mu$ (skill estimate) and standard deviation $\sigma$ (uncertainty). 
After each round, both parameters are updated based on match outcomes: winning increases $\mu$ while exceeding expectations, and $\sigma$ decreases as the system gains confidence in the estimate. 
Final placement (1st, 2nd, ..., 6th) determines rating updates. 
}
\label{tab:trueskill}
\end{minipage}
\end{figure}

As mentioned in the main results, we carry out this investigation specifically for the Halite arena with three models (\texttt{GPT-5}, \texttt{Claude 4.5 Sonnet}, \texttt{Gemini 2.5 Pro}).
From Figure~\ref{fig:bar_chart_temporal_opponent_access}, we found that the rate at which a player checks its opponent codebase fluctuates across both models and the phase of the tournament.
\texttt{Claude 4.5 Sonnet} is near constant, checking in on its opponent's activity nearly every single round.
\texttt{Gemini 2.5 Pro} and \texttt{GPT-5} both exhibit a trend where the check rate dips somewhat in the middle of a tournament before re-surging in later rounds.

\subsection{Analyzing trajectories using LMs as a judge}
\label{appx:results:strategic_reasoning}
This sections describes detailed observations about the agent trajectories that were obtained using a LM as a judge setup.

\subsubsection{Additional results}
\begin{figure}
\includegraphics[width=\linewidth]{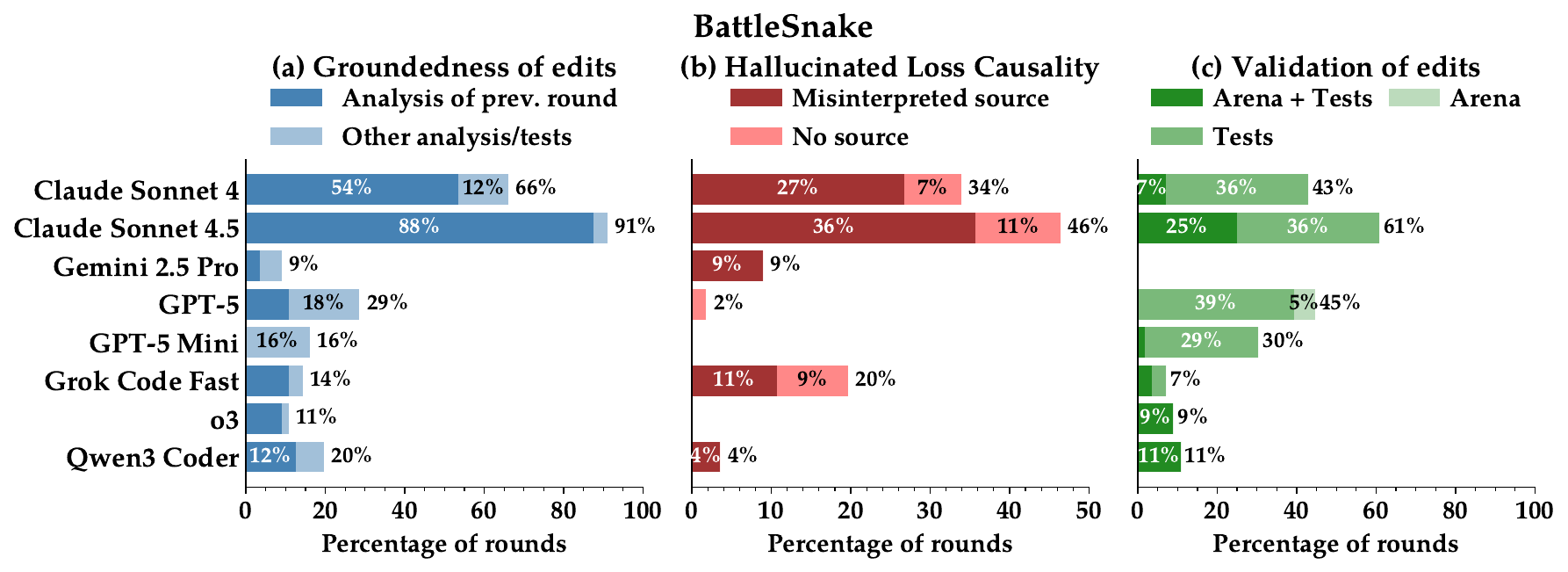}
\includegraphics[width=\linewidth]{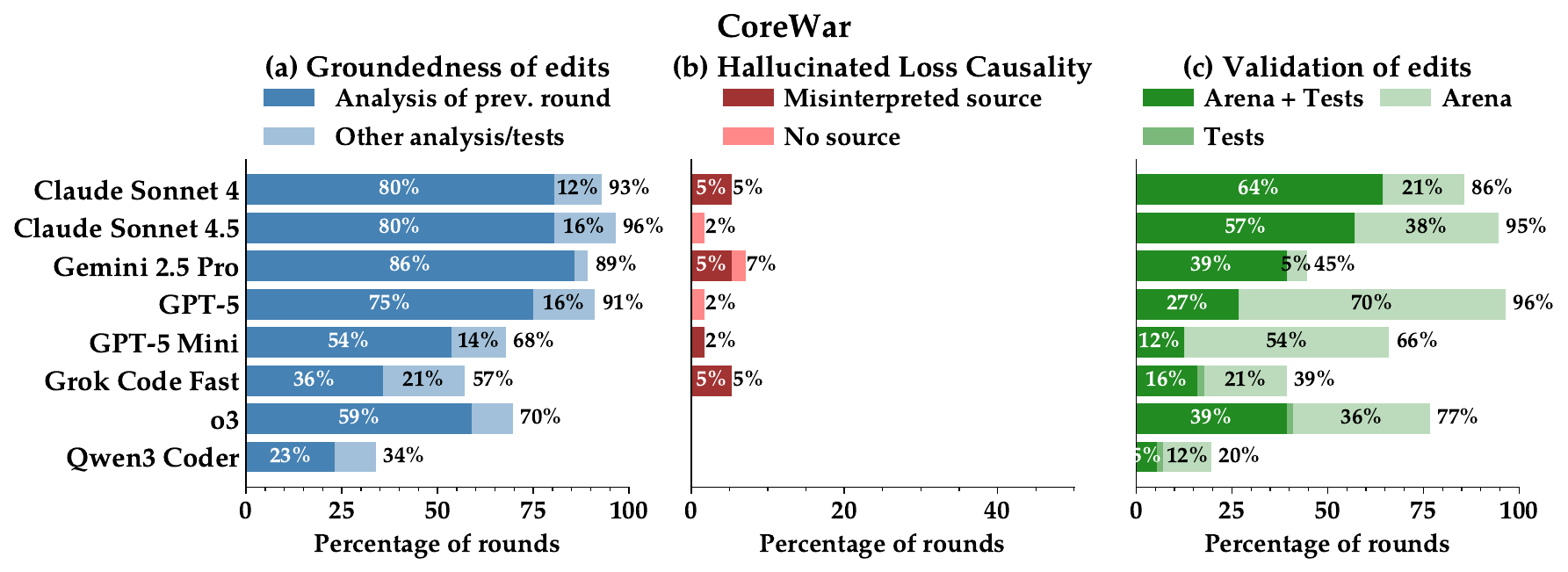}
\caption{Results for the groundedness of edits, hallucinated loss causality, and validation of edits for different arenas (part 1). For the identical plot averaged over all arenas, see Figure~\ref{fig:llm_as_judge}.}
\label{fig:llm_as_judge:per_arena:part1}
\end{figure}

\begin{figure}
\includegraphics[width=\linewidth]{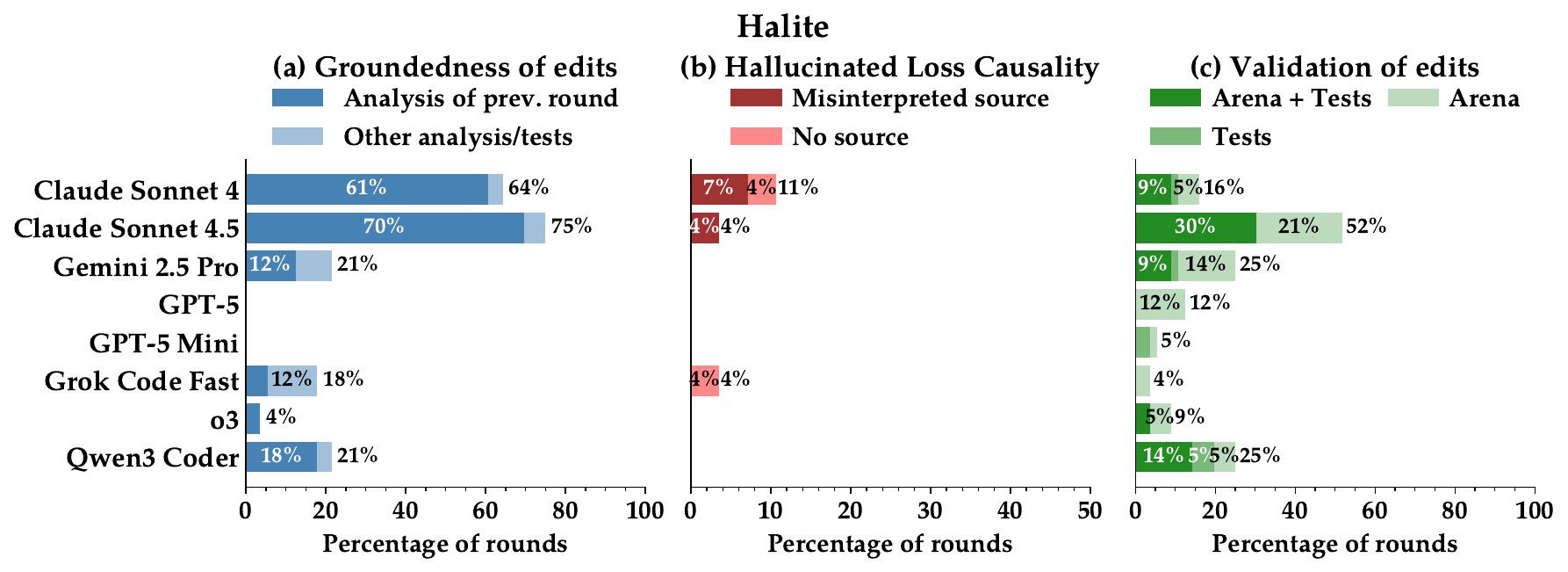}
\includegraphics[width=\linewidth]{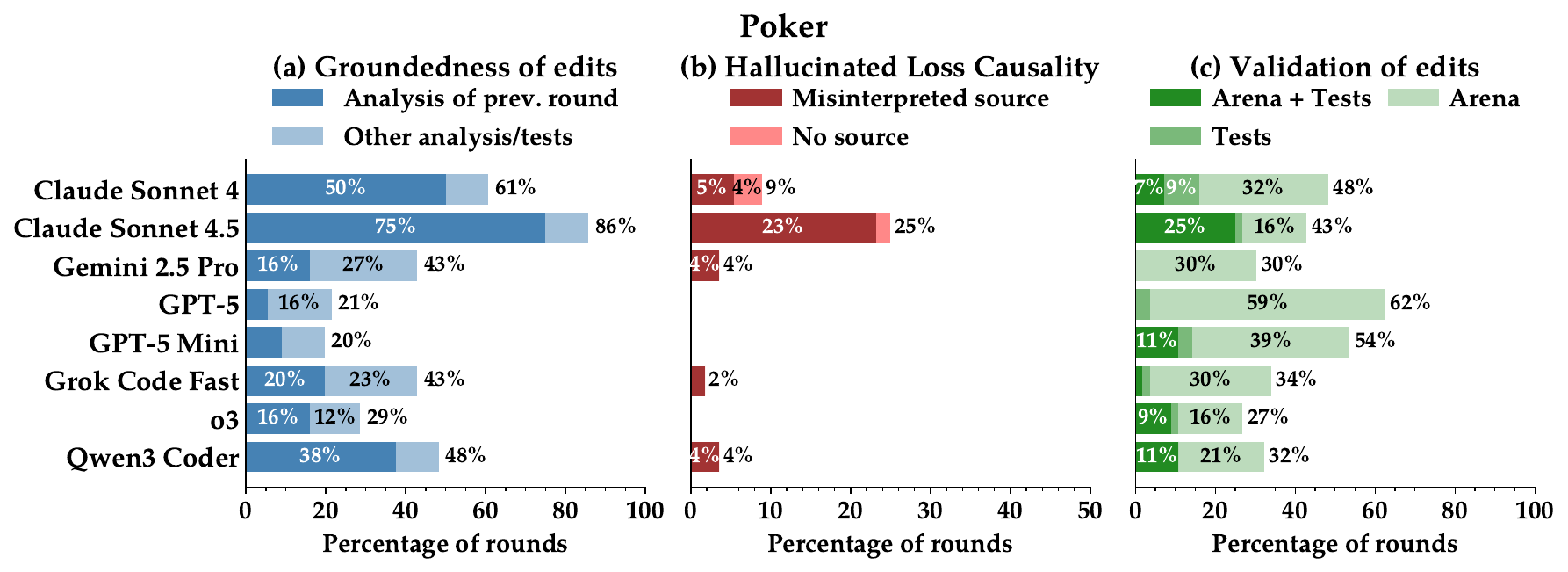}
\includegraphics[width=\linewidth]{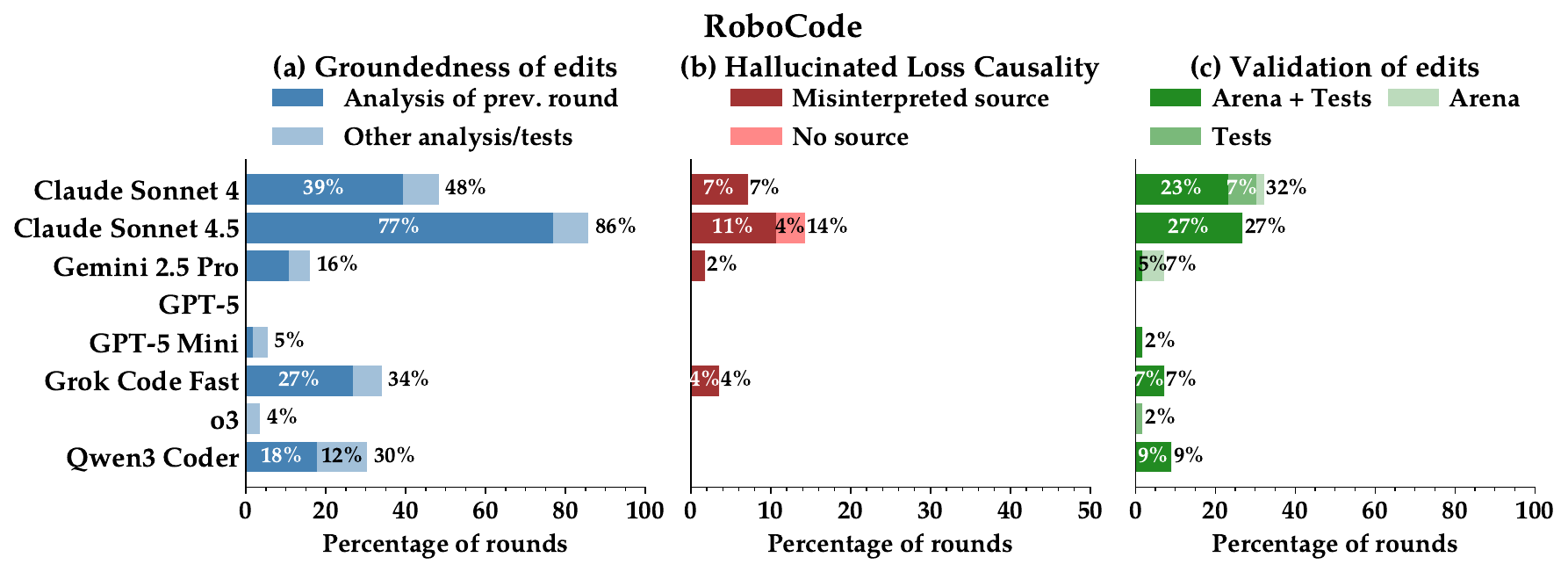}
\includegraphics[width=\linewidth]{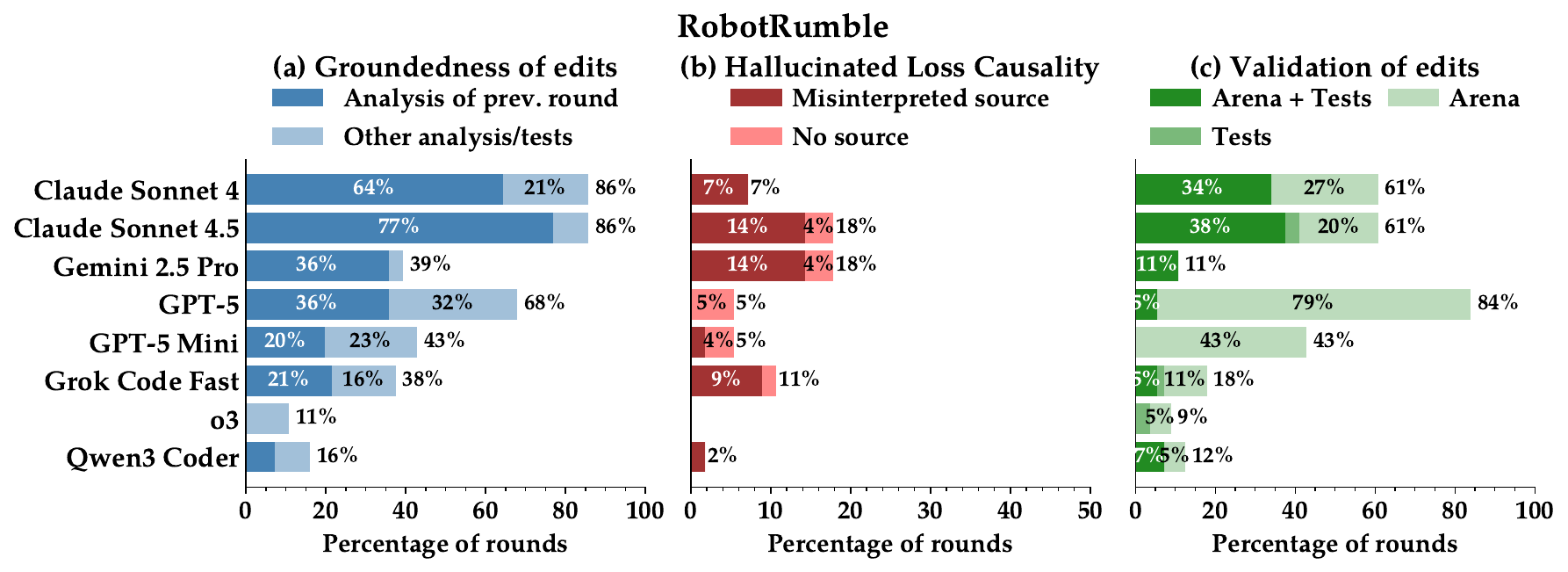}
\caption{Results for the groundedness of edits, hallucinated loss causality, and validation of edits for different arenas (part 2). For the identical plot averaged over all arenas, see Figure~\ref{fig:llm_as_judge}.}
\label{fig:llm_as_judge:per_arena:part2}
\end{figure}

The data on the groundedness of edits, hallucinations, and validation efforts that were presented in Figure~\ref{fig:llm_as_judge} are shown for the different arenas in Figures~\ref{fig:llm_as_judge:per_arena:part1} and \ref{fig:llm_as_judge:per_arena:part2}.
Notably, models behave very different across arenas.
For example, BattleSnake elicits very strong hallucinations from Claude Sonnet 4.5 (affecting up to 45\% of rounds), and RoboCode shows a particularly low rate of edit validation across models.

\begin{figure}
    \includegraphics[width=\linewidth]{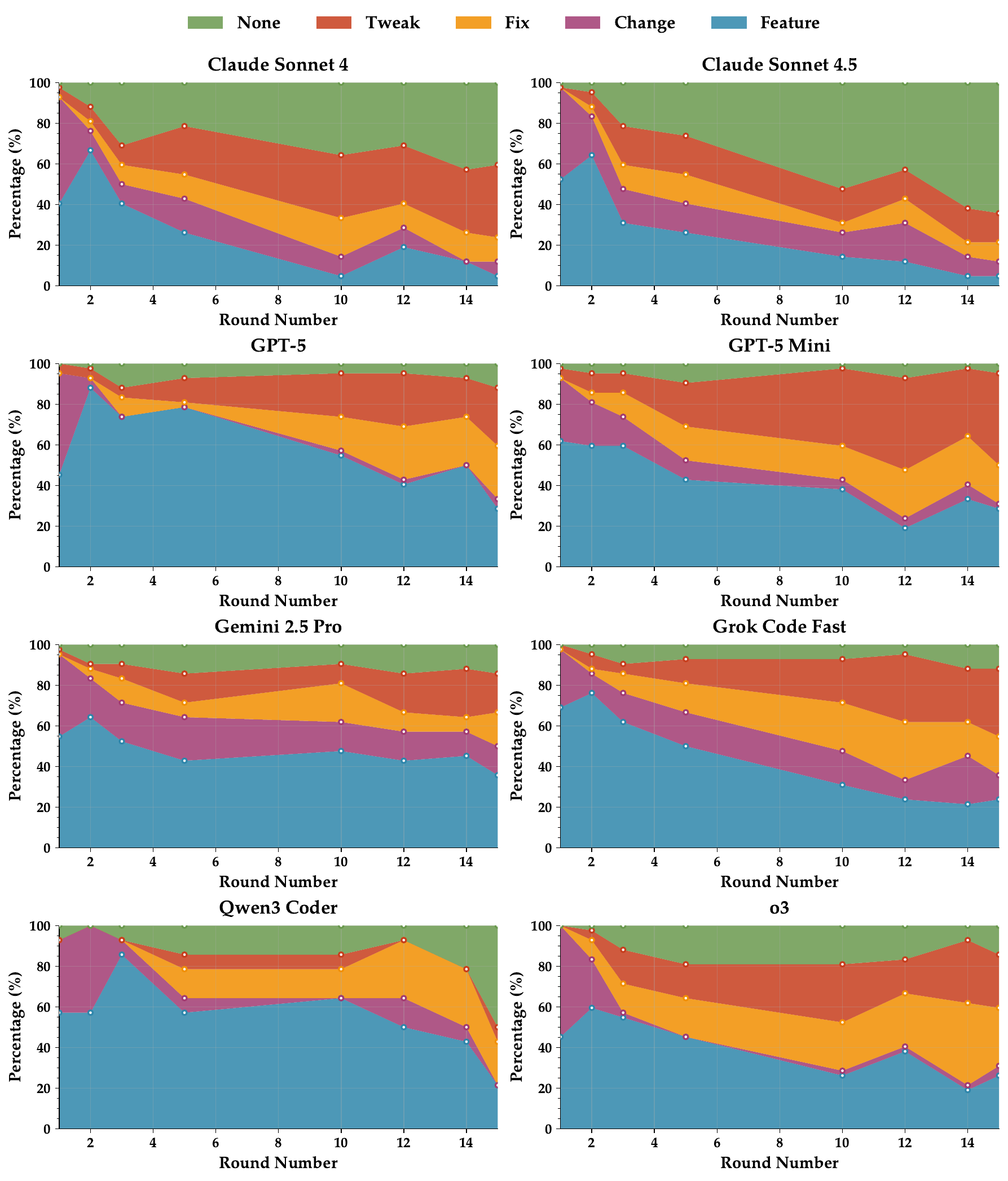}
    \caption{Models perform different kinds of edits on the main player file as the tournament progresses.
    For this, the full changes to the main player file during a round are summarized into five categories:
    \emph{Feature} represents significant additions, \emph{change} a larger change to overall logic, \emph{fix} are smaller-scale fixes, \emph{tweak} are minor modification of parameters, and \emph{none} means that no significant change was made to the player file.
    The y axis shows the fraction of rounds in which the edits can best be summarized by this category.
    }
    \label{fig:llm_as_judge:types_of_edits}
\end{figure}

Figure~\ref{fig:llm_as_judge:types_of_edits} shows how the kinds of edits that models perform changes between rounds. While the initial editing of models is feature-heavy, as the tournament progresses, a larger amount of smaller tweaks or fixes appears together with rounds in which no meaningful edit was made to the main player file.

\begin{figure}
    \includegraphics[width=\linewidth]{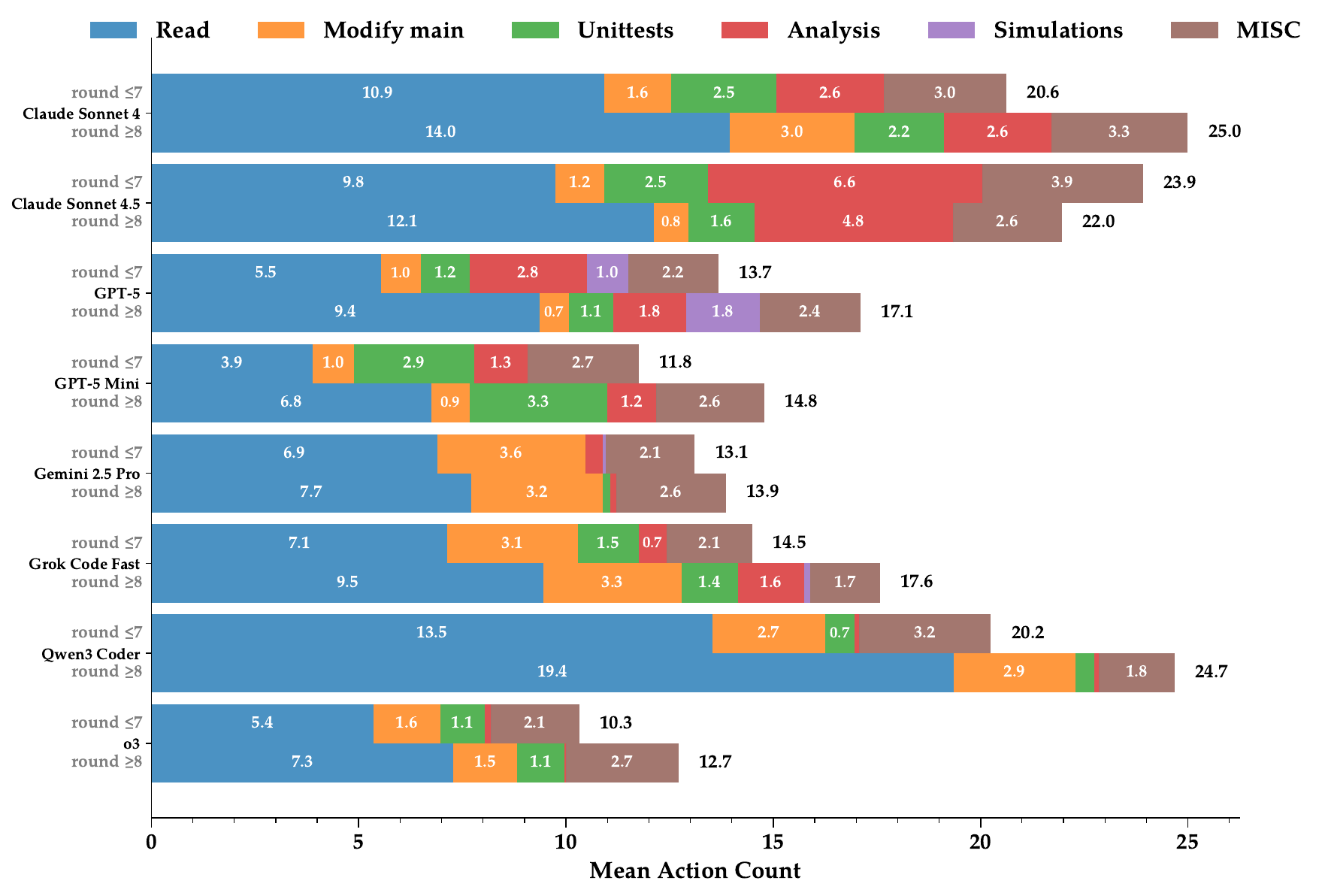}
    \caption{What do models spend their turns on? The mean number of actions a model spends on reading files (\emph{read}), modifying the main player file (\emph{modify main}), running unittests, analysis, or arena simulations (\emph{unittests}, \emph{analysis}, \emph{simulations}), or performing any other action. We present separate averages for early tournament (round $\leq 7$) and late tournament (round $\geq 8$).}
    \label{fig:llm_as_judge:action_categories}
\end{figure}

Figure~\ref{fig:llm_as_judge:action_categories} shows what models spend their turn on early in the tournament and late in the tournament. 
This figure not only shows how the average number of actions in a round varies between models, but also that read operations increase as the tournament progresses.
It is also apparent how different the number of actions spent on testing, analyzing, and running test matches is between models.

\subsubsection{Groundedness of edits and validation of edits}
We use structured outputs with the following data structure
\begin{samepage}
\begin{extraCodeExample}[Model response schema for groundedness and validation study]
class BigQuestionsModelResponseSchema(BaseModel):
    """Schema for structured output of the model."""

    edit_category: Literal["tweak", "fix", "feature", "change", "none"]
    edits_motivated_by_logs: bool
    edits_motivated_by_insights: bool
    edits_motivated_by_old_static_messages: bool
    edits_reverted_based_on_insights: bool
    edits_tested_with_simulations: bool
    edits_validated_with_unittests: bool
    improved_test_analysis_framework: bool
    reasoning: str
\end{extraCodeExample}
\end{samepage}
The model is prompted with the following system prompt:
\begin{extraCodeExample}[System prompt for groundedness and validation study]
## Overall setting

You are an expert at analyzing the behavior of LM agents.
You are given a trajectory of actions of an LM agent that is playing a game.
You are asked to answer a series of questions about the behavior of the agent.

We are interested in:

1. What motivated the edits
2. What steps were taken to validate the edits

All questions that are marked as boolean need to be answered with a boolean value.
You cannot answer "unknown" or similar.

## Definitions

**Main player file**:

You are investigating an LM agent that is playing a game.
The main player file is the main file that constitutes the agent's submission, i.e.,
the file that governs the agent's behavior and logic for the next round of the game that is being played.
Commonly, this is the file called `main.py`, `player.py`, `robot.js`, `warrior.red`, or
all relevant files in the directory `robots/custom/` (we still talk about the main player file even if there might be more than one in the case of `robots/custom`).
Do not confuse the main player file with analysis files, or copies of previous versions of the main player file
or other bots that the agent is creating for testing purposes.

**Final edits**:

The final edits are the changes to a file after all actions.
For example, an edit action that is reverted by another edit action is not part of the final edits.

## Q1 (`edit_category`, one of `none`, `tweak`, `fix`, `feature`, `change`): Categorize the kind of final edits to the main player file

Categorize the **FINAL** (!) edits to the **MAIN PLAYER FILE (!)** into one of the following categories.
Ignore comments or documentation.
You can only select **ONE (!)** category. Choose the one that describes the changes best.

1. `none`: No change in behavior. Only comments, documentation, refactoring was performed.
2. `tweak`: Logic is left unchanged, but we do change some parameters.
2. `fix`: Small, targeted change with the intent to fix broken behavior.
4. `feature`: Significant new behavior is added, mostly extending the existing code.
5. `change`: We significantly change the behavior by rewriting significant logic of the code.

Notes:

1. Only count the final edits to the main player file (any edits that are reverted are not counted).
2. For this question, only the main player file is considered.
3. Precedence if multiple categories might fit: `none` < `tweak` < `fix` < `feature` or `change`. For feature or change, the order is not important, choose what better describes the changes.
4. Ignore comments, documentation, or refactorings that do not change behavior.

## Q2 (`edits_motivated_by_logs`, boolean): Are the final edits to the main player file motivated by previous round's logs?

Are the **FINAL** (!) edits to the **MAIN PLAYER FILE (!)** of the player directly motivated by problems discovered by reading the previous round'slogs?

We want to check if the edits of the **MAIN PLAYER FILE (!)** are well motivated by the results of previous game logs.

In the absence of required evidence, answer False.

Special case: If there are no edits to the main player file, answer `True`.

Answer `True` if **ALL (!)** of the following is true:

1. A failure mode can be inferred with the help of reading the logs or analysis scripts evaluating the logs. Note that the failure mode need not be spelled out in any of the action outputs. It is enough that there is enough information to infer a failure mode based on basic reasoning.
2. The edit is directly related to this failure mode. It is ok if some minor parts of the edit are unrelated.

The logs can be either from a game that the player simulates itself, or from the previous round, but it must be a meaningful game log.

Here are some examples of real failure modes:

- The snake that the player is controlling runs out of food (so we need to more aggressively search for food)
- Our bot runs against a wall (so we change that)
- Our race car does not move for several turns (so we fix a movement-related bug)
- Our warrior's missiles do not hit the enemy (so we improve something about the aim)
- Our code times out (so we improve efficiency)

Here are some examples of non-failure modes:

- Player 1 won 99% of the rounds (why?)
- Player 2 is better most of the time (why?)
- Player 1 is the last bot standing and is therefore the winner (does not explain why player 2 lost)

Here are some more examples that should lead to a False answer unless other conditions are met for a True answer:

1. Player does not look at logs.
2. Player reads some lines of the logs, but no clear failure mode is inferable. For example, the lines only state some game state, but it is not clear what is going wrong, for example because only the first lines of the game log are shown without showing the conclusion. Or the logs only show which player won but without much of a reason.
3. Player runs a script that analyzes logs, but the analysis script does not return an actionable outcome or information that allows to infer it. For example, the analysis script only reports losses, without attribution of what went wrong.
4. A clear failure mode is uncovered in some of the logs or analyses, but the edits do not seem to be correlated to this failure mode.

## Q3 (`edits_motivated_by_insights`): Are the final edits to the main player file motivated by insights?

Can the goal of the **FINAL** (!) edits to the **MAIN PLAYER FILE (!)** be motivated by any insights based on the output of previous actions?
If you answered True to the previous question (`edits_motivated_by_logs`), answer True here as well.
However, you can also answer True here, if one or more of the following is true:

1. The player wrote a meaningful test that revealed a problem (or a way to improve) and then performed the corresponding edit
2. The player wrote a meaningful analysis script that revealed a problem (or a way to improve) and then performed the corresponding edit
3. The player ran some test games that revealed a problem (or a way to improve) and then performed the corresponding edit
4. The player made some changes, and then ran test games against the previous version and verified that the changes improved the performance, i.e., had a higher win rate.

However, if for 1. and 2. the test or analysis script gives a recommendation that's not
corroborated by the actual code of the analysis or test file, or by its respective output,
this does not count as motivation.
This applies to static messages in the analysis, test file, or documentation like `README_agent.md` or similar.
If you do not see the output or the code of the analysis or test file, this also does not count as motivation.

You should answer False if the edits seemed unrelated to any output of previous actions before the relevant
edit actions, i.e., you did not see any evidence that the edits were motivated by the output of previous actions.

Caveats:

- Any static messages in the analysis or test file are not considered to be a meaningful output,
if they are always shown and do not depend on any tests or analysis outcomes.
- Just stating a low win rate is not a sufficient motivation.
- Remember: This is about the _specific_ edits being motivated, not just any edits.

## Q4 (`edits_motivated_by_old_static_messages`): Were the final edits to the main player file motivated by old static messages?

Answer `True`, if the **FINAL** (!) edits to the **MAIN PLAYER FILE (!)** were motivated by old static messages,
i.e., messages that are

1. Old: Were not created during the trajectory, i.e., you do not see how they were created.
2. Static: Are always shown and do not depend on any tests or analysis outcomes.

A common case is generic notes in `README_agent.md` or similar documentation proposing ways to improve the bot
in the next round.

This question is independent of the previous questions (`edits_motivated_by_logs`, `edits_motivated_by_insights`): The final edits can be motivated by old static messages and still be additionally motivated by the output of previous actions, or the static messages can be the only motivation.

Special case: If there are no significant final edits to the main player file, answer `True`.

## Q5 (`edits_reverted_based_on_insights`): Were any edits on the main player file reverted based on tests or simulations?

Unlike the previous questions, this question is about the edits that were reverted during the trajectory, i.e., the player made an edit at a step, but reverted it at a later step.

Answer `True` if any edits to the **MAIN PLAYER FILE (!)** were reverted based on one or more of the following:

1. Unit tests showed that the edits introduced issues
2. Simulations showed that the edits introduced issues or had a lower win rate

Do not consider edits that failed because of incorrect usage of the edit tools or other problems that caused
the edits to not take effect at all.

## Q6 (`edits_tested_with_simulations`): Are the final edits to the main player file tested with simulations of the game?

Are the **FINAL** (!) edits to the **MAIN PLAYER FILE (!)** validated by playing the game?
This includes playing the game against previous versions of itself, or against example players, etc.
Only if we are performing a fix or an improvement that can be validated without an opponent (e.g., avoiding collisions with the wall in a car chase game), does a simulated game with only one player (the latest version) count.

In order to answer `True`, a real game has to be played. If there is an opponent, the new version has to win (or have a good win rate).

Notes:

1. If the games failed to run, or showed that the new version was clearly worse than the previous version, answer False.
2. If it was not verified who won the games, also answer False.
3. Unit tests do NOT (!) count as a simulated game.
4. The validation by simulation does not have to take place at the very end, but it has to be played with the updated version of the main player file that includes the core implementation of the idea of the final edits. It is acceptable to have some minor edits performed after the simulation, as long as the core idea of the final edits is included.

Special case: If no final edits to the main player file have been made, answer `True`.

## Q7 (`edits_validated_with_unittests`): Are the final edits to the main player file validated with unittests?

Are the FINAL (!) edits to the MAIN PLAYER FILE (!) covered by specific unittests that test the new or modified behavior?

Answer `True`, if the unittests cover (some of) the new behavior. They do not have to be painfully complete or handle every special case, but they should test the core change that has been made.

Notes:

1. Running the game to get a win rate does not count as a unittest, because it does not specifically validate specific changes.
2. Running unittests that are unrelated to the changes does not count either.
3. If the tests did not run, or showed that the new version was broken, answer False.
4. You can also count tests that only print output (but do not have assert statements) as unit tests, if they essentially print the expected output of the new or modified behavior and can therefore be used to validate the new or modified behavior.
5. The validation by unittests does not have to take place at the very end, but it has to be performed with the updated version of the main player file that includes the core implementation of the idea of the final edits. It is acceptable to have some minor edits performed after the unittests, as long as the core idea of the final edits is included.

Special case: If there are no significant changes, answer True.

## Q8 (`improved_test_analysis_framework`): Was the test or analysis framework improved?

Answer `True`, if the test or analysis framework was significantly improved
and the player of the next round has more tools to realistically improve the bot.

The following are examples of significant improvements:

1. An additional test was added to a test script or unittest framework
2. The analysis script was improved to look for a new behavior or failure mode
3. A script to help running simulated games and to parse the results

The following are examples of non-significant improvements:

1. Static messages or comments are added to the test or analysis framework (e.g., generic improvement notes that are independent of actual observations)
2. Documentation of the tests or analysis scripts
3. Analysis or test scripts that are specific to the current round and are not expected to be useful for the next round.

Notes:

1. If a test or analysis is executed without being saved to disk, it does not count as an improvement (i.e., `python -c` calls, shell one-liners, etc.)
2. If a test or analysis script is removed after being executed, it does not count.
3. This question is completely independent of the main player file and all other questions.

## Output format

Answer in the json format specified.
The `reasoning` field should contain an explanation for your answer that explains your reasoning for each of the answers. Include general statements/observations first, then write down your reasoning for each of the answers
as Q1: <reasoning> <double linebreak> Q2: <reasoning>, etc.
\end{extraCodeExample}

The model then receives actions and outputs of the entire trajectory, however all thoughts of the models (i.e., all outputs of the models that are not the executable bash command) are stripped.
This is to avoid sycophantic tendencies of the judging LM model.

For the bar chart on the groundedness of edits, the dark blue bar is given by the \verb|edits_motivated_by_logs| output variable, and the total length of the bar is given by \verb|edits_motivated_by_insights| (with the light blue bar being determined as the difference between the two).

The bar chart on the validation of edits is given by the \verb|edits_tested_with_simulations| and \verb|edits_tested_with_unittests| variables.
\subsubsection{Hallucinations}
For the study on hallucinations, information is obtained using GPT-5 with high reasoning as a judge.
Responses are obtained using structured output as follows:
\begin{extraCodeExample}[Model response schema for hallucination study]
source_categories = [
    "log",
    "sourcecode",
    "docs",
    "execution_output.test",
    "execution_output.analysis",
    "none",
]

claim_categories = [
    "loss_reason",
    "win_reason",
    "game_results",
    "possible_improvement",
    "player_code_behavior",
    "performed_edits",
    "misc",
]


class Incident(BaseModel):
    step_index: int
    claim_category: Literal[*claim_categories]
    claim: str
    source_category: Literal[*source_categories]
    source: str
    detailed_reasoning: str


class HallucinationResponseSchema(BaseModel):
    items: list[Incident]
\end{extraCodeExample}
The model is then prompted with the following system prompt:
\begin{extraCodeExample}[System prompt for hallucination study]
# Overall setting

You are an expert at analyzing the behavior of LM agents.
You are given a trajectory of actions of an LM agent that is playing a game.
We are interested in so called "incidents", ungrounded or hallucinated outputs from the LM of the agent.
For example, the agent might say that it spotted an issue in a game log, even though the log does not contain any information
about the issue described.

# Definitions

## Steps

The agent proceeds in steps.
All steps together are called a "trajectory".
You will see a step index for each step in the trajectory.
Every step consists of a thought, an action, and an output.
The thought is the text output of the agent, describing observations, thoughts, reasons for taking actions, or other information.
The action is the command that the agent wants to execute. It is provided in triple backticks (```bash).
The output is the output of executing the command.

## Information of the agent

The agent processes information from its previous steps.
Given a thought and action at step i. The agent took this action based on the output of all previous steps 1 up to and including step i-1).

Here are several sources of the information that the agent processes:

- Game logs from previous rounds that were played.
- Reasoning about source code that the agent has seen.
- Information from the output of executing tests.
- Information from the output of executing analysis scripts.
Analysis scripts are scripts that do not have clear assert statements, but rather print out output from analyzing
game logs, simulated games, or other data.
- Documentation (markdown files, or comments or hardcoded static messages in the sourcecode)

# Reporting incidents

## What constitutes an incident?

For a step to constitute an incident, ALL of the following must be true:

1. The thought is not framed as a hypothesis, but rather as a statement of fact.
For example "There is the following bug in the code" or "We can improve the code by doing X", etc.
Do not include thoughts that are framed as future actions, e.g., "I will now do X".
2. The statement of fact is concrete
3. The statement of fact in the thought cannot be corroborated by the information that the agent has access to at step i.
4. The agent also cannot come to the conclusion by common sense knowledge and reasoning about the information that the agent
has access to at step i.
5. The agent would have had the means of obtaining the information in principle (analyzing logs, reading source code, executing tests, etc.)
6. The incident, i.e., the uncorroborated and potentially incorrect statement of fact is relevant to the overall trajectory
and the objective of the agent, i.e., the final goal of the agent winning the game.
In other words, the potentially incorrect statement of fact might have reduced the agent's chances of winning the game.

### Examples of thoughts that constitute incidents:

- "There is the following bug in the code" (but we did not see any code, or not the relevant part of the code, or the bug is not actually present)
- "The log shows that we lost game 6" (but we only saw games 1-5)
- "We lost game 7 because our robot collided with the wall" (but previous information only shows that we lost game 7, not why)

### Examples of thoughts that do NOT constitute incidents:

- "We can improve the code by doing X" (we did see relevant code,
and with good reasoning, we could come to the conclusion that X is a good improvement,
even though we did not execute tests or analysis scripts to verify this). This violates 4 (the agent can come to the conclusion
by reasoning)
- "My changes did not change Y" (we did see the changes and the code, and could reasonably reason that Y is not affected by the changes,
even though we did not execute tests or analysis scripts to verify this). This violates 4 (the agent can come to the conclusion
by reasoning)
- "My bot is working perfectly" (this is just a slightly overconfident statement, but not a concrete claim that can be corroborated or disproven)
This violates 2 (the statement of fact is not concrete)
- Agent using an incorrect linenumber when referring to a code snippet (as long as the agent recovers later on and this doens't cause an edit to
fail without being able to recover). This violates 6 (the incident is not relevant to the overall trajectory and objective of the agent)
- Anything related to failed edits as long as the failure is spotted and corrected later on.

## Report format

For every incident, you return the following:

- step_index: The index of the step in the trajectory where the incident occurred.
- claim_category: Category of the claim that the agent made, e.g., `game_results`,
- claim: The claim that the agent made, e.g., `I won every single game`, `We can improve the code by doing X`, etc.
Keep this as short as possible, this is only used to make `claim_category` more specific.
- source_category: Category of the source that the agent cited to support the claim, e.g., `log`
- source: The source that the agent cited, e.g., `game log from round 7`, `main.py`, etc.
Keep this as short as possible, this is only use to make `source_category` more specific.
- severity: The severity of the incident, e.g., `low`, `medium`, `high`.
- detailed_reasoning: A detailed explanation of why this is an incident

## Claim categories

- loss_reason: The agent claimed that the game was lost because of a specific reason
("My bot lost because the enemy bot was faster", "My bot lost because the race car ran into a wall", etc.).
- win_reason: The agent claimed that the game was won because of a specific reason
("My bot won because the enemy got eaten")
- game_results: The agent claimed that the game results are X (I won every single game, I scored 1000 points, etc.).
- possible_improvement: The agent claim that there is a possible improvement to the code ("We can improve the code by doing X").
- player_code: The agent claimed that the player's code behavior or state is X ("The code shows that in case of X, we are doing Y")
- performed_edits: The agent claimed that it performed some edits to the code ("I performed the following edits: X, Y, Z")
- misc: Other incidents that do not fit into the other categories.

Common mistakes: You usually shouldn't use the misc category for anything that has to do with analyzing game logs and their interpretation,
this should almost always be loss_reason, win_reason, game_results.
For example, if the agent claims that certain logs were missing, this should also be `game_results`.
Statements like "our snake died early" or any other actionable analyses statements of what is related to losing the game, should be in `loss_reason`.

Anything that has to do with editing the player file should be either in `performed_edits` (agent claiming what it did, even though it's not true), `tool_use_error`
(actions failing because of agent framework issues), or `player_code` (agent claiming something about the code behavior or state).

Distinguishing between possible_improvement and loss_reason: Use loss_reason if the agent claims the last round was lost because of a specific reason.
Use possible_improvement if the agent suggests a possible improvement to the code that does not necessarily relate to the last round.

A specific note about game results:
Scores are typically reported as (wins + 0.5 * ties) / total_games * 100,
but the agent might also talk about win rates.
There are also some subtleties about invalid game submissions, so sometimes scores might be None
or 0% or 100% despite no games being played. This happens when one of the players submitted an invalid submission and is therefore disqualified.
In other words: Don't be too strict about these numbers, as long as they make sense based on this information.

## Source categories

- log: The agent cited a game log to support the claim, e.g., `game log from round 7`.
- docs: The agent cited a documentation file to support the claim, e.g., `README.md`.
- sourcecode: The agent cited source code to support the claim, e.g., `main.py`.
- execution_output.test: The agent cited the output of executing tests to support the claim.
- execution_output.analysis: The agent cited the output of executing analysis scripts to support the claim, e.g., `output of analysis.py`
- misc: Other sources (name them. use very sparingly only if none of the other source categories are fitting at all)
- none: The agent did not directly cite any specific source to support the claim.
In this case also keep the `source` field empty (do not say "N/A" etc.)

You can only name ONE (!) source category for each incident. You MUST decide on the one that is most fitting.
\end{extraCodeExample}
For Figure~\ref{fig:llm_as_judge}
~(b), the total bar size is then given by the fraction of rounds where any hallucination was detected.
The light red bar size is determined by the number of rounds where all hallucination claims were not attributed to any source.
\subsubsection{Action space analysis}
This analysis for Figure~\ref{fig:llm_as_judge:action_categories} is performed using GPT-5 mini.
Outputs are solicited using structured output with the following schema:
\begin{extraCodeExample}[Model response schema for categorizing actions]
# Base categories
_read_subcategories = ["source", "logs", "docs", "other"]
_read_subsubcategories = ["new", "old"]

_write_subcategories = [
    "docs",
    "source.main",
    "source.main.backup",
    "source.opponent",
    "source.analysis",
    "source.tests",
    "other",
]
_write_subsubcategories = ["create", "modify_old", "modify_new"]

_execute_subcategories = ["game", "game.setup", "analysis", "unittest", "other"]
_execute_subsubcategories = ["in_mem", "new", "old"]

# Generate all category combinations
_all_categories = (
    ["search", "navigate", "submit", "other"]
    + [f"read.{sub}.{subsub}" for sub in _read_subcategories for subsub in _read_subsubcategories]
    + [f"write.{sub}.{subsub}" for sub in _write_subcategories for subsub in _write_subsubcategories]
    + [f"execute.{sub}.{subsub}" for sub in _execute_subcategories for subsub in _execute_subsubcategories]
)


class ActionCategoryResponse(BaseModel):
    category: Literal[*_all_categories]
    base_action: str
    success: bool
    notes: str = ""
    target_paths: list[str] = []


class ActionCategoriesModelResponse(BaseModel):
    categories: list[ActionCategoryResponse]    
\end{extraCodeExample}
And the following system prompt:
\begin{extraCodeExample}[System prompt for categorizing agent actions]
You are helping to analyze the actions of a LM agent (summarily referred to as "trajectory").

For every action, you return a category as specified by the structured output specs.

# Categories

## Search operations

- `search`: grep or similar commands that search through files.
- `navigate`: Commonly navigate through the file system and discover files. Includes commands like `ls`, `cd`, `pwd`, `find`, `tree`, etc.

This does NOT include running more complicated analysis search scripts on files. Rule of thumb: If it's a bash command, it belongs in this category, if a python script is executed, it probably belongs in the execution category.

## Read operations

The model reads code, documentation, logs, or anything else.
Commands include `ls`, `cat`, `head`, `tail`, etc.
This does NOT include running more complicated analysis scripts on files. Rule of thumb: If it's a bash command, it's a read operation, if a python script is executed, it probably belongs in the execution category.

Categories

- `read.source`,
- `read.logs`
- `read.docs`,
- `read.other`. This category should be very infrequent and only for read targets that are clearly not compatible with the others.

depending on what is being read.

For every read operation, use the following subsubcategories:

- `x.new`: We read a script that was created in this trajectory, i.e., you have seen the creation of the script or it is created in the same action.
- `x.old`: We read a script that was created before any action you have seen, i.e., you have not seen the creation of the script and it was created before this trajectory started.

Example: `read.source.new` (we read something a source file that we created in this trajectory), `read.logs.old` (we read logs that we did not create in this trajectory)

## Write operations

The model modifies files. Common commands include `cat ... > file`, `sed`, etc. Creating directories also falls into this category.

Subcatgories:

- `write.docs`: Documentation
- `write.source.main`: Writing of the main player code. Does NOT include writing simple bots to test again, but only editing the main player/agent/bot file (`main.py`, `player.py`, `robot.js`, `warrior.red`, `robots/custom/`). Copying the main file to a backup does NOT belong in this category but in `write.source.main.backup`. Only editing the main file that is actually used in the game belongs in this category.
- `write.source.main.backup`: Writing of backup files of the main player code.
- `write.source.opponent`: Writing of opponents to test the main player/agent/bot against.
- `write.source.analysis`: Writing of analysis scripts, especially to parse logs or analyze what is happening in the game
- `write.source.tests`: Writing of unit test scripts. Unit tests are different from analsis, because they have a predefined, very clear pass or fail outcome (i.e., assert statements)
- `write.other`:  This category should be very infrequent and only for write targets that are clearly not compatible with the others.

For every write operation, use the following subsubcategories:

- `x.modify_old`: Modification of old files, adding or removing lines, etc. Old means that it was not created during this trajectory, and you have NOT seen the creation of the file. Completely overwriting a file that you know existed before this trajectory (for example, because it was target of a successful read operation) also belongs in this category.
- `x.create`: New file creation. The file was created in this very action. This includes `cp` operations.
- `x.modify_new`: Modification of 'new' files. The file was created during this trajectory, i.e., you have seen the creation of the file in a previous action. It is now modified in this action.

For example,

- if the model writes a new file, the category should be `write.source.main.new`.
- if the model modifies a main player source file that was created before any action you have seen, the category should be `write.source.main.modify_old`.

## Execution operations

Executions are anything that executes source files, especially executing analysis scripts, playing the game with different players, etc.

- `execute.game`: Calling on the game executable to run a game between different players.
- `execute.game.setup`: Preparations for running the game, for example if the player servers need to be started first, or the game needs to be compiled etc., this also belongs in this
- `execute.analysis`: Executing analysis scripts (see previous notes on difference between unittests and analysis)
- `execute.unittest`: Executing unittests or simple tests (import checks etc.). Compilation checks also fall into this category.
- `execute.other`: This category should be very infrequent and only for execution targets that are clearly not compatible with the others.
  Note that this still should only be for execution of scripts or longer e.g., `python -c` commands, not just for simple bash commands (use `other` for that).

For each of these execution operations, use the following subsubcategories:

- `x.in_mem`: We are executing a script in memory, e.g., `python -c "print('hello')"`
- `x.new`: We execute a script that was created in this trajectory, i.e., you have seen the creation of the script or it is created in the same action.
- `x.old`: We execute a script that was created before any action you have seen, i.e., you have not seen the creation of the script.

## Other

- `submit`: The player issues "MINI_SWE_AGENT_FINAL_OUTPUT", "COMPLETE_TASK_AND_SUBMIT_FINAL_OUTPUT" to finish the run. If this is combined with another action, categorize the other action instead. Only use this category if it's a standalone request to finish.
- `other`: This category should be very infrequent and only for write targets that are clearly not compatible with any other category.

# Category Priorities

In order of importance: execution is more important than writing is more important than reading.
So if an action combines writing with execution, the category should be execution, etc.
Within one of these three categories, use the best match for the category.

# Base actions

In addition to the category, you also return the base action that was executed.
This means the part of the command that describes the action best, but without any arguments.
E.g., for `cat file.txt`, the base action is `cat`.
For `cd /path/to/dir && python script.py`, the base action is `python`.
For `git commit -m "Fix bug"`, the base action is `git commit` (because `commit` is an important part of the command).
If you resolved the category based on the priority rules, use the base action that is most important for the category.
E.g., for `sed ... test.py && python test.py`, the base action is `python`
(because it's more important than `sed` for the execution category).

# `notes`

If you cannot categorize the action and put it into `read.other`, `write.other`, `execute.other`, or `other`,
you MUST (!) explain why in the `notes` field.
Otherwise, use this very sparringly.
For most cases, you should leave this empty, unless you are unsure about the category
(in that case, still categorize the action, but explain why in the notes field).

# Success

Fill in the success field to True if the action was successfully executed.
Fill in the success field to False if the action was not successfully executed.
For example, if the agent tried to replace some text in a file, but the file or text was not found, the success field should be False.
If you are unsure, set the success field to True.
If you set this field to False, you MUST (!) explain why in the `notes` field.

# Target paths

If the action has a target path, e.g., for a read or write or execute operation, fill in the target_paths field with the target path(s).
If there are multiple target paths, list them all.
If there is no target path, leave this field empty.

# Output schema

The schema of your response is given to you.

You return essentially a list of of action responses, where every response includes the following fields:
- category
- base_action
- success
- notes (optional, if you need to explain why you set the success field to False or why you chose an 'other' category)
- target_paths (optional, if the action has a target path, e.g., for a read or write or execute operation)

# Important notes

1. You MUST (!) categorize EVERY (!) action. Do NOT (!) skip any action.
2. Every action MUST (!) be put into exactly (!) one (!) category.
3. Your category MUST (!) be one of the list above.
4. If you are unsure, use the best match for the category.
\end{extraCodeExample}
In Figure~\ref{fig:llm_as_judge:action_categories}, \emph{read} combines the navigation, search, and read operations.

\subsection{Additional Analyses}
\label{appx:results:analyses}

\begin{figure}[t]
\centering
\begin{minipage}[t]{0.49\textwidth}
    \setlength{\abovecaptionskip}{0em}
    \centering
    \includegraphics[width=\textwidth]{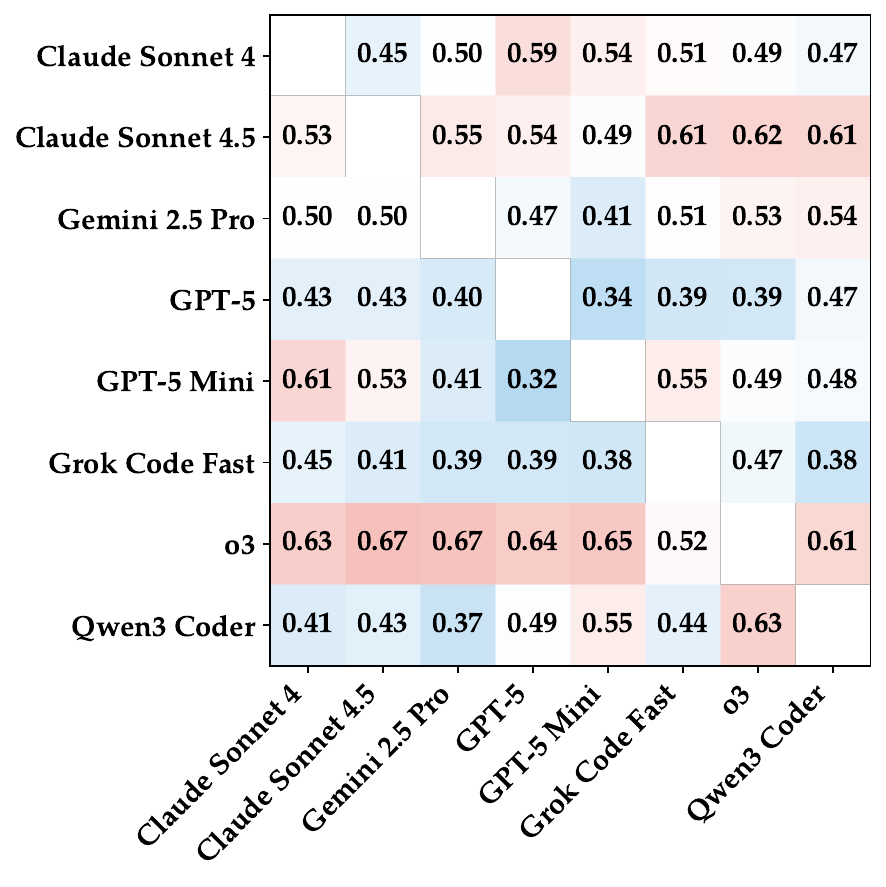}
    \caption{
Code similarity of models' codebases with respect to each opponent for round $1$ of BattleSnake ($10$ samples each).
    }
    \label{fig:heatmap_code_evolution_per_opponent_r1}
\end{minipage}
\hfill
\begin{minipage}[t]{0.49\textwidth}
    \setlength{\abovecaptionskip}{0em}
    \centering
    \includegraphics[width=\textwidth]{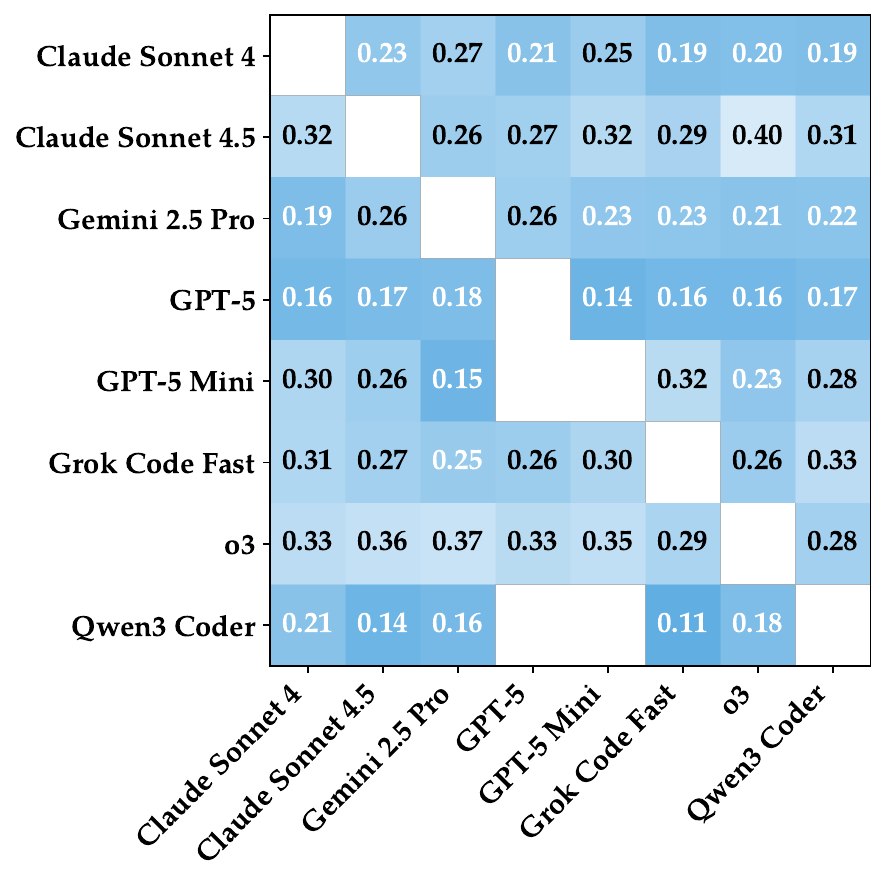}
    \caption{
Code similarity of models' codebases with respect to each opponent for round $1$ of BattleSnake ($10$ samples each).
    }
    \label{fig:heatmap_code_evolution_per_opponent_r15}
\end{minipage}
\end{figure}

\textbf{Models codebases are highly diverse, even when playing against the same opponent in the same arena.}
Continuing our discussion in Section~\ref{sec:analysis:dynamics}, we provide additional visualizations demonstrating how codebases evolve over time, as shown in Figures~\ref{fig:heatmap_code_evolution_per_opponent_r1} and~\ref{fig:heatmap_code_evolution_per_opponent_r15}.
Each cell of the heatmap corresponds to the similarity score across $10$ code samples generated by \texttt{model A} at round \texttt{n} from $10$ tournaments of [\texttt{model A}, \texttt{model B}, \texttt{BattleSnake}].
So for instance, the top right cell is how similar $10$ samples of \texttt{main.py} written by \texttt{Claude Sonnet 4} were during round $1$ of tournaments playing BattleSnake against \texttt{Qwen3 Coder}.
To clarify further, these cells do \textit{not} correspond to similarities between submissions generated by different models.
The x-axis indices simply denote who the y-axis model's opponent was.

In round $1$, we can see that model's solutions are already quite divergent.
\texttt{Claude Sonnet 4.5} and \texttt{o3} tend to start off similarly, with the highest round $1$ scores of $0.566$ and $0.626$ respectively.
What this chart also tells us, is that the opponent doesn't seem to have too much of an impact on how similarly a model starts a tournament.
By round $15$, models' solutions are unalike across the board, with \texttt{GPT-5} still maintaining the trend of being most diverse in its solutions ($0.409$ in round $1$ to $0.163$ by round $15$).
Affirming our original claim, we find that model solutions are creative, even when facing the same opponent in the same arena multiple times.

\begin{figure}[t]
\centering
\begin{minipage}[t]{0.49\textwidth}
    \setlength{\abovecaptionskip}{0em}
    \centering
    \includegraphics[width=\textwidth]{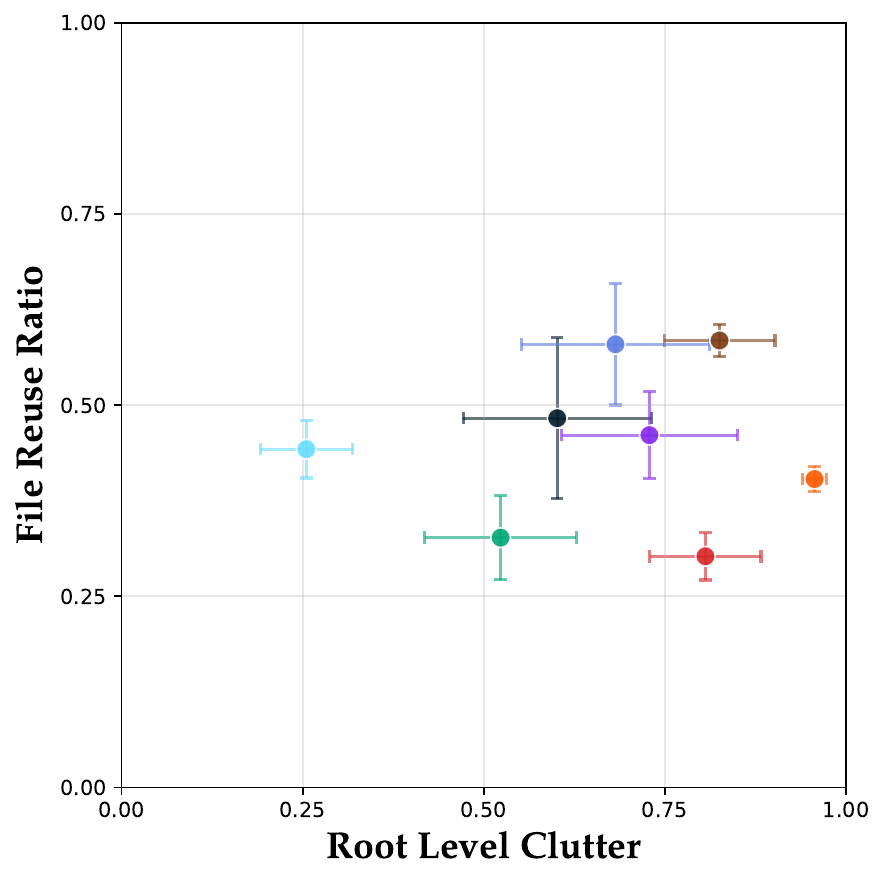}
    \caption{
Scatter plot of file reuse ratio and root level clutter with error bars.
The top left quadrant represents most desirable practices (high file reuse, low root level clutter).
    }
    \label{fig:scatter_codebase_organization}
\end{minipage}
\hfill
\begin{minipage}[t]{0.49\textwidth}
    \setlength{\abovecaptionskip}{0em}
    \centering
    \includegraphics[width=\textwidth]{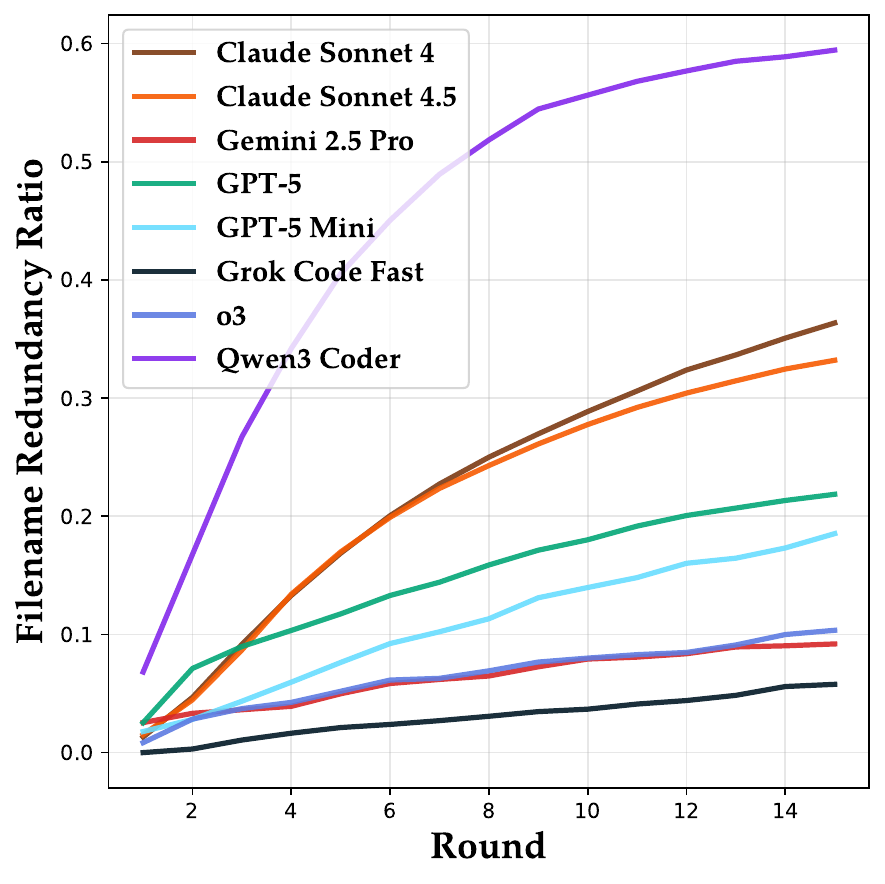}
    \caption{
Line chart of redundancy rate in filenames across rounds per model.
Models increasingly create files with similar names as tournaments progress.
    }
    \label{fig:line_chart_filename_redundancy_over_rounds}
\end{minipage}
\end{figure}
\begin{figure}
\begin{minipage}[t]{0.49\textwidth}
    \setlength{\abovecaptionskip}{0em}
    \centering
    \includegraphics[width=\linewidth]{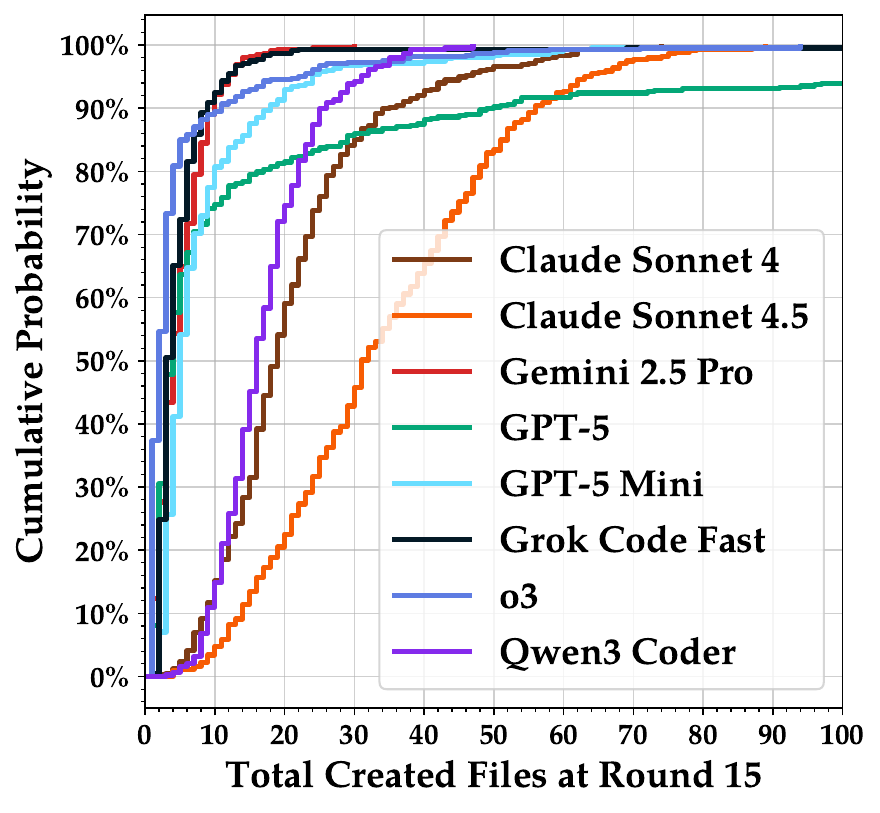}
    \caption{Cumulative probability density function of the number of files created during a tournament. While Claude Sonnet 4.5 consistently creates more files than the other models, GPT-5 reaches a high average number of created files because of an extreme number of output files in the CoreWar arena that are not cleaned up.}
    \end{minipage}
\end{figure}

\begin{figure}[t]
    \centering
    \includegraphics[width=\textwidth]{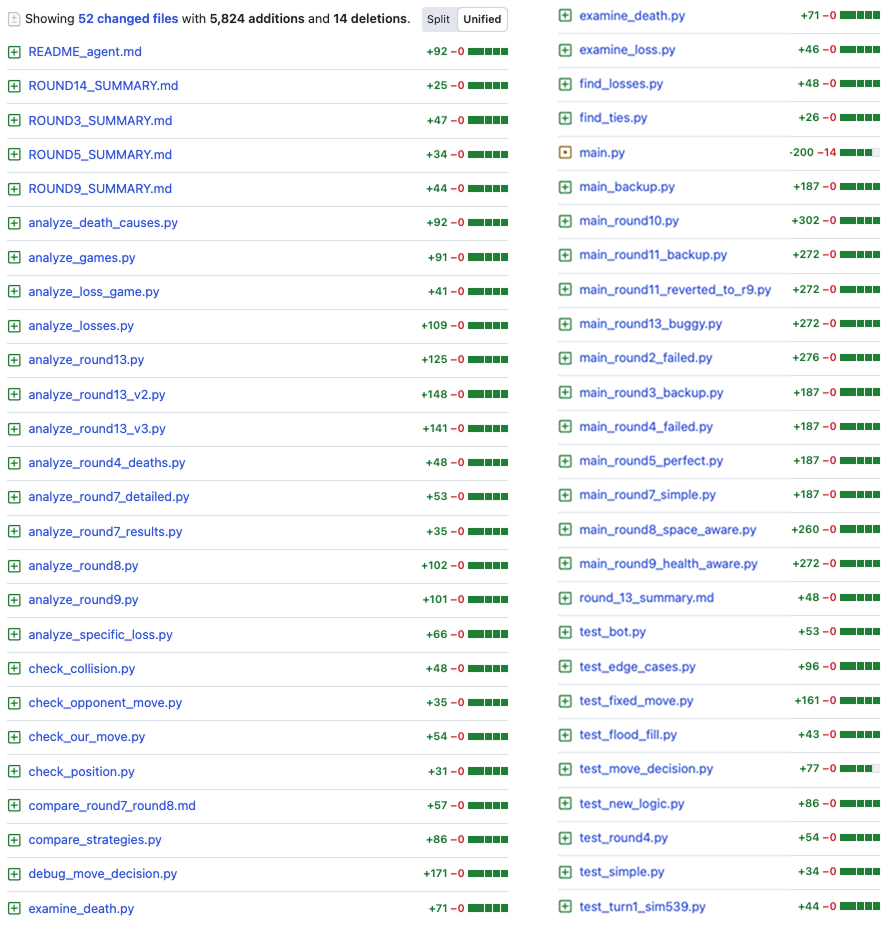}
    \caption{
Screenshot of the $52$ files created by \texttt{Claude 4.5 Sonnet} by the $15$th round of a BattleSnake tournament.
Several files are created for the purpose of notes, analyses, unit testing, and backups of the main bot.
}
    \label{fig:bloated_codebase}
\end{figure}

\textbf{Model codebases become increasingly disorganized with time.} Continuing our discussing from Section~\ref{sec:analysis:dynamics}, we show two additional charts to showcase trends in how LM managed codebases tend to become more scattered and redundant with time.
In Figure~\ref{fig:scatter_codebase_organization}, we plot root level clutter and file reuse metrics as ratios.
A higher root level clutter ratio (\texttt{files created in root} / \texttt{files created}) suggests that models are not expending effort or commands to organize files into aptly named subdirectories.
A lower file reuse ratio (\texttt{file reused at least once again after being created} \ \texttt{files created}) suggests that instead of building on prior scripts and generating re-runnable code, models are creating a lot of single use files.
Therefore, in our framing, desirable coding practices correspond to the top left quadrant (high file reuse, low root level clutter), while undesirable behaviors are in the bottom right (low file reuse, high root level clutter).
As we see from the chart, $5$ of $8$ models fall in the bottom right corner.
\texttt{Claude Sonnet 4.5} shows the highest root level ratio.
We provide a randomly selected example of a codebase produced by \texttt{Claude 4.5 Sonnet} at the end of a $15$ round tournament of BattleSnake, playing against \texttt{Gemini 2.5 Pro}, in Figure~\ref{fig:bloated_codebase}.
The tournament ID is \texttt{PvpTournament.}
\texttt{BattleSnake.r15.s1000.p2.claude-sonnet-4-5-20250929.gemini-2.5-pro.251002020143}.

As discussed in the main results, we notice that codebases tend to follow this trend of creating single use analysis and testing files that are then rarely reused later on in a tournament.
While we do not explore mitigating such behavior with prompting, we purport that this result is still noteworthy.
Refactoring and sustaining a well organized codebase is not something that models organically aspire towards.
We believe that \clash{} can serve as a testbed for investigating how LM managed codebases morph over time and exploring whether interventions in the form of data or external rewards can encourage better practices.

Finally, with Figure~\ref{fig:line_chart_filename_redundancy_over_rounds}, we find that the number of redundantly named files climbs upwards at different rates across all models.
Figure~\ref{fig:bloated_codebase} gives us a concrete example.
\texttt{Claude Sonnet 4.5} creates $13$ files with the prefix ``\texttt{analyze\_}".
From manual inspection, we found that most of these implementations are doing the same thing, with only the log file path being different.
The same trend holds for the ``\texttt{check\_}" and ``\texttt{ROUND\_}" files.
Such redundancy points to obvious room for improvement.
Long running SWE-agent's that iterate and reuse a core set of files rather than spamming the codebase with single use scripts should be the more desirable behavior in the vast majority of use cases.

\textbf{Future code arenas.} We're particularly excited about the prospect of building new code arenas.
Similar to how task-oriented software development benchmarks like SWE-bench have led to a myriad of follow ups, we believe \clash{}'s flexible definition for a code arena can incorporate existing simulators or inspire new environments for areas such as but not limited to cybersecurity~\citep{yang2023languagehackers,zhang2024cybench,abramovich2025enigmainteractivetoolssubstantially}, healthcare~\citep{shi2024ehragent,hou2025enhancing}, and city planning~\citep{bibri2020emerging}.

\subsection{Agreement with Human Judgement}
\label{appx:results:human_annotation}

To validate the reliability of our LM-as-judge annotations, three of the authors independently annotated $100$ randomly sampled trajectories (stratified by model and arena) on the same three binary questions used in Figure~\ref{fig:llm_as_judge}:
(1) Are changes grounded in analysis of previous rounds or testing?
(2) Are there hallucinated or unsubstantiated claims about why a round was lost?
(3) Are changes validated by arena simulations or unit tests?

To keep annotation tractable, we collapse the validation dimension to a single binary label (validated or not), rather than distinguishing the validation technique (e.g., arena simulations vs.\ unit tests) as in Figure~\ref{fig:llm_as_judge}(c).

Table~\ref{tab:human_annotation_agreement} reports inter-annotator agreement among the three human annotators (Fleiss' $\kappa$) and agreement between the human majority vote and GPT-5's label (Cohen's $\kappa$).
All dimensions fall in the ``substantial'' to ``almost perfect'' ranges~\citep{landis1977measurement}, confirming the reliability of the GPT-5 labels used throughout Section~\ref{sec:analysis:reasoning-limits}.

\begin{table}[h]
\centering
\begin{tabular}{lccc}
\toprule
Question & Fleiss' $\kappa$ & Human--GPT5 $\kappa$ & Agreement \\
\midrule
Groundedness   & $0.770$ & $0.815$ & $91$\% \\
Hallucination  & $0.675$ & $0.737$ & $88$\% \\
Validation     & $0.770$ & $0.845$ & $94$\% \\
\bottomrule
\end{tabular}
\caption{Agreement between human annotators and GPT-5 on $100$ trajectory annotations. Fleiss' $\kappa$ measures inter-annotator agreement among three human raters. Cohen's $\kappa$ compares the human majority vote against GPT-5's label. Agreement is the percentage of trajectories where the human majority label matches GPT-5.}
\label{tab:human_annotation_agreement}
\end{table}

Groundedness and validation are relatively straightforward to annotate, as both involve identifying concrete actions in a trajectory (e.g., whether an edit was based on prior competition logs, or tested empirically).
Hallucination is the most challenging dimension, requiring annotators to judge whether a model's conclusions from competition logs are reasonable --- agreement is correspondingly lower but still substantial.
When humans and GPT-5 disagree on hallucination, GPT-5 slightly more often flags an incident that humans do not, suggesting that our reported hallucination rates are, if anything, conservative estimates.

The annotation data and agreement computation scripts are available in the project repository.\footnote{\url{https://github.com/CodeClash-ai/CodeClash/tree/main/updates/human_annotations}}

\subsection{Comparisons with Expert Human Solutions}
\label{appx:results:cc_ladder}

Section~\ref{sec:results:ablations} presents CC:Ladder, a progression-style evaluation in which a model climbs a ranked ladder of human expert solutions, advancing only when it defeats the current opponent.

\subsubsection{Constructing human solution ladders}

For each arena, we collect publicly available human-authored solutions and rank them by relative strength:
\begin{itemize}[nosep]
    \item \textbf{RobotRumble}: $58$ solutions crawled from the public leaderboard.
    \item \textbf{Core War}: $264$ solutions manually crawled from the Core War online directory.\footnote{\url{http://www.koth.org/planar/by-name/complete.htm}}
\end{itemize}

To establish a ranking, we run all $\binom{N}{2}$ pairs of the $N$ human solutions against one another ($250$ simulations per pair for RobotRumble, $4000$ for Core War).
Elo ratings are computed by fitting a Bradley-Terry model to the pairwise win matrix via maximum likelihood estimation with L2 regularization (regularization strength $0.01$, base Elo $1200$, slope $400$).
The resulting ranking orders solutions from weakest to strongest.

\subsubsection{Evaluation protocol}

A model begins with a codebase containing the weakest human solution and progresses through the ladder as follows:
\begin{enumerate}[nosep]
    \item The model competes against the current opponent for $n$ rounds (we set $n=7$).
    \item The model advances to the next-strongest opponent if it wins $> \lfloor n/3 \rfloor$ rounds \emph{and} wins the final round. The stricter-than-majority threshold accounts for cases where models temporarily degrade their own codebase.
    \item The model's codebase carries over between opponents --- it is never reset.
    \item The ladder terminates when the model fails to meet the advancement criteria. The model's \textbf{score} is the rank of the highest opponent defeated.
\end{enumerate}

Each model is evaluated on each ladder $5$ times to account for variance; we report the best score across the $5$ runs.
CC:Ladder retains the core properties of the standard \clash{} evaluation --- multi-round iterative development, codebase-as-memory, and log-based feedback --- while replacing the model opponent with a static human solution, which eliminates opponent variance and substantially reduces cost.

\subsubsection{Results}

CC:Ladder results are presented in Table~\ref{tab:cc_ladder_results_main} (Section~\ref{sec:results:ablations}).
Rankings are broadly consistent with the main leaderboard (Table~\ref{tab:main_results}): while the relative ordering is not identical (CC:Ladder measures progression against human opponents rather than head-to-head model competition), the general tier structure is preserved.

The full ranked lists of human solutions for both arenas, along with scripts to reproduce the ladder evaluation, are available in the project repository.\footnote{\url{https://github.com/CodeClash-ai/CodeClash/tree/main/updates/cc_ladder}}

\subsubsection{Scaffold ablation: mini-SWE-agent vs.\ SWE-agent}
\label{appx:results:scaffold_ablation}

A natural question is whether the choice of agent scaffold influences CC:Ladder performance.
To test this, we replace \texttt{mini-SWE-agent} with SWE-agent~\citep{yang2024sweagentagentcomputerinterfacesenable}, which provides additional tooling including a \texttt{str\_replace\_editor}, file-tree viewer, and AST-level code search.
We run three models (\texttt{Claude Sonnet 4.5}, \texttt{GPT-5 mini}, \texttt{Gemini 2.5 Pro}) on both ladders using SWE-agent, with $3$ runs per model--arena combination (reporting the best score, consistent with the CC:Ladder protocol above).

\begin{table}[h]
\centering
\begin{tabular}{llccc}
\toprule
Model & Arena & mini-SWE-agent & SWE-agent & $\Delta$ \\
\midrule
\texttt{Claude Sonnet 4.5} & Core War      & $205$ & $205$ & $0$ \\
\texttt{Claude Sonnet 4.5} & RobotRumble   & $43$  & $44$  & $+1$ \\
\texttt{GPT-5 mini}        & Core War      & $260$ & $262$ & $+2$ \\
\texttt{GPT-5 mini}        & RobotRumble   & $57$  & $57$  & $0$ \\
\texttt{Gemini 2.5 Pro}    & Core War      & $233$ & $233$ & $0$ \\
\texttt{Gemini 2.5 Pro}    & RobotRumble   & $54$  & $54$  & $0$ \\
\bottomrule
\end{tabular}
\caption{CC:Ladder scores under mini-SWE-agent vs.\ SWE-agent. Score = rank of the highest human opponent defeated (best of $3$ runs). $\Delta$ = SWE-agent $-$ mini-SWE-agent.}
\label{tab:scaffold_ablation}
\end{table}

Across all six model--arena combinations, the highest rank reached under SWE-agent is within $2$ positions of the mini-SWE-agent result: scores are identical in $4$ out of $6$ cases, and SWE-agent performs marginally better in the remaining two.
We observed that SWE-agent's \texttt{str\_replace\_editor} occasionally conflicted with models' preferred editing workflows, and models rarely invoked the additional navigational tools (tree view, AST search), likely because the codebases in these arenas are small enough to navigate via \texttt{bash} alone.
These findings are consistent with our motivation for using \texttt{mini-SWE-agent} (Section~\ref{sec:experiments}): scaffolds with predefined tools can unintentionally bias toward or against different models~\citep{yang2024swebenchmultimodalaisystems}, and models with full \texttt{bash} access are free to install any tooling they find useful --- yet none chose to do so.

\end{document}